\documentclass[nofootinbib, reprint, amsmath, pra, amssymb, ]{revtex4-2}

\usepackage[utf8]{inputenc}
\usepackage{amsmath}
\usepackage{graphicx}
\usepackage{hyperref}
\usepackage{xcolor}
\usepackage{gensymb}
\usepackage{lipsum}
\usepackage{upgreek} 

\usepackage{array}[=2016-10-06] 
\usepackage{tabularx}

\newcommand{\upd}{\mathrm{d}} 

\DeclareMathOperator{\sech}{sech}

\newcommand{\D}{\mathrm{d}} 

\begin{document}

\title{Solitary waves in a phononic integrated circuit }

\author{Timothy M.F. Hirsch$^{1,3}$}
\author{Xiaoya Jin$^{1}$}
\author{Nicolas P. Mauranyapin$^{1}$} 
\author{Nishta Arora$^{1}$}
\author{Erick Romero$^{1}$}
\author{Matthew Reeves$^{1,3}$}
\author{Glen I. Harris$^{1,3}$}
\author{Warwick P. Bowen$^{1,2}$}
\email{Corresponding author: w.bowen@uq.edu.au}
\author{Christopher G. Baker$^{1,3}$}
\affiliation{$^{1}$School of Mathematics and Physics, The University of Queensland, Brisbane, Australia}
\affiliation{$^{2}$ARC Centre of Excellence in Quantum Biotechnology,
     School of Mathematics and Physics, University of Queensland, Brisbane, Australia.}
\affiliation{$^{3}$Cortisonic Pty Ltd, Brisbane, Australia}

\date{\today}

\maketitle

{\bfseries 

Solitons are universal nonlinear excitations that appear in settings as varied as optics, water waves, and quantum gases~\cite{suhMicroresonatorSolitonDualcomb2016,braschPhotonicChipBased2016,zabuskyInteractionSolitonsCollisionless1965,remoissenetWavesCalledSolitons1999,streckerFormationPropagationMatterwave2002}. While reduced models of soliton dynamics are well established, their validity and dynamical behaviour in strongly nonlinear regimes with frequent interactions remain largely unexplored experimentally. Progress has been constrained by the difficulty of simultaneously achieving precise control of dispersion and nonlinearity, together with the temporal and spatial resolution required for dynamical observations. Here we overcome these difficulties by producing acoustic solitons in integrated phononic waveguides. We exploit the interplay between waveguide dispersion and mechanical Kerr nonlinearity to generate `dark' solitons that persist over metre-scale propagation distances. The slow phonon velocity allows direct imaging of hundreds of dark soliton collisions---two orders of magnitude more than have previously been accessible~\cite{foursaInvestigationBlackGraySoliton1996,wellerExperimentalObservationOscillating2008}---as well as soliton fission and the melting of a soliton Wigner crystal. Furthermore, the unprecedented dynamical resolution allows us to verify two long-predicted aspects of dark soliton behaviour: the existence of a collisional phase shift and two depth-dependent collision regimes~\cite{kivsharDarkOpticalSolitons1998,huangDarkSolitonsTheir2001}. These results not only illuminate fundamental nonlinear energy transport processes, but also show a path towards acoustic versions of soliton-enabled technologies such as frequency combs and mode-locked lasers~\cite{suhMicroresonatorSolitonDualcomb2016,braschPhotonicChipBased2016,greluDissipativeSolitonsModelocked2012}.

}

\begin{figure*}
   \centering
   \includegraphics[width=\textwidth]{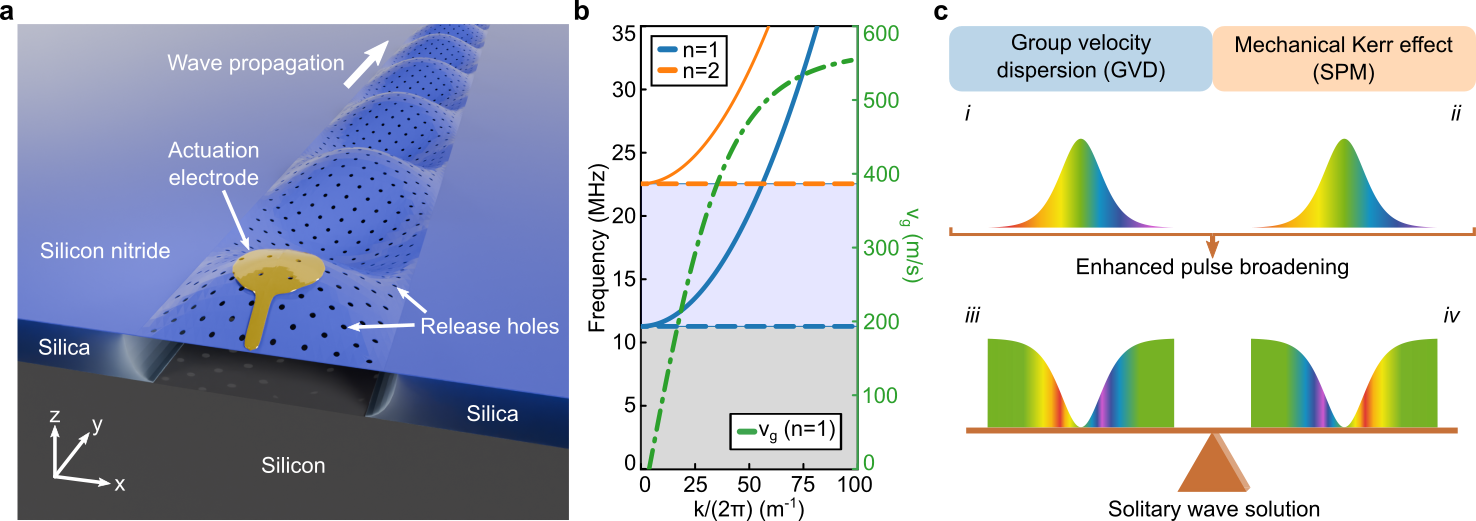}
   \caption{\textbf{Dispersion and mechanical Kerr nonlinearity in an on-chip acoustic waveguide}.  \textbf{a}, Schematic illustration of out-of plane guided acoustic waves in a membrane waveguide. Note the section view is for illustrative purposes; in the physical device the extremity of the waveguide is also clamped (see Fig.~\ref{fig:Fig2}\textbf{a}). \textbf{b},  Waveguide dispersion relation for the fundamental (n=1; solid blue), and second order (n=2; solid orange) acoustic modes (left axis). Grey shading: below cut-off; blue shading: single mode region~\cite{romeroPropagationImagingMechanical2019,mauranyapinTunnelingTransverseAcoustic2021}. Group velocity $v_g$ (green, dash-dot --- right axis). \textbf{c}, Pulse chirp arising from GVD and self-phase modulation (SPM) arising from the mechanical Kerr nonlinearity. In our system, GVD and SPM combine to broaden bright pulses ($i \,\&\, ii$), and counterbalance in dark pulses leading to solitary wave solutions ($iii \,\&\, iv$)~\cite{agrawalNonlinearFiberOptics2019}.}
   \label{fig:Fig1}
\end{figure*}

Solitons are nonlinear waves that preserve their shape throughout propagation and collisions~\cite{zabuskyInteractionSolitonsCollisionless1965}. Formed by a balance of dispersion and nonlinearity, solitons are a natural structural basis for understanding nonlinear systems~\cite{remoissenetWavesCalledSolitons1999}.
They are found in diverse settings including shallow~\cite{remoissenetWavesCalledSolitons1999, chabchoubExperimentalObservationDark2013, onoratoRogueShockWaves2016} and deep~\cite{remoissenetWavesCalledSolitons1999} water, normal 
and superconducting 
circuits~\cite{remoissenetWavesCalledSolitons1999}, plasmas~\cite{zabuskyInteractionSolitonsCollisionless1965}, 
Bose-Einstein condensates (BECs)~\cite{burgerDarkSolitonsBoseEinstein1999,streckerFormationPropagationMatterwave2002,beckerOscillationsInteractionsDark2008}, superfluid helium~\cite{ancilottoFirstObservationBright2018, reevesNonlinearWaveDynamics2025}, and nonlinear optics~\cite{kivsharDarkOpticalSolitons1998,herrTemporalSolitonsOptical2014}.
In optics especially, solitons have shown not just scientific but also engineering value: optical solitons now underpin technologies such as frequency combs~\cite{suhMicroresonatorSolitonDualcomb2016,braschPhotonicChipBased2016} and mode-locked lasers~\cite{greluDissipativeSolitonsModelocked2012}, and techniques like pulse compression~\cite{chernikovSolitonPulseCompression1993} and supercontinuum generation~\cite{dudleySupercontinuumGenerationPhotonic2006}. However, despite this success in optics, solitons have yet to achieve a comparable technological maturity in other physical systems. Furthermore, some features of the behaviour of solitons in their `dark' form have not been demonstrated, due to limited spatial resolution in BEC experiments~\cite{wellerExperimentalObservationOscillating2008} and difficulty generating and sustaining the required background excitation in optics~\cite{kivsharDarkOpticalSolitons1998}. 

Over the past few decades, integrated phononic circuits have been actively developed, boasting features similar to integrated photonics but benefiting from the low dissipation and slow velocity of acoustic phonons~\cite{schmidFundamentalsNanomechanicalResonators2016, xiSoftclampedTopologicalWaveguide2025}.  
There have been major efforts to observe solitons in these phononic circuits~\cite{hatanakaPhononWaveguidesElectromechanical2014, kurosuOnchipTemporalFocusing2018,kurosuMechanicalKerrNonlinearity2020}, due to their rich nonlinear dynamics and potential applications in nonlinear wave technologies.
The challenge has been fundamental: acoustic nonlinearities are typically weak, and dispersion and dissipation are difficult to simultaneously engineer at the nanoscale~\cite{kurosuMechanicalKerrNonlinearity2020}.
For instance, popular designs such as Rayleigh-like and Love-like phononic rib waveguides~\cite{fuPhononicIntegratedCircuitry2019,mayorGigahertzPhononicIntegrated2021} and phononic crystal waveguides~\cite{fangOpticalTransductionRouting2016}  
have weak quadratic nonlinearities that require impractically large stresses to achieve useful nonlinear behaviour.
Alternatively, transverse waves in suspended structures~\cite{midtvedtNonlinearPhononicsUsing2014,chaElectricalTuningElastic2018, hatanakaPhononWaveguidesElectromechanical2014, kurosuMechanicalKerrNonlinearity2020, romeroPropagationImagingMechanical2019} provide access to larger cubic Duffing-type geometric nonlinearities, which have been leveraged to achieve four-wave mixing~\cite{kurosuMechanicalKerrNonlinearity2020} and nanomechanical logic~\cite{romeroAcousticallyDrivenSinglefrequency2024,jinNanomechanicalErrorCorrection2025}. However, due to dissipation-limited propagation times, insufficient actuation amplitudes, and the need for dispersion engineering to balance the defocussing geometric nonlinearity, on-chip acoustic solitons have remained out of reach~\cite{kurosuMechanicalKerrNonlinearity2020}.
Investigations of acoustic solitons have therefore been largely restricted to macroscopic model systems, including bead chains~\cite{costeSolitaryWavesChain1997,daraioTunabilitySolitaryWave2006}, granular media~\cite{patilReviewExploitingNonlinearity2022} and 
3D-printed acoustic metamaterials~\cite{zhangProgrammableRobustStatic2019}, or numerical studies~\cite{charalampidisPhononicRogueWaves2018,miyazawaRogueSolitaryWaves2022}.

Our solution to the above challenges, illustrated in Fig.~\ref{fig:Fig1}\textbf{a}, uses transverse acoustic waves in a hard-clamped, high-tensile stress membrane, suspended via an array of sub-wavelength release holes~\cite{romeroAcousticallyDrivenSinglefrequency2024}. Unlike previous works~\cite{chaElectricalTuningElastic2018,kurosuMechanicalKerrNonlinearity2020}, we employ thin high tensile-stress silicon nitride ($\sigma=1$ GPa)  operating in the membrane limit~\cite{schmidFundamentalsNanomechanicalResonators2016}. This provides a more than order-of-magnitude reduction in dissipation~\cite{chaElectricalTuningElastic2018,kurosuMechanicalKerrNonlinearity2020}, enabling the observation of acoustic pulse 
evolution over several meters. We combine this low dissipation with an electrostatic actuation scheme~\cite{schmidFundamentalsNanomechanicalResonators2016} that allows us to produce wave amplitudes on the scale of tens of nanometres (exceeding earlier demonstrations by close to two orders of magnitude~\cite{kurosuMechanicalKerrNonlinearity2020}). Together, these features allow strong nonlinear behaviour to be observed within the dissipation lengthscale.

Our approach allows us to achieve solitary waves (the counterpart to solitons in non-idealised media~\cite{remoissenetWavesCalledSolitons1999}) in integrated phononics. 
We take advantage of the slow phonon velocity to directly image dark solitary wave collisions, achieving unprecedented observational detail unavailable to optical or BEC experiments. This resolution enables us to confirm the prediction that dark solitons experience a phase shift upon collision through direct imaging of their interaction~\cite{huangDarkSolitonsTheir2001}. Additionally, the low phonon frequencies allow us to use radiofrequency electronics to precisely initialise one or multiple dark solitary waves with varying relative velocities, such that the waveguide represents a programmable soliton collider capable of producing head-on and overtaking collisions between solitons of different depths. We use this functionality to confirm the topologically protected nature of dark solitons, and that they exhibit two kinds of collision---an avoided crossing or the formation of a transient composite structure---depending on their relative depths~\cite{huangDarkSolitonsTheir2001}. These results illustrate the potential of phononics for studies of nonlinear wave physics, and show a path towards integrated phononic circuits with strongly nonlinear functionality.

\section{Background}

Soliton formation requires a balance of nonlinearity and dispersion. The dispersion here is provided by the waveguide geometry:
the clamped boundary conditions on the sides of the waveguide introduce a cutoff frequency ($\Omega_c=2\pi\times11.3$ MHz) below which waves are exponentially attenuated (Fig.~\ref{fig:Fig1}\textbf{b}), with the group velocity $v_g$ rising from 0 at cutoff  towards $c=\sqrt{\frac{\sigma}{\rho}}\sim 570\,\mathrm{m/s}$ at large wavenumbers~\cite{romeroPropagationImagingMechanical2019}. This corresponds to a regime of anomalous group velocity dispersion (GVD), where higher frequencies propagate faster.
Nonlinearity arises from stretching of the membrane at large amplitudes of motion, increasing its tension (i.e. a hardening Duffing nonlinearity~\cite{schmidFundamentalsNanomechanicalResonators2016}) and thereby increasing the group velocity.
This so-called mechanical Kerr effect~\cite{kurosuMechanicalKerrNonlinearity2020} is of opposite sign to that commonly encountered in optics~\cite{kivsharDarkOpticalSolitons1998}, and induces a pulse chirp through self phase modulation (SPM) (Fig.~\ref{fig:Fig1}\textbf{c})~\cite{kurosuMechanicalKerrNonlinearity2020,agrawalNonlinearFiberOptics2019}. 
The combined dispersive and nonlinear dynamics are well captured by the nonlinear Schr\"{o}dinger equation (NLSE)
widely used to describe envelope solitons~\cite{kivsharDarkOpticalSolitons1998,kurosuMechanicalKerrNonlinearity2020,chabchoubExperimentalObservationDark2013, agrawalNonlinearFiberOptics2019}:
\begin{equation}
    \frac{\partial A}{\partial y}=-\frac{\alpha}{2}A-\frac{i k_2}{2} \frac{\partial^2 A}{\partial T^2}+i \xi\left|A\right|^2 A 
    \label{eqn:NLSE}
\end{equation}
where $A$ is the slowly varying pulse amplitude; $T$ the `fast time' coordinate measured in a frame moving with the group velocity $v_g$ of the carrier waves; $\alpha$ the energy loss rate; $k_2=\frac{\partial^2 k_y}{\partial \Omega^2}$ the GVD coefficient ($k_2$ is evaluated at the pulse centre frequency); and $\xi$ the nonlinear coefficient. In the Supplementary Information we derive Eq.~\eqref{eqn:NLSE} for our system (including an additional third-order dispersion term) and determine values for $\alpha$, $k_2$, and $\xi$ using ab-initio theory and experimental measurements.

In this work we follow the convention of reserving the word `soliton' for exact solutions to Eq.~\eqref{eqn:NLSE} in the absence of loss, which have perfectly stable shape and propagation speed~\cite{zabuskyInteractionSolitonsCollisionless1965}. As is the case for any experimental platform, 
our solitary waves experience loss and are therefore not idealised solitons; however, our losses are sufficiently small that the waves demonstrate metastable soliton-like behaviour, adiabatically maintaining the balance between nonlinearity and dispersion throughout the propagation~\cite{remoissenetWavesCalledSolitons1999}.

\begin{figure*}
    \centering
        \includegraphics[width=\textwidth]{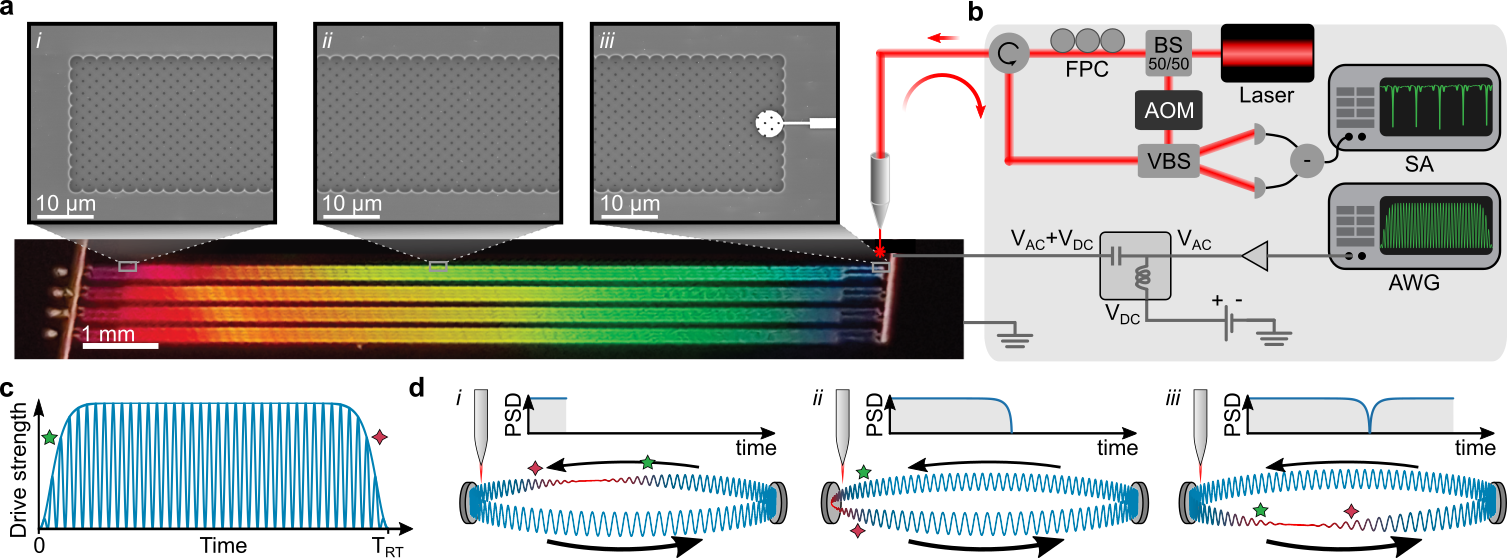}
        \caption{\textbf{Experimental setup and actuation protocol.}   \textbf{a}, Bottom: color optical photograph of the fabricated device, showing four groups of four acoustic waveguides of length $L=1$ cm and width $W=25$ $\mu$m. The iridescence is caused by the periodic array of sub-wavelength release holes~\cite{mauranyapinTunnelingTransverseAcoustic2021,romeroAcousticallyDrivenSinglefrequency2024}. Top: zoom-ins show scanning electron microscope images of the top waveguide. (\textit{i}): waveguide termination; (\textit{ii}): mid-section of waveguide; (\textit{iii}): waveguide extremity with electrostatic actuation electrode. \textbf{b}, Optical and electronic drive and readout scheme. AOM: acousto-optic modulator; VBS: variable beam-splitter; SA: spectrum analyzer; AWG: arbitrary waveform generator; FPC: fiber polarization controller. See Supplementary Information for more details. \textbf{c}, Illustration of the actuation pulse that produces a dark solitary wave.  \textbf{d}, Diagram of the soliton's progression through the waveguide and resulting measured power spectrum: (\textit{i}) before, (\textit{ii}) during, and (\textit{iii}) after passing the optical fiber probe.}
    \label{fig:Fig2}
\end{figure*}

In our architecture, the chirp arising from GVD and SPM add constructively in the case of bright pulses (Fig.~\ref{fig:Fig1}\textbf{c} i\&ii), resulting in enhanced pulse broadening. Previous works have responded to this by engineering a band gap in the waveguide that flips the sign of the GVD near the band edge, allowing bright solitons to be supported in principle~\cite{kurosuOnchipTemporalFocusing2018,kurosuMechanicalKerrNonlinearity2020}. Here, we instead note that for dark pulses, GVD and SPM can counterbalance and enable solitary wave solutions~\cite{kivsharDarkOpticalSolitons1998,kivsharDarkOpticalSolitons1998,agrawalNonlinearFiberOptics2019} (Fig.~\ref{fig:Fig1}\textbf{c} iii\&iv)---these dark solitary waves are what we experimentally realise. This combination of GVD and SPM corresponds to a regime akin to that found in optical fibers, where dark solitons were first observed~\cite{emplitPicosecondStepsDark1987, weinerExperimentalObservationFundamental1988} (albeit achieved here through a combination of anomalous dispersion and defocussing nonlinearity, compared to normal dispersion and focussing nonlinearity in those works).

The realisation of dark solitary waves in this phononic architecture provides the opportunity to probe soliton collisions, which have been of extensive interest across diverse nonlinear media~\cite{zabuskyInteractionSolitonsCollisionless1965,kivsharDarkOpticalSolitons1998,agrawalNonlinearFiberOptics2019,chenLaboratoryExperimentsCounterpropagating2014,maitreDarkSolitonMoleculesExcitonPolariton2020,hoeferDarkdarkSolitonsModulational2011,streckerFormationPropagationMatterwave2002,wellerExperimentalObservationOscillating2008}. Unlike their bright counterparts, which can either attract or repel each other depending on their relative phase~\cite{chenLaboratoryExperimentsCounterpropagating2014}, dark solitons only experience mutually repulsive interactions that result in a positive collisional phase shift in the direction of travel~\cite{kivsharDarkOpticalSolitons1998,huangDarkSolitonsTheir2001}. Direct measurement of this phase shift has been impossible, precluded for instance in BECs by limited spatial resolution and condensate lifetime~\cite{ wellerExperimentalObservationOscillating2008}, and in optics by the short interaction time, limiting measurement of the phase shift to inference from measurements of arrival times~\cite{foursaInvestigationBlackGraySoliton1996}. By comparison, our setup permits direct visualisation of these collisions in the time domain, facilitated by the slow phonon group velocity, low dissipation, and ability to initialise a near-uniform background (see next section).

\subsection{Solitary wave initialisation}

Figure~\ref{fig:Fig2}\textbf{a} shows our fabricated devices:  $L=1$ cm-long  straight acoustic waveguides capacitively actuated at one end, with the other extremity forming a reflective boundary.  Previous works had released the membrane by directly etching the silicon substrate~\cite{mauranyapinTunnelingTransverseAcoustic2021}. Here, we instead use a silica sacrificial layer enabling the fabrication of a controllable nanometer-scale electrode gap (Fig.~\ref{fig:Fig1}\textbf{a}). This, along with the use of high actuation DC voltage (up to 150 V, see Fig.~\ref{fig:Fig2}\textbf{b}), allows us to reach very large wave amplitudes ($A_0\sim 60$ nm) without relying on resonant enhancement in an acoustic cavity (see Supplementary Information). Acoustic wave readout is performed  using a custom-built laser Doppler vibrometer~\cite{romeroAcousticallyDrivenSinglefrequency2024} shown in Fig.~\ref{fig:Fig2}\textbf{b} (see Supplementary Information).

\begin{figure*}[ht]
   \centering
    \includegraphics[width=\textwidth]{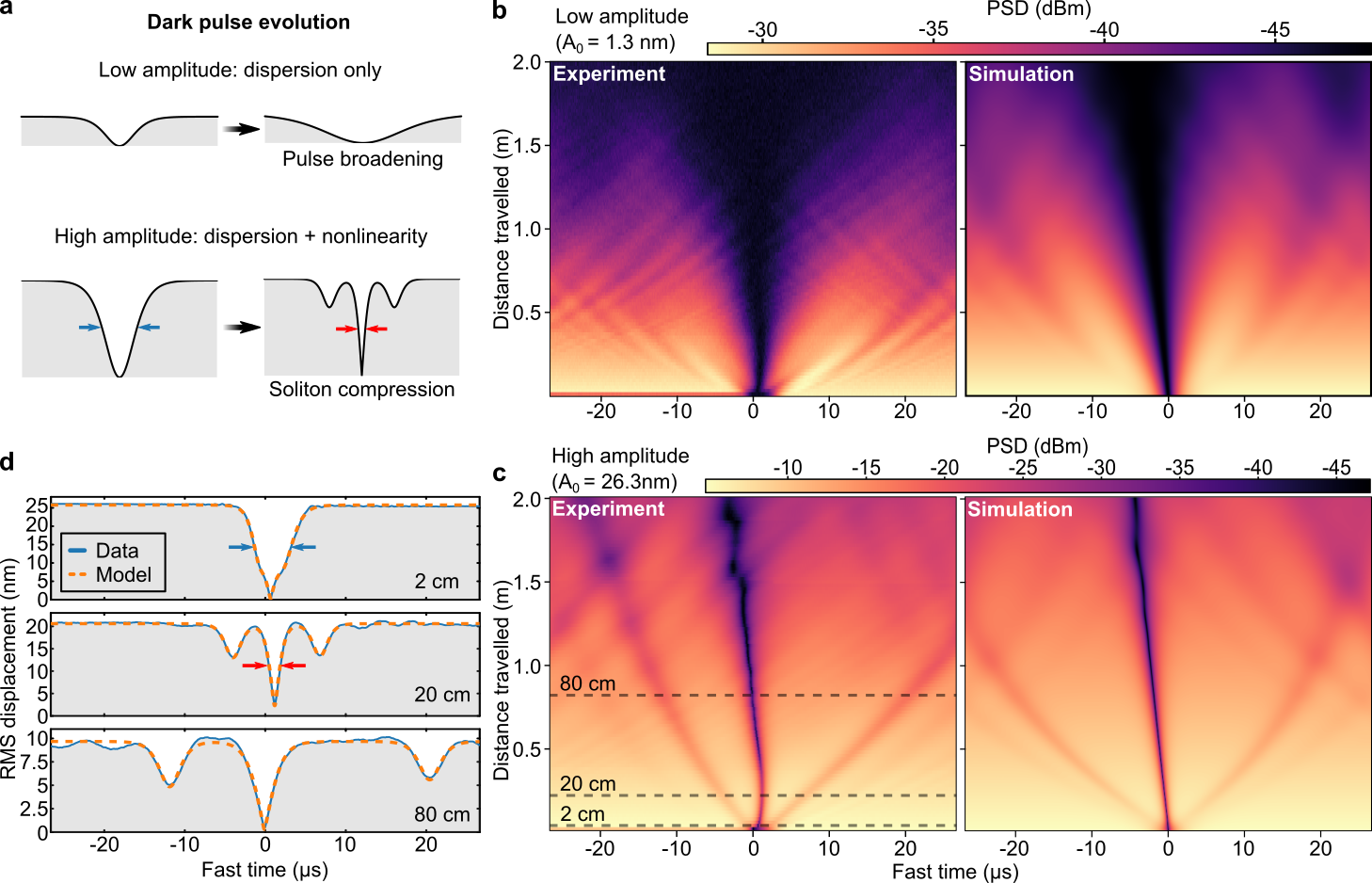}
    \caption{\textbf{Acoustic dark solitary waves.}  \textbf{a}, Sketch of predicted dark pulse behaviour in the low amplitude (linear)  and high amplitude (nonlinear) regimes. Note the soliton compression highlighted by the red and blue arrows, and the shedding of grey solitary waves on either side of the pulse. \textbf{b}, Experimental pulse evolution (left panel)  and NLSE simulation (right panel) in the low amplitude regime ($A_0=1.3$ nm). \textbf{c}, Experimental pulse evolution (left panel)  and NLSE simulation (right panel) in the high amplitude regime ($A_0=26.3$ nm).  \textbf{d}, Cross-sections of the  measured RMS displacement at the coordinates shown in (\textbf{c}) (blue), along with a black and grey soliton model fit (dashed orange) - see Supplementary Information. }
    \label{fig:Fig3}
\end{figure*}

A dark solitary wave requires a bright, high-amplitude background, whose amplitude we denote $A_0$.  A natural way to initialise the wave is as a dark notch within a bright envelope, as has been done using pulse shaping in fiber optics~\cite{weinerExperimentalObservationFundamental1988}.  This however leads to issues with the non-uniformity of the bright pulse and its interaction with the waveguide boundaries within a finite-sized structure, limiting the evolution time before the background intensity becomes chaotic (see  extended Fig.~\ref{fig:tanh-in-exp})~\cite{thurstonCollisionsDarkSolitons1991}. Instead, we develop an alternate approach illustrated in Fig.~\ref{fig:Fig2}\textbf{c}, whereby we drive the acoustic waveguide for a time equal to its round trip time $T_{RT}=2L/v_g$. The ramp-up and ramp-down portions on opposite sides of the drive (red and green stars) form, when recombined, a hyperbolic tangent `tanh' profile. This approach is conceptually similar to experiments in optics that sandwiched a dark region between two bright pulses~\cite{emplitPicosecondStepsDark1987,weinerExperimentalObservationFundamental1988}; however, here we avoid using pulses altogether, creating a uniform background that significantly extends the soliton lifetime.

Driving for the round trip time fully fills the waveguide with both forward and backward travelling waves, creating a standing-wave background modulated only by the discrete soliton dip travelling back and forth, as illustrated in Fig.~\ref{fig:Fig2}\textbf{d}. This aligns well with models of dark solitons in trapped BECs~\cite{burgerDarkSolitonsBoseEinstein1999}, where the soliton acts as a transient, travelling defect on the uniform standing-wave condensate. The nonlinear interactions within the soliton wave packet remain significant, while the cross-interaction with the counter-propagating background component can be abstracted to an effective renormalization of the sound speed, 
justified by the high relative speed ($2 v_g \sim $ 1 km/s), as we will see later in the case of soliton collisions.  Use of an arbitrary waveform generator (Fig.~\ref{fig:Fig2}\textbf{b}) lets us simultaneously control the carrier frequency (and hence the dispersion), the background amplitude, the initial dip width and the dip depth (greyness). We additionally control the phase shift across the dip, a crucial parameter responsible for a dark soliton's topologically protected nature which ensures its robust propagation~\cite{kivsharDarkOpticalSolitons1998}---as in early BEC experiments where a $\pi$ phase slip was optically imprinted upon the condensate to initialise dark solitons~\cite{burgerDarkSolitonsBoseEinstein1999}.

\section{Results}

\subsection{Dark pulse evolution}

We first verify a hallmark of solitary wave physics---namely, amplitude-dependent pulse propagation. A dark pulse on a low amplitude background $A_0$ will evolve solely under the influence of dispersion, leading to gradual pulse broadening (Fig.~\ref{fig:Fig3}\textbf{a}, top). In contrast, the same dark pulse on a high amplitude background, evolving under the combined influence of dispersion and nonlinearity, is predicted to decompose along the new solutions of the nonlinear system, namely black (full extinction) and grey (partial extinction) solitary waves~\cite{kivsharDarkOpticalSolitons1998,agrawalNonlinearFiberOptics2019}. In this example, a pulse whose temporal width $T_0$  weakly exceeds the stable soliton width $T_s=\sqrt{\frac{|k_2|}{\xi A_0^2}}$ will exhibit soliton compression~\cite{agrawalNonlinearFiberOptics2019,kurosuMechanicalKerrNonlinearity2020}, as well as shed radiation and grey solitary waves travelling outward at a relative speed of $c_s\sin(\theta)$, where $c_s$ is the Bogoliubov speed of sound~\cite{kivsharDarkOpticalSolitons1998,onoratoRogueShockWaves2016}:
\begin{equation}
    \label{eqn:c_s}
    c_s=\sqrt{\xi A_0^2 |k_2|}.
\end{equation}
The parameter $\theta$ characterises the solitary wave depth; $\theta=0$ corresponds to black, full extinction solitons which are stationary in the moving frame, while $0<\theta<\frac{\pi}{2}$ corresponds to grey solitons which separate in the moving frame (see bottom of Fig.~\ref{fig:Fig3}\textbf{a})~\cite{kivsharDarkOpticalSolitons1998}.

\begin{figure*}
    \centering
    \includegraphics[width=1.0\textwidth]{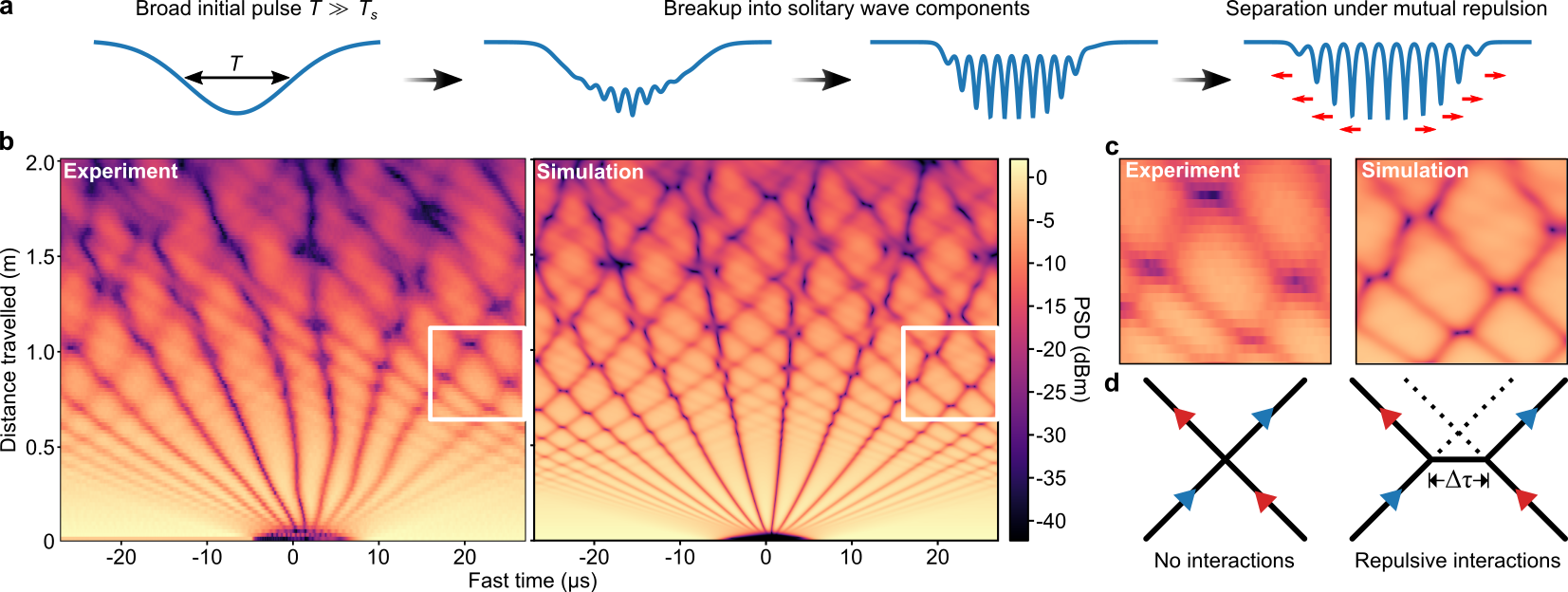}
    \caption{\textbf{Soliton fission and solitary wave repulsion.}  \textbf{a}, Schematic illustration of the break-up of a broad ($T\gg T_s$) initial dark pulse into its solitary wave components.  \textbf{b} Experiment (left) and NLSE simulation (right). Here $A_0=59.5\,\mathrm{nm}$. \textbf{c}, Zoom-in of the wave crossings (white boxes in \textbf{b}).  \textbf{d}, Sketch of wave crossings in the case of no interactions (left) and repulsive interactions (right). The collisional phase shift is apparent, as each soliton emerges ahead of its pre-collision trajectory~\cite{huangDarkSolitonsTheir2001,thurstonCollisionsDarkSolitons1991}.}
    \label{fig:Fig4}
\end{figure*}

The left panels of Figs.~\ref{fig:Fig3}\textbf{b} and ~\ref{fig:Fig3}\textbf{c} show waterfall plots of the experimentally measured dark pulse evolution, using the initialisation procedure shown in Fig.~\ref{fig:Fig2}, for low and high  amplitude backgrounds (respectively $A_0=1.3$ nm and $A_0=26.3$ nm) (see Supplementary Information for experimental details). These plots are composed by slicing the photocurrent time series data into regular intervals of the round trip time $T_\mathrm{RT}$, such that the $k^\mathrm{th}$ row of pixels from the bottom of the plot represents the pulse evolution after $k-1$ round trips around the waveguide. The data reveal the striking amplitude dependence of the pulse evolution highlighted in Fig.~\ref{fig:Fig3}\textbf{a}.  The right panels of Figs~\ref{fig:Fig3}\textbf{b} and {\textbf{c}} plot the simulated evolution of the injected dark pulse, modelled by the NLSE (Eq.\eqref{eqn:NLSE}) with no free parameters, exhibiting excellent agreement with experiment in both cases (see Supplementary Information for simulation details, and extended Fig.~\ref{fig:ExtFig-DarkSolAllVoltages+widthxamp} for pulse evolution at 4 additional intermediate amplitudes). The evolution of bright pulses, where nonlinearity at higher amplitudes leads to enhanced pulse broadening is shown in extended Fig.~\ref{fig:ExtFig-BrightPulse}. Figure~\ref{fig:Fig3}\textbf{d} shows cross-sections of the RMS displacement after 2~cm, 20~cm and 80~cm of pulse propagation (blue trace). These highlight soliton compression and the nucleation of a black solitary wave---characterized by a complete extinction of the transverse vibration and a characteristic tanh profile---accompanied by the shedding of two grey solitary waves as the broad initial pulse decomposes onto the basis of soliton solutions (dashed orange traces are fits to the data with one black and two grey profiles; for details of the fit see Eq.~\eqref{eqn:soliton-fission-N-soliton-fit} in the Supplementary Information). The cross-sections further highlight how the solitary wave adiabatically evolves throughout the propagation, gradually broadening as the background amplitude decreases due to acoustic losses, so as to continuously maintain the balance between dispersion and nonlinearity. This is evidenced by a constant amplitude-width product (a conserved quantity for NLSE solitons) maintained over 1.5m of propagation ($i.e.\sim$40 000 wavelengths)---see Supplementary Fig.~\ref{fig:ExtFig-DarkSolAllVoltages+widthxamp} for detailed analysis.

\subsection{Soliton fission and soliton repulsion}

To further confirm the solitary nature of the generated waves, we next initialise a dark pulse (on a high amplitude background) whose initial width $T_0$ is much greater than the soliton width $T_s$, such that it represents a nonlinear superposition of multiple separate fundamental dark solitons~\cite{zakharovExactTheoryTwoDimensional1972,agrawalNonlinearFiberOptics2019}. The expected outcome, termed soliton fission, is a key process in supercontinuum generation in optical fibers~\cite{dudleySupercontinuumGenerationPhotonic2006} as well as water and superfluid waves~\cite{trilloExperimentalObservationTheoretical2016,reevesNonlinearWaveDynamics2025}, and is illustrated in Fig.~\ref{fig:Fig4}\textbf{a}.
The broad pulse initially narrows, reaching a maximal soliton compression at a distance $L_{\mathrm{fiss}}=\sqrt{L_{NL}L_{D}}$, where $L_{NL}=1/(\xi A_0^2)$ and $L_D=T_0^2/|k_2|$ are respectively the nonlinear and dispersive lengths~\cite{agrawalNonlinearFiberOptics2019}.
With our parameters ($T_0=6.88\,\mathrm{\upmu s}$, $A_0=59.5\,\mathrm{nm}$) $L_D = 1.29\,\mathrm{m}$, $L_{NL}=3.9\,\mathrm{mm}$ 
and $L_\mathrm{fiss}=7.1\,\mathrm{cm}$.

In this highly nonlinear regime where $L_D\gg L_\mathrm{NL}$, theory predicts that after maximal soliton compression, the pulse will break up into a combination of dispersive radiation and a train of $N_\mathrm{fiss}=\sqrt{\frac{L_D}{L_{NL}}}=\sqrt{\frac{T_0^2 \xi A_0^2}{|k_2|}}\simeq 18$ fundamental solitons~\cite{agrawalNonlinearFiberOptics2019}, which then spatially separate under the effect of mutual interactions~\cite{dudleySupercontinuumGenerationPhotonic2006,kivsharDarkOpticalSolitons1998}. This is observed in our experimental results, shown in Fig.~\ref{fig:Fig4}\textbf{b}, left panel. Once again, the NLSE simulation presents excellent agreement with the observed physics (Fig.~\ref{fig:Fig4}\textbf{b}, right panel). The lower phonon velocity enables us to observe this fission process while it is occurring and with exceptional resolution, something particularly challenging to do in optics due to the much faster wave speed~\cite{yiImagingSolitonDynamics2018}, or in BECs due to limited experimental lifetimes and spatial resolution~\cite{burgerDarkSolitonsBoseEinstein1999,wellerExperimentalObservationOscillating2008}.
The central solitary wave travels at $v_g$ (and therefore nearly vertically in Fig.~\ref{fig:Fig4}\textbf{b}) while the fastest (slowest) solitary waves move approximately 2\% faster (slower), performing one extra (fewer) round-trip after 1 m or 50 round trips. Due to this speed difference, the solitary waves will eventually collide, with the faster waves overtaking the slowest. These crossings create a hexagonal-looking lattice (Fig.~\ref{fig:Fig4}\textbf{c}), indicative of repulsive soliton-soliton interactions (Fig.~\ref{fig:Fig4}\textbf{d})~\cite{huangDarkSolitonsTheir2001,thurstonCollisionsDarkSolitons1991}. Because of these repulsive interactions, dark solitons are predicted to exit collisions with a positive phase shift in their travelling directions, which we observe here as a temporal shift $\Delta\tau$.

\subsection{Programmable soliton collisions}

\begin{figure*}
    \centering
    \includegraphics[width=1.0\textwidth]{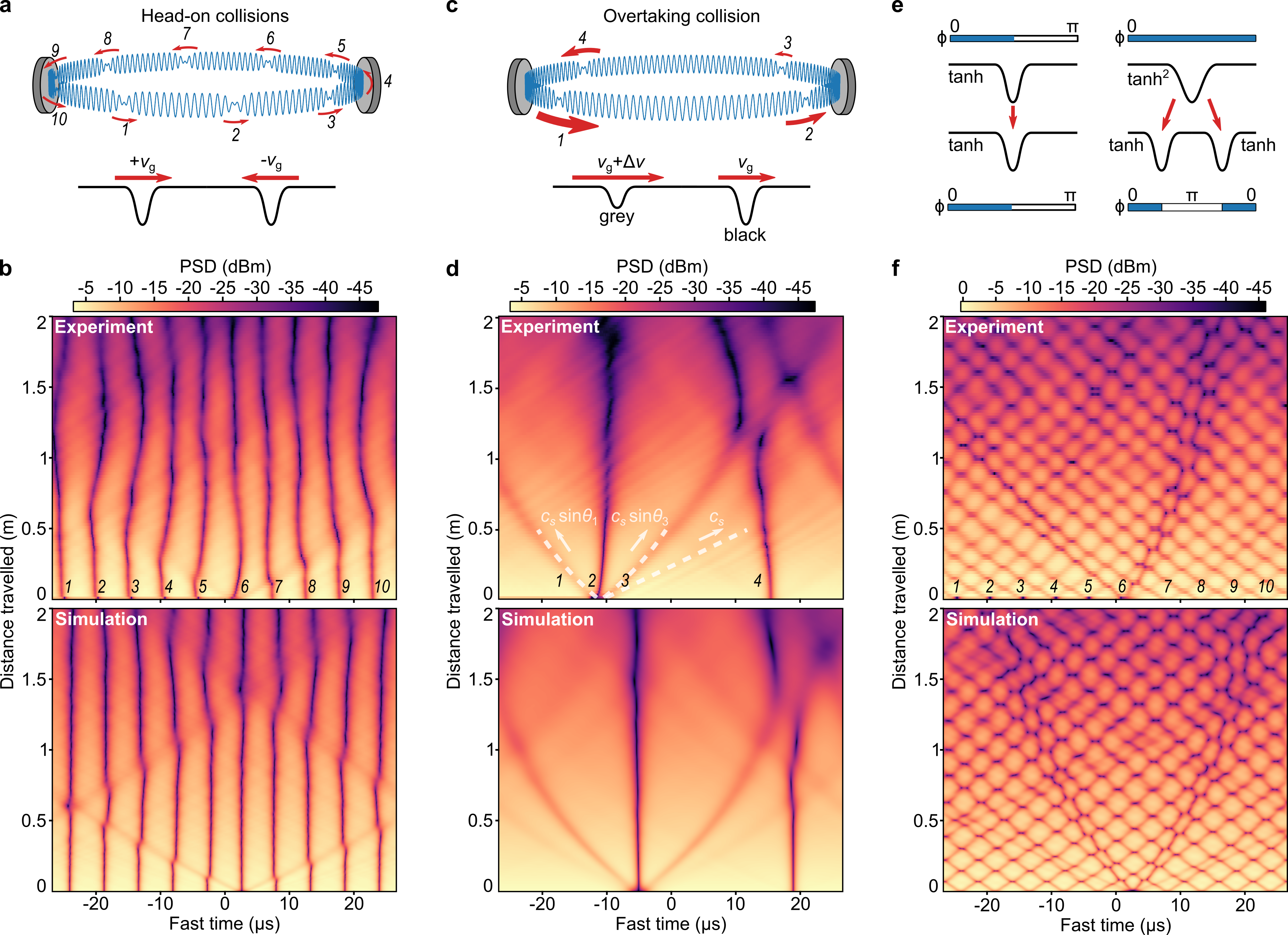}
    \caption{\textbf{Programmable soliton-soliton collider}.   \textbf{a}, Sketch of the waveguide initialisation to perform head-on collisions. \textbf{b}, Experimental data (top) and NLSE simulation (bottom). \textbf{c}, Sketch of the waveguide initialisation to perform overtaking collisions.  \textbf{d}, Experimental data (top) and NLSE simulation (bottom). \textbf{e}, Role of the parity of the injected pulse: an appropriately-sized tanh pulse (odd) forms a single solitary wave, while an even $\tanh^2$  pulse forms a pair of dark solitary waves.  \textbf{f}, Experimental data (top) and NLSE simulation (bottom). All NLSE simulations are performed with no free parameters (see Supplementary Information).}
    \label{fig:Fig5}
\end{figure*}

To systematically study both head-on and overtaking dark solitary wave collisions, and the resulting temporal shifts, we use the arbitrary waveform generator (AWG) (Fig.~\ref{fig:Fig2}\textbf{b}) to controllably inject multiple solitary waves into our chip. This strategy, similar to the phase imprinting used to initialise solitons in BECs~\cite{burgerDarkSolitonsBoseEinstein1999, denschlagGeneratingSolitonsPhase2000}, allows us programmatically engineer different types of soliton collisions. We first study head-on collision. To do so, we use the AWG to initialise 10 black solitary waves equally distributed within the waveguide, whose tanh profile and width $T_s(A_0)$ matches that of an ideal black soliton at this amplitude. These are all nominally identical such that they all propagate with identical speed $v_g$, as illustrated in Fig.~\ref{fig:Fig5}\textbf{a}. Each solitary wave collides with every other wave twice per round-trip (and with itself upon reflection at the waveguide extremities), such that, after a propagation distance of 1 meter (i.e. 50 round trips), it has experienced 1000 head-on collisions. Despite these collisions, the solitary waves propagate broadly unperturbed, as shown experimentally in Fig.~\ref{fig:Fig5}\textbf{b} (top panel) and well reproduced by NLSE simulations (bottom panel). These results validate our earlier assumption of separability of forward and backward travelling wave components, and highlight a hallmark of solitary waves: their ability to preserve shape upon collisions~\cite{kivsharDarkOpticalSolitons1998,chabchoubExperimentalObservationDark2013,reevesNonlinearWaveDynamics2025,dudleySupercontinuumGenerationPhotonic2006, agrawalNonlinearFiberOptics2019}.

In the experiment, all solitary waves are fully imprinted into the actuation waveform, except for the wave labelled \#6 in Fig.~\ref{fig:Fig5}\textbf{b}, which is formed by the temporal wrapping of the waveform's beginning and end (as illustrated in Fig.~\ref{fig:Fig2}\textbf{c}). It therefore departs slightly from the ideal soliton eigenmode shape, and sheds dispersive radiation and sound waves travelling at oblique angles (as also shown in Fig.~\ref{fig:Fig3}). Despite these perturbations, the generated solitary waves do not cross, as their long-range mutually repulsive interactions help them self-organize and remain in a distribution that uniformly fills the resonator. This is well captured by the NLSE simulations (bottom panel). 
This configuration  can be naturally interpreted as a finite-temperature one-dimensional  Wigner crystal~\cite{deshpande_one-dimensional_2008, shapir_imaging_2019}  of solitons on a closed ring (Fig.~\ref{fig:Fig5}), in which our high-finesse acoustic waveguide provides the confinement needed to stabilise the repulsive interactions and form a stable lattice. 
Over time, the sound waves introduced by the imperfect imprinting of the wrapped soliton provide the energy to melt the lattice into a soliton‑liquid state.
We confirm this by computing the average pair-correlation function $g(\Delta t)$ over the two meters of evolution, which reveals the decaying positional order characteristic of a liquid state~\cite{coleSolitonCrystalsKerr2017} (see Supplementary Information for more details).

In optical microcavities, crystals of bright dissipative Kerr solitons have recently been demonstrated, and promise to enhance the repetition rate and power efficiency of microcomb sources~\cite{coleSolitonCrystalsKerr2017}. Those systems require a careful choice of resonator and laser parameters to counteract the natural attractive interaction between bright solitons. By contrast, in our undriven NLSE regime, dark solitons exhibit robust mutual repulsion without fine‑tuning, and the initial conditions can be programmed with high precision. This level of control opens the door to systematic studies of the thermodynamics of soliton crystals~\cite{coleSolitonCrystalsKerr2017}, liquids, and gases~\cite{redor2019experimental}, as well as direct measurements of the phonon dispersion relation of the soliton lattice itself.

Due to the large relative speed, head-on collisions last only on the order of microseconds and produce indiscernible phase shifts, as in  earlier BEC experiments~\cite{wellerExperimentalObservationOscillating2008}.
To resolve the shifts we switch to programming overtaking solitary wave collisions, as illustrated in Fig.~\ref{fig:Fig5}\textbf{c}. As we have seen, dark solitary waves naturally repel. To overcome this, we program one `ideal' tanh solitary wave (wave \#4), as well as one broad dip, similar to that used to generate the results in Fig.~\ref{fig:Fig3}. This broad dip will shed two grey solitary waves (waves \#1 and \#3), launched at relative speeds $c_s\sin\theta_1$ and $c_s\sin\theta_3$, which we use to overcome the intrinsic repulsion and engineer an overtaking collision. Experimental results are shown in Fig.~\ref{fig:Fig5}\textbf{d}, where black solitary wave \#4 overtakes the slower grey wave \#3. The grey solitary waves' relative velocities closely match the predicted values (dashed white lines in experimental plot) calculated from Eq.~\eqref{eqn:c_s}; we estimate $\theta_1\approx36^\circ$ and $\theta_3\approx39^\circ$ from the relationship $\sin\theta=A_\mathrm{dip}/A_0$, where $A_\mathrm{dip}$ is the amplitude in the center of the solitary wave~\cite{kivsharDarkOpticalSolitons1998}. Additionally, the relative velocity of the sound waves shed by imperfectly initialised solitary wave closely match the predicted value of $c_s$. Compared to the head-on collision case, the interaction time of this overtaking collision is enhanced by a factor of $(2v_g)/(v_g^2c_s(\sin\theta_3))\gtrsim500$, leading to a clear avoided crossing and observation of the predicted temporal shift arising due to repulsive interactions~\cite{huangDarkSolitonsTheir2001,wellerExperimentalObservationOscillating2008}.

Finally, we seek to validate the existence of a $\pi$ phase shift across the centre of the dark soliton---a topologically protected parameter responsible for its robust propagation~\cite{kivsharDarkOpticalSolitons1998}---by varying the parity of the injected pulse (Fig.~\ref{fig:Fig5}\textbf{e}). We perform the same experiment as in Fig.~\ref{fig:Fig5}\textbf{b}, simply replacing the injected odd parity tanh profiles (with $\pi$ phase shift across the minimum) by even parity $\tanh^2$ profiles (with no accompanying phase shift). The inverse scattering transform predicts that an even input pulse always produces (at least) a pair of dark solitary waves, with equal amplitudes and opposite velocities~\cite{zakharovExactTheoryTwoDimensional1972,denschlagGeneratingSolitonsPhase2000}. This is indeed what we observe (Fig.~\ref{fig:Fig5}\textbf{f}), where each of the ten input pulses quickly bifurcates, filling the waveguide with twenty solitary waves in good agreement with NLSE simulations (see also extended Fig.~\ref{fig:tanh-in-exp} for the evolution of a single $\tanh^2$ pulse). The hexagonal lattice formed by the trajectories of these solitary waves represents hundreds of grey soliton collisions, in each of which the solitary waves form a composite structure before separating with positive spatial shifts~\cite{huangDarkSolitonsTheir2001,thurstonCollisionsDarkSolitons1991}.

Fig.~\ref{fig:Fig5}\textbf{d} and~\textbf{f} validate that dark solitons exhibit two kinds of collision, namely an avoided crossing for darker solitons and the formation of a composite structure for more grey solitons~\cite{thurstonCollisionsDarkSolitons1991,huangDarkSolitonsTheir2001}. It has been predicted that in either kind of collision, the size of the temporal shift $\Delta\tau$ increases when the collision involves darker solitons~\cite{thurstonCollisionsDarkSolitons1991,huangDarkSolitonsTheir2001}---we confirm this, measuring $\Delta\tau\approx3.1\,\mathrm{\upmu s}$ in Fig.~\ref{fig:Fig5}\textbf{d} and $\Delta\tau\approx1.6\,\mathrm{\upmu s}$ in Fig.~\ref{fig:Fig5}\textbf{f}. In theory, $\Delta\tau$ depends only on the depth parameters of the colliding solitons, which are conserved quantities of the collision~\cite{thurstonCollisionsDarkSolitons1991,huangDarkSolitonsTheir2001}. This explains why we see the same temporal shift across the hundreds of collisions in Fig~\ref{fig:Fig5}\textbf{f}, even as the background amplitude slowly decays by a factor of $\sim10$.

\section{Conclusion}

This work demonstrates the first realisation of acoustic solitary waves on a phononic chip. By implementing a programmable soliton collider, we achieve unprecedented direct observation of dark soliton collisional dynamics. We demonstrate soliton compression, soliton fission, as well as Kerr crystal-like self-organisation.   
Furthermore, we map the collisional trajectories to experimentally verify a positive collisional phase shift and two depth-dependent collision regimes~\cite{huangDarkSolitonsTheir2001}. Whereas previous dark soliton experiments have captured single or few collisions~\cite{wellerExperimentalObservationOscillating2008, foursaInvestigationBlackGraySoliton1996}, our platform achieves a resolution and collision count competitive with current experiments on bright solitons in optics~\cite{yiImagingSolitonDynamics2018,copieSpaceTimeObservation2023}.

Beyond fundamental observations, our results introduce soliton engineering to integrated phononics, marking the transition from passive linear functionality to sophisticated nonlinear signal processing. The simplicity of our waveguide architecture facilitates low-loss topological phononics~\cite{xiSoftclampedTopologicalWaveguide2025}, adiabatic soliton compression~\cite{chernikovSolitonPulseCompression1993}, and band gap engineering~\cite{eggletonBraggGratingSolitons1996,kurosuMechanicalKerrNonlinearity2020}. Additionally, electrostatic tuning offers a path for dynamic tuning of the dispersion and nonlinear Kerr effect~\cite{chaElectricalTuningElastic2018,samantaTuningGeometricNonlinearity2018}. These capabilities open the door to bright soliton pulses and frequency combs integrated within membrane phononic circuits~\cite{hirschTutorialMembranePhononic2026}. Variations on this setup could access the Lugiato-Lefever regime associated with dissipative Kerr solitons~\cite{lugiatoSpatialDissipativeStructures1987}, or serve as a phononic `wave tank' for exploring nonlinear hydrodynamics such as rogue waves~\cite{reevesNonlinearWaveDynamics2025}.

\section*{Acknowledgements}
This research was primarily funded by the Australian Research Council and Lockheed Martin Corporation through the Australian Research Council Linkage Grant No. LP190101159. Support was also provided by the Australian Research Council Centre of Excellence for Engineered Quantum Systems (No. CE170100009). G.I.H. (No. DE210100848), C.G.B. (No. FT240100405) and M.T.R (No. DE220101548) acknowledge their Australian Research Council Fellowships. This work was performed in part at the Queensland node of the Australian National Fabrication Facility, a company established under the National Collaborative Research Infrastructure Strategy to provide nano and micro-fabrication facilities for Australia’s researchers. The authors acknowledge the facilities, and the scientific and technical assistance, of the Microscopy Australia Facility at the Centre for Microscopy and Microanalysis (CMM), the University of Queensland.

\section*{Author contributions}

T.M.F.H. and C.G.B. acquired the experimental data. T.M.F.H. fabricated the chip, with assistance from N.A. and E.R.. T.M.F.H. built the experimental setup, with assistance from N.P.M., X. J. and G.I.H.. C.G.B, T.M.F.H. and M.T.R.  developed the theory and performed the numerical simulations. C.G.B. and W.P.B. jointly conceived and supervised the project.  C.G.B. and  T.M.F.H.  wrote the manuscript, with support from W.P.B..  Funding Acquisition was led by W.P.B., with support from C.G.B. and G.I.H.. All authors discussed and analyzed the results and provided feedback on the manuscript.

\section*{Competing interests}
T.M.F.H, C.G.B, M.T.R, and G.I.H. are employed by Cortisonic Pty Ltd, a company developing low-power nanomechanical computing solutions.

\section*{Data availability}
The data that support the findings of this study will be made available in a public, open-access repository. 

\vspace{1cm}

\bibliography{bib}

@book{agrawalNonlinearFiberOptics2019,
  title = {Nonlinear Fiber Optics},
  author = {Agrawal, G. P.},
  year = 2019,
  edition = {Sixth edition},
  publisher = {Academic Press},
  address = {London, United Kingdom ; San Diego, CA, United States},
  isbn = {978-0-12-817042-7},
  lccn = {QC448 .A38 2019}
}

@article{ancilottoFirstObservationBright2018,
  title = {First {{Observation}} of {{Bright Solitons}} in {{Bulk Superfluid 4He}}},
  author = {Ancilotto, Francesco and Levy, David and Pimentel, Jessica and Eloranta, Jussi},
  year = 2018,
  month = jan,
  journal = {Physical Review Letters},
  volume = {120},
  number = {3},
  pages = {035302},
  publisher = {American Physical Society},
  doi = {10.1103/PhysRevLett.120.035302},
  urldate = {2026-01-13}
}

@article{bakerHighBandwidthOnchip2016,
  title = {High Bandwidth On-Chip Capacitive Tuning of Microtoroid Resonators},
  author = {Baker, Christopher G. and Bekker, Christiaan and McAuslan, David L. and Sheridan, Eoin and Bowen, Warwick P.},
  year = 2016,
  month = sep,
  journal = {Optics Express},
  volume = {24},
  number = {18},
  pages = {20400},
  issn = {1094-4087},
  doi = {10.1364/OE.24.020400},
  urldate = {2025-02-25},
  copyright = {https://doi.org/10.1364/OA\_License\_v1\#VOR-OA},
  langid = {english}
}

@article{beckerOscillationsInteractionsDark2008,
  title = {Oscillations and Interactions of Dark and Dark--Bright Solitons in {{Bose}}--{{Einstein}} Condensates},
  author = {Becker, Christoph and Stellmer, Simon and {Soltan-Panahi}, Parvis and Baumert, Mathis and Richter, Eva-Maria and Kronj{\"a}ger, Jochen and Bongs, Kai and Sengstock, Klaus},
  year = 2008,
  month = jun,
  journal = {Nature Physics},
  volume = {4},
  number = {6},
  pages = {496--501},
  publisher = {Nature Publishing Group},
  issn = {1745-2481},
  doi = {10.1038/nphys962},
  urldate = {2025-12-16},
  copyright = {2008 Springer Nature Limited},
  langid = {english}
}

@article{beliaevOpticalStructuralComposition2022,
  title = {Optical, Structural and Composition Properties of Silicon Nitride Films Deposited by Reactive Radio-Frequency Sputtering, Low Pressure and Plasma-Enhanced Chemical Vapor Deposition},
  author = {Beliaev, Leonid Yu. and Shkondin, Evgeniy and Lavrinenko, Andrei V. and Takayama, Osamu},
  year = 2022,
  month = dec,
  journal = {Thin Solid Films},
  volume = {763},
  pages = {139568},
  issn = {0040-6090},
  doi = {10.1016/j.tsf.2022.139568},
  urldate = {2025-01-22}
}

@article{braschPhotonicChipBased2016,
  title = {Photonic Chip--Based Optical Frequency Comb Using Soliton {{Cherenkov}} Radiation},
  author = {Brasch, V. and Geiselmann, M. and Herr, T. and Lihachev, G. and Pfeiffer, M. H. P. and Gorodetsky, M. L. and Kippenberg, T. J.},
  year = 2016,
  month = jan,
  journal = {Science},
  volume = {351},
  number = {6271},
  pages = {357--360},
  publisher = {American Association for the Advancement of Science},
  doi = {10.1126/science.aad4811},
  urldate = {2026-02-05}
}

@article{burgerDarkSolitonsBoseEinstein1999,
  title = {Dark {{Solitons}} in {{Bose-Einstein Condensates}}},
  author = {Burger, S. and Bongs, K. and Dettmer, S. and Ertmer, W. and Sengstock, K. and Sanpera, A. and Shlyapnikov, G. V. and Lewenstein, M.},
  year = 1999,
  month = dec,
  journal = {Physical Review Letters},
  volume = {83},
  number = {25},
  pages = {5198--5201},
  publisher = {American Physical Society},
  doi = {10.1103/PhysRevLett.83.5198},
  urldate = {2025-12-16}
}

@article{chabchoubExperimentalObservationDark2013,
  title = {Experimental {{Observation}} of {{Dark Solitons}} on the {{Surface}} of {{Water}}},
  author = {Chabchoub, A. and Kimmoun, O. and Branger, H. and Hoffmann, N. and Proment, D. and Onorato, M. and Akhmediev, N.},
  year = 2013,
  month = mar,
  journal = {Physical Review Letters},
  volume = {110},
  number = {12},
  pages = {124101},
  publisher = {American Physical Society},
  doi = {10.1103/PhysRevLett.110.124101},
  urldate = {2024-10-08}
}

@article{chabchoubRogueWaveObservation2011,
  title = {Rogue {{Wave Observation}} in a {{Water Wave Tank}}},
  author = {Chabchoub, A. and Hoffmann, N. P. and Akhmediev, N.},
  year = 2011,
  month = may,
  journal = {Physical Review Letters},
  volume = {106},
  number = {20},
  pages = {204502},
  publisher = {American Physical Society},
  doi = {10.1103/PhysRevLett.106.204502},
  urldate = {2026-01-21}
}

@article{chaElectricalTuningElastic2018,
  title = {Electrical Tuning of Elastic Wave Propagation in Nanomechanical Lattices at {{MHz}} Frequencies},
  author = {Cha, Jinwoong and Daraio, Chiara},
  year = 2018,
  month = nov,
  journal = {Nature Nanotechnology},
  volume = {13},
  number = {11},
  pages = {1016--1020},
  publisher = {Nature Publishing Group},
  issn = {1748-3395},
  doi = {10.1038/s41565-018-0252-6},
  urldate = {2023-09-12},
  copyright = {2018 The Author(s), under exclusive licence to Springer Nature Limited},
  langid = {english}
}

@article{charalampidisPhononicRogueWaves2018,
  title = {Phononic Rogue Waves},
  author = {Charalampidis, E. G. and Lee, J. and Kevrekidis, P. G. and Chong, C.},
  year = 2018,
  month = sep,
  journal = {Physical Review E},
  volume = {98},
  number = {3},
  pages = {032903},
  publisher = {American Physical Society},
  doi = {10.1103/PhysRevE.98.032903},
  urldate = {2025-12-16}
}

@article{chenLaboratoryExperimentsCounterpropagating2014,
  title = {Laboratory Experiments on Counter-Propagating Collisions of Solitary Waves. {{Part}} 1. {{Wave}} Interactions},
  author = {Chen, Yongshuai and Yeh, Harry},
  year = 2014,
  month = jun,
  journal = {Journal of Fluid Mechanics},
  volume = {749},
  pages = {577--596},
  issn = {0022-1120, 1469-7645},
  doi = {10.1017/jfm.2014.231},
  urldate = {2025-12-16},
  langid = {english}
}

@article{chernikovSolitonPulseCompression1993,
  title = {Soliton Pulse Compression in Dispersion-Decreasing Fiber},
  author = {Chernikov, S. V. and Dianov, E. M. and Richardson, D. J. and Payne, D. N.},
  year = 1993,
  month = apr,
  journal = {Optics Letters},
  volume = {18},
  number = {7},
  pages = {476--478},
  publisher = {Optica Publishing Group},
  issn = {1539-4794},
  doi = {10.1364/OL.18.000476},
  urldate = {2025-12-16},
  copyright = {\copyright{} 1993 Optical Society of America},
  langid = {english}
}

@article{coleSolitonCrystalsKerr2017,
  title = {Soliton Crystals in {{Kerr}} Resonators},
  author = {Cole, Daniel C. and Lamb, Erin S. and Del'Haye, Pascal and Diddams, Scott A. and Papp, Scott B.},
  year = 2017,
  month = oct,
  journal = {Nature Photonics},
  volume = {11},
  number = {10},
  pages = {671--676},
  publisher = {Nature Publishing Group},
  issn = {1749-4893},
  doi = {10.1038/s41566-017-0009-z},
  urldate = {2026-01-12},
  copyright = {2017 The Author(s)},
  langid = {english}
}

@article{copieSpaceTimeObservation2023,
  title = {Space--Time Observation of the Dynamics of Soliton Collisions in a Recirculating Optical Fiber Loop},
  author = {Copie, Fran{\c c}ois and Suret, Pierre and Randoux, St{\'e}phane},
  year = 2023,
  month = oct,
  journal = {Optics Communications},
  volume = {545},
  pages = {129647},
  issn = {0030-4018},
  doi = {10.1016/j.optcom.2023.129647},
  urldate = {2026-01-16}
}

@article{costeSolitaryWavesChain1997,
  title = {Solitary Waves in a Chain of Beads under {{Hertz}} Contact},
  author = {Coste, C. and Falcon, E. and Fauve, S.},
  year = 1997,
  month = nov,
  journal = {Physical Review E},
  volume = {56},
  number = {5},
  pages = {6104--6117},
  publisher = {American Physical Society},
  doi = {10.1103/PhysRevE.56.6104},
  urldate = {2025-12-16}
}

@article{daraioTunabilitySolitaryWave2006,
  title = {Tunability of Solitary Wave Properties in One-Dimensional Strongly Nonlinear Phononic Crystals},
  author = {Daraio, C. and Nesterenko, V. F. and Herbold, E. B. and Jin, S.},
  year = 2006,
  month = feb,
  journal = {Physical Review E},
  volume = {73},
  number = {2},
  pages = {026610},
  publisher = {American Physical Society},
  doi = {10.1103/PhysRevE.73.026610},
  urldate = {2025-12-16}
}

@article{denschlagGeneratingSolitonsPhase2000,
  title = {Generating {{Solitons}} by {{Phase Engineering}} of a {{Bose-Einstein Condensate}}},
  author = {Denschlag, J. and Simsarian, J. E. and Feder, D. L. and Clark, Charles W. and Collins, L. A. and Cubizolles, J. and Deng, L. and Hagley, E. W. and Helmerson, K. and Reinhardt, W. P. and Rolston, S. L. and Schneider, B. I. and Phillips, W. D.},
  year = 2000,
  month = jan,
  journal = {Science},
  volume = {287},
  number = {5450},
  pages = {97--101},
  publisher = {American Association for the Advancement of Science},
  doi = {10.1126/science.287.5450.97},
  urldate = {2026-01-12}
}

@book{dingemansWaterWavePropagation1997,
  title = {Water {{Wave Propagation Over Uneven Bottoms}} ({{In}} 2 {{Parts}})},
  author = {Dingemans, Maarten W.},
  year = 1997,
  month = jan,
  publisher = {World Scientific},
  googlebooks = {r57sCgAAQBAJ},
  isbn = {978-981-4506-58-8},
  langid = {english}
}

@article{dudleySupercontinuumGenerationPhotonic2006,
  title = {Supercontinuum Generation in Photonic Crystal Fiber},
  author = {Dudley, John M. and Genty, Go{\"e}ry and Coen, St{\'e}phane},
  year = 2006,
  month = oct,
  journal = {Reviews of Modern Physics},
  volume = {78},
  number = {4},
  pages = {1135--1184},
  publisher = {American Physical Society},
  doi = {10.1103/RevModPhys.78.1135},
  urldate = {2025-12-16}
}

@article{eggletonBraggGratingSolitons1996,
  title = {Bragg {{Grating Solitons}}},
  author = {Eggleton, Benjamin J. and Slusher, R. E. and {de Sterke}, C. Martijn and Krug, Peter A. and Sipe, J. E.},
  year = 1996,
  month = mar,
  journal = {Physical Review Letters},
  volume = {76},
  number = {10},
  pages = {1627--1630},
  publisher = {American Physical Society},
  doi = {10.1103/PhysRevLett.76.1627},
  urldate = {2025-12-16}
}

@article{emplitPicosecondStepsDark1987,
  title = {Picosecond Steps and Dark Pulses through Nonlinear Single Mode Fibers},
  author = {Emplit, P. and Hamaide, J. P. and Reynaud, F. and Froehly, C. and Barthelemy, A.},
  year = 1987,
  month = jun,
  journal = {Optics Communications},
  volume = {62},
  number = {6},
  pages = {374--379},
  issn = {0030-4018},
  doi = {10.1016/0030-4018(87)90003-4},
  urldate = {2026-02-12}
}

@article{fangOpticalTransductionRouting2016,
  title = {Optical Transduction and Routing of Microwave Phonons in Cavity-Optomechanical Circuits},
  author = {Fang, Kejie and Matheny, Matthew H. and Luan, Xingsheng and Painter, Oskar},
  year = 2016,
  month = jul,
  journal = {Nature Photonics},
  volume = {10},
  number = {7},
  pages = {489--496},
  publisher = {Nature Publishing Group},
  issn = {1749-4893},
  doi = {10.1038/nphoton.2016.107},
  urldate = {2023-10-04},
  copyright = {2016 Springer Nature Limited},
  langid = {english}
}

@article{foursaInvestigationBlackGraySoliton1996,
  title = {Investigation of {{Black-Gray Soliton Interaction}}},
  author = {Foursa, Dmitri and Emplit, Philippe},
  year = 1996,
  month = nov,
  journal = {Physical Review Letters},
  volume = {77},
  number = {19},
  pages = {4011--4014},
  publisher = {American Physical Society},
  doi = {10.1103/PhysRevLett.77.4011},
  urldate = {2026-01-19}
}

@article{fuPhononicIntegratedCircuitry2019,
  title = {Phononic Integrated Circuitry and Spin--Orbit Interaction of Phonons},
  author = {Fu, Wei and Shen, Zhen and Xu, Yuntao and Zou, Chang-Ling and Cheng, Risheng and Han, Xu and Tang, Hong X.},
  year = 2019,
  month = jun,
  journal = {Nature Communications},
  volume = {10},
  number = {1},
  pages = {2743},
  issn = {2041-1723},
  doi = {10.1038/s41467-019-10852-3},
  urldate = {2019-07-18},
  copyright = {2019 The Author(s)},
  langid = {english}
}

@article{gardeniersLPCVDSiliconrichSilicon1996,
  title = {{{LPCVD}} Silicon-rich Silicon Nitride Films for Applications in Micromechanics, Studied with Statistical Experimental Design*},
  author = {Gardeniers, J. G. E. and Tilmans, H. A. C. and Visser, C. C. G.},
  year = 1996,
  month = sep,
  journal = {Journal of Vacuum Science \& Technology A},
  volume = {14},
  number = {5},
  pages = {2879--2892},
  issn = {0734-2101},
  doi = {10.1116/1.580239},
  urldate = {2025-01-22}
}

@article{greluDissipativeSolitonsModelocked2012,
  title = {Dissipative Solitons for Mode-Locked Lasers},
  author = {Grelu, Philippe and Akhmediev, Nail},
  year = 2012,
  month = feb,
  journal = {Nature Photonics},
  volume = {6},
  number = {2},
  pages = {84--92},
  publisher = {Nature Publishing Group},
  issn = {1749-4893},
  doi = {10.1038/nphoton.2011.345},
  urldate = {2022-03-29},
  copyright = {2012 Nature Publishing Group, a division of Macmillan Publishers Limited. All Rights Reserved.},
  langid = {english}
}

@article{hatanakaPhononWaveguidesElectromechanical2014,
  title = {Phonon Waveguides for Electromechanical Circuits},
  author = {Hatanaka, D. and Mahboob, I. and Onomitsu, K. and Yamaguchi, H.},
  year = 2014,
  month = jul,
  journal = {Nature Nanotechnology},
  volume = {9},
  number = {7},
  pages = {520--524},
  issn = {1748-3395},
  doi = {10.1038/nnano.2014.107},
  urldate = {2019-07-19},
  copyright = {2014 Nature Publishing Group},
  langid = {english}
}

@article{herrTemporalSolitonsOptical2014,
  title = {Temporal Solitons in Optical Microresonators},
  author = {Herr, T. and Brasch, V. and Jost, J. D. and Wang, C. Y. and Kondratiev, N. M. and Gorodetsky, M. L. and Kippenberg, T. J.},
  year = 2014,
  month = feb,
  journal = {Nature Photonics},
  volume = {8},
  number = {2},
  pages = {145--152},
  publisher = {Nature Publishing Group},
  issn = {1749-4893},
  doi = {10.1038/nphoton.2013.343},
  urldate = {2025-04-09},
  copyright = {2013 Springer Nature Limited},
  langid = {english}
}

@article{hirschDirectionalEmissionOnchip2024,
  title = {Directional Emission in an On-Chip Acoustic Waveguide},
  author = {Hirsch, T. M. F. and Mauranyapin, N. P. and Romero, E. and Jin, X. and Harris, G. and Baker, C. G. and Bowen, W. P.},
  year = 2024,
  month = jan,
  journal = {Applied Physics Letters},
  volume = {124},
  number = {1},
  pages = {013504},
  issn = {0003-6951},
  doi = {10.1063/5.0180794},
  urldate = {2024-01-08}
}

@article{hirschTutorialMembranePhononic2026,
  title = {Tutorial: {{Membrane}} Phononic Integrated Circuits},
  shorttitle = {Tutorial},
  author = {Hirsch, Timothy M. F. and Mauranyapin, Nicolas P. and Romero, Erick and Harris, Glen I. and Jin, Xiaoya and Arora, Nishta and Bekker, Christiaan J. and Meng, Chao and Bowen, Warwick P. and Baker, Christopher G.},
  year = 2026,
  month = feb,
  journal = {Journal of Applied Physics},
  volume = {139},
  number = {8},
  pages = {081102},
  issn = {0021-8979},
  doi = {10.1063/5.0304976},
  urldate = {2026-02-25}
}

@article{hoeferDarkdarkSolitonsModulational2011,
  title = {Dark-Dark Solitons and Modulational Instability in Miscible Two-Component {{Bose-Einstein}} Condensates},
  author = {Hoefer, M. A. and Chang, J. J. and Hamner, C. and Engels, P.},
  year = 2011,
  month = oct,
  journal = {Physical Review A},
  volume = {84},
  number = {4},
  pages = {041605},
  publisher = {American Physical Society},
  doi = {10.1103/PhysRevA.84.041605},
  urldate = {2026-01-20}
}

@article{huangDarkSolitonsTheir2001,
  title = {Dark Solitons and Their Head-on Collisions in {{Bose-Einstein}} Condensates},
  author = {Huang, Guoxiang and Velarde, Manuel G. and Makarov, Valeri A.},
  year = 2001,
  month = jun,
  journal = {Physical Review A},
  volume = {64},
  number = {1},
  pages = {013617},
  publisher = {American Physical Society},
  doi = {10.1103/PhysRevA.64.013617},
  urldate = {2025-12-16}
}

@misc{jinNanomechanicalErrorCorrection2025,
  title = {Nanomechanical {{Error Correction}}},
  author = {Jin, Xiaoya and Baker, Christopher G. and Romero, Erick and Arora, Nishta and Mauranyapin, Nicolas P. and Hirsch, Timothy M. F. and Harris, Glen I. and Bowen, Warwick P.},
  year = 2025,
  month = sep,
  number = {arXiv:2509.11560},
  eprint = {2509.11560},
  primaryclass = {physics},
  publisher = {arXiv},
  doi = {10.48550/arXiv.2509.11560},
  urldate = {2025-09-23},
  archiveprefix = {arXiv}
}

@misc{karjantoNonlinearSchrodingerEquation2019,
  title = {The Nonlinear {{Schr\"odinger}} Equation: {{A}} Mathematical Model with Its Wide-Ranging Applications},
  shorttitle = {The Nonlinear {{Schr\"odinger}} Equation},
  author = {Karjanto, N.},
  year = 2019,
  month = dec,
  number = {arXiv:1912.10683},
  eprint = {1912.10683},
  primaryclass = {nlin},
  publisher = {arXiv},
  doi = {10.48550/arXiv.1912.10683},
  urldate = {2026-02-16},
  archiveprefix = {arXiv}
}

@article{khaykovichFormationMatterWaveBright2002,
  title = {Formation of a {{Matter-Wave Bright Soliton}}},
  author = {Khaykovich, L. and Schreck, F. and Ferrari, G. and Bourdel, T. and Cubizolles, J. and Carr, L. D. and Castin, Y. and Salomon, C.},
  year = 2002,
  month = may,
  journal = {Science},
  volume = {296},
  number = {5571},
  pages = {1290--1293},
  publisher = {American Association for the Advancement of Science},
  doi = {10.1126/science.1071021},
  urldate = {2026-01-21}
}

@article{kivsharDarkOpticalSolitons1998,
  title = {Dark Optical Solitons: Physics and Applications},
  shorttitle = {Dark Optical Solitons},
  author = {Kivshar, Yuri S. and {Luther-Davies}, Barry},
  year = 1998,
  month = may,
  journal = {Physics Reports},
  volume = {298},
  number = {2},
  pages = {81--197},
  issn = {0370-1573},
  doi = {10.1016/S0370-1573(97)00073-2},
  urldate = {2025-04-11}
}

@article{kurosuMechanicalKerrNonlinearity2020,
  title = {Mechanical {{Kerr Nonlinearity}} of {{Wave Propagation}} in an {{On-Chip Nanoelectromechanical Waveguide}}},
  author = {Kurosu, M. and Hatanaka, D. and Yamaguchi, H.},
  year = 2020,
  month = jan,
  journal = {Physical Review Applied},
  volume = {13},
  number = {1},
  pages = {014056},
  issn = {2331-7019},
  doi = {10.1103/PhysRevApplied.13.014056},
  urldate = {2021-04-27},
  langid = {english}
}

@article{kurosuOnchipTemporalFocusing2018,
  title = {On-Chip Temporal Focusing of Elastic Waves in a Phononic Crystal Waveguide},
  author = {Kurosu, M. and Hatanaka, D. and Onomitsu, K. and Yamaguchi, H.},
  year = 2018,
  month = apr,
  journal = {Nature Communications},
  volume = {9},
  number = {1},
  pages = {1331},
  publisher = {Nature Publishing Group},
  issn = {2041-1723},
  doi = {10.1038/s41467-018-03726-7},
  urldate = {2020-07-01},
  copyright = {2018 The Author(s)},
  langid = {english}
}

@incollection{lifshitzNonlinearDynamicsNanomechanical2008,
  title = {Nonlinear {{Dynamics}} of {{Nanomechanical}} and {{Micromechanical Resonators}}},
  booktitle = {Review of {{Nonlinear Dynamics}} and {{Complexity}}},
  author = {Lifshitz, Ron and Cross, M.C.},
  year = 2008,
  publisher = {WILEY-VCH Verlag GmbH \& Co. KGaA},
  address = {Weinheim},
  isbn = {978-3-527-40729-3}
}

@article{lugiatoSpatialDissipativeStructures1987,
  title = {Spatial {{Dissipative Structures}} in {{Passive Optical Systems}}},
  author = {Lugiato, L. A. and Lefever, R.},
  year = 1987,
  month = may,
  journal = {Physical Review Letters},
  volume = {58},
  number = {21},
  pages = {2209--2211},
  publisher = {American Physical Society},
  doi = {10.1103/PhysRevLett.58.2209},
  urldate = {2026-01-17}
}

@article{maitreDarkSolitonMoleculesExcitonPolariton2020,
  title = {Dark-{{Soliton Molecules}} in an {{Exciton-Polariton Superfluid}}},
  author = {Ma{\^i}tre, Anne and Lerario, Giovanni and Medeiros, Adri{\`a} and Claude, Ferdinand and Glorieux, Quentin and Giacobino, Elisabeth and Pigeon, Simon and Bramati, Alberto},
  year = 2020,
  month = nov,
  journal = {Physical Review X},
  volume = {10},
  number = {4},
  pages = {041028},
  issn = {2160-3308},
  doi = {10.1103/PhysRevX.10.041028},
  urldate = {2025-12-16},
  langid = {english}
}

@article{schulzWignerCrystalOne1993,
  title = {Wigner Crystal in One Dimension},
  author = {Schulz, H. J.},
  year = 1993,
  month = sep,
  journal = {Physical Review Letters},
  volume = {71},
  number = {12},
  pages = {1864--1867},
  publisher = {American Physical Society},
  doi = {10.1103/PhysRevLett.71.1864},
  urldate = {2026-03-13}
}

@article{landigMeasuringDynamicStructure2015,
  title = {Measuring the Dynamic Structure Factor of a Quantum Gas Undergoing a Structural Phase Transition},
  author = {Landig, Renate and Brennecke, Ferdinand and Mottl, Rafael and Donner, Tobias and Esslinger, Tilman},
  year = 2015,
  month = may,
  journal = {Nature Communications},
  volume = {6},
  number = {1},
  pages = {7046},
  publisher = {Nature Publishing Group},
  issn = {2041-1723},
  doi = {10.1038/ncomms8046},
  urldate = {2026-03-13},
  copyright = {2015 The Author(s)},
  langid = {english}
}

@article{karpovDynamicsSolitonCrystals2019,
  title = {Dynamics of Soliton Crystals in Optical Microresonators},
  author = {Karpov, Maxim and Pfeiffer, Martin H. P. and Guo, Hairun and Weng, Wenle and Liu, Junqiu and Kippenberg, Tobias J.},
  year = 2019,
  month = oct,
  journal = {Nature Physics},
  volume = {15},
  number = {10},
  pages = {1071--1077},
  publisher = {Nature Publishing Group},
  issn = {1745-2481},
  doi = {10.1038/s41567-019-0635-0},
  urldate = {2026-03-04},
  copyright = {2019 The Author(s), under exclusive licence to Springer Nature Limited},
  langid = {english}
}

@article{mauranyapinTunnelingTransverseAcoustic2021,
  title = {Tunneling of {{Transverse Acoustic Waves}} on a {{Silicon Chip}}},
  author = {Mauranyapin, Nicolas P. and Romero, Erick and Kalra, Rachpon and Harris, Glen and Baker, Christopher G. and Bowen, Warwick P.},
  year = 2021,
  month = may,
  journal = {Physical Review Applied},
  volume = {15},
  number = {5},
  pages = {054036},
  publisher = {American Physical Society},
  doi = {10.1103/PhysRevApplied.15.054036},
  urldate = {2021-06-02}
}

@article{mayorGigahertzPhononicIntegrated2021,
  title = {Gigahertz {{Phononic Integrated Circuits}} on {{Thin-Film Lithium Niobate}} on {{Sapphire}}},
  author = {Mayor, Felix M. and Jiang, Wentao and Sarabalis, Christopher J. and McKenna, Timothy P. and Witmer, Jeremy D. and {Safavi-Naeini}, Amir H.},
  year = 2021,
  month = jan,
  journal = {Physical Review Applied},
  volume = {15},
  number = {1},
  pages = {014039},
  publisher = {American Physical Society},
  doi = {10.1103/PhysRevApplied.15.014039},
  urldate = {2023-09-13}
}

@article{midtvedtNonlinearPhononicsUsing2014,
  title = {Nonlinear Phononics Using Atomically Thin Membranes},
  author = {Midtvedt, Daniel and Isacsson, Andreas and Croy, Alexander},
  year = 2014,
  month = sep,
  journal = {Nature Communications},
  volume = {5},
  number = {1},
  pages = {4838},
  publisher = {Nature Publishing Group},
  issn = {2041-1723},
  doi = {10.1038/ncomms5838},
  urldate = {2026-02-03},
  copyright = {2014 Springer Nature Limited},
  langid = {english}
}

@article{miyazawaRogueSolitaryWaves2022,
  title = {Rogue and Solitary Waves in Coupled Phononic Crystals},
  author = {Miyazawa, Y. and Chong, C. and Kevrekidis, P. G. and Yang, J.},
  year = 2022,
  month = mar,
  journal = {Physical Review E},
  volume = {105},
  number = {3},
  pages = {034202},
  publisher = {American Physical Society},
  doi = {10.1103/PhysRevE.105.034202},
  urldate = {2026-01-12}
}

@article{mollenauerExperimentalObservationPicosecond1980,
  title = {Experimental {{Observation}} of {{Picosecond Pulse Narrowing}} and {{Solitons}} in {{Optical Fibers}}},
  author = {Mollenauer, L. F. and Stolen, R. H. and Gordon, J. P.},
  year = 1980,
  month = sep,
  journal = {Physical Review Letters},
  volume = {45},
  number = {13},
  pages = {1095--1098},
  publisher = {American Physical Society},
  doi = {10.1103/PhysRevLett.45.1095},
  urldate = {2026-01-21}
}

@book{nayfehNonlinearOscillations2004,
  title = {Nonlinear Oscillations},
  author = {Nayfeh, Ali Hasan and Mook, Dean T.},
  year = 2004,
  series = {Physics Textbook},
  edition = {Wiley classics library ed. 1995, [Nachdr.]},
  publisher = {Wiley},
  address = {Weinheim},
  isbn = {978-0-471-12142-8},
  langid = {english}
}

@phdthesis{norteNanofabricationOnChipOptical2014,
  title = {Nanofabrication for {{On-Chip Optical Levitation}}, {{Atom-Trapping}}, and {{Superconducting Quantum Circuits}}},
  author = {Norte, Richard Alexander},
  year = 2014,
  month = nov,
  doi = {10.7907/Z9WS8R61},
  urldate = {2024-06-19},
  collaborator = {{Institute for Quantum Information and Matter} and {IQIM} and {Kavli Nanoscience Institute}},
  copyright = {No commercial reproduction, distribution, display or performance rights in this work are provided.},
  langid = {english},
  school = {California Institute of Technology}
}

@article{okawachiOctavespanningFrequencyComb2011,
  title = {Octave-Spanning Frequency Comb Generation in a Silicon Nitride Chip},
  author = {Okawachi, Yoshitomo and Saha, Kasturi and Levy, Jacob S. and Wen, Y. Henry and Lipson, Michal and Gaeta, Alexander L.},
  year = 2011,
  month = sep,
  journal = {Optics Letters},
  volume = {36},
  number = {17},
  pages = {3398--3400},
  publisher = {Optica Publishing Group},
  issn = {1539-4794},
  doi = {10.1364/OL.36.003398},
  urldate = {2026-01-21},
  copyright = {\copyright{} 2011 Optical Society of America},
  langid = {english}
}

@book{onoratoRogueShockWaves2016,
  title = {Rogue and {{Shock Waves}} in {{Nonlinear Dispersive Media}}},
  editor = {Onorato, Miguel and Resitori, Stefania and Baronio, Fabio},
  year = 2016,
  series = {Lecture {{Notes}} in {{Physics}}},
  volume = {926},
  publisher = {Springer International Publishing},
  address = {Cham},
  doi = {10.1007/978-3-319-39214-1},
  urldate = {2025-10-11},
  copyright = {http://www.springer.com/tdm},
  isbn = {978-3-319-39212-7 978-3-319-39214-1},
  langid = {english}
}

@article{patilReviewExploitingNonlinearity2022,
  title = {Review of Exploiting Nonlinearity in Phononic Materials to Enable Nonlinear Wave Responses},
  author = {Patil, Ganesh U. and Matlack, Kathryn H.},
  year = 2022,
  month = jan,
  journal = {Acta Mechanica},
  volume = {233},
  number = {1},
  pages = {1--46},
  issn = {1619-6937},
  doi = {10.1007/s00707-021-03089-z},
  urldate = {2025-12-16},
  langid = {english}
}

@article{reevesNonlinearWaveDynamics2025,
  title = {Nonlinear Wave Dynamics on a Chip},
  author = {Reeves, Matthew T. and Wasserman, Walter W. and Harrison, Raymond A. and Marinkovi{\'c}, Igor and Luu, Nicole and Sawadsky, Andreas and Sfendla, Yasmine L. and Harris, Glen I. and Bowen, Warwick P. and Baker, Christopher G.},
  year = 2025,
  month = oct,
  journal = {Science},
  volume = {390},
  number = {6771},
  pages = {371--376},
  publisher = {American Association for the Advancement of Science},
  doi = {10.1126/science.ady3042},
  urldate = {2025-12-16}
}

@book{remoissenetWavesCalledSolitons1999,
  title = {Waves {{Called Solitons}}},
  author = {Remoissenet, Michel},
  year = 1999,
  series = {Advanced {{Texts}} in {{Physics}}},
  publisher = {Springer Berlin Heidelberg},
  address = {Berlin, Heidelberg},
  doi = {10.1007/978-3-662-03790-4},
  urldate = {2026-01-13},
  copyright = {http://www.springer.com/tdm},
  isbn = {978-3-642-08519-2 978-3-662-03790-4}
}

@article{meyerWignerCrystalPhysics2008,
  title = {Wigner Crystal Physics in Quantum Wires},
  author = {Meyer, Julia S and Matveev, K A},
  year = 2008,
  month = dec,
  journal = {Journal of Physics: Condensed Matter},
  volume = {21},
  number = {2},
  pages = {023203},
  issn = {0953-8984},
  doi = {10.1088/0953-8984/21/2/023203},
  urldate = {2026-03-13},
  langid = {english}
}

@article{romeroAcousticallyDrivenSinglefrequency2024,
  title = {Acoustically Driven Single-Frequency Mechanical Logic},
  author = {Romero, Erick and Mauranyapin, Nicolas P. and Hirsch, Timothy M.F. and Kalra, Rachpon and Baker, Christopher G. and Harris, Glen I. and Bowen, Warwick P.},
  year = 2024,
  month = may,
  journal = {Physical Review Applied},
  volume = {21},
  number = {5},
  pages = {054029},
  publisher = {American Physical Society},
  doi = {10.1103/PhysRevApplied.21.054029},
  urldate = {2024-06-03}
}

@article{romeroPropagationImagingMechanical2019,
  title = {Propagation and {{Imaging}} of {{Mechanical Waves}} in a {{Highly Stressed Single-Mode Acoustic Waveguide}}},
  author = {Romero, E. and Kalra, R. and Mauranyapin, N.P. and Baker, C.G. and Meng, C. and Bowen, W.P.},
  year = 2019,
  month = jun,
  journal = {Physical Review Applied},
  volume = {11},
  number = {6},
  pages = {064035},
  publisher = {American Physical Society},
  doi = {10.1103/PhysRevApplied.11.064035},
  urldate = {2021-03-22}
}

@article{samantaTuningGeometricNonlinearity2018,
  title = {Tuning of Geometric Nonlinearity in Ultrathin Nanoelectromechanical Systems},
  author = {Samanta, Chandan and Arora, Nishta and Naik, A. K.},
  year = 2018,
  month = sep,
  journal = {Applied Physics Letters},
  volume = {113},
  number = {11},
  pages = {113101},
  issn = {0003-6951},
  doi = {10.1063/1.5026775},
  urldate = {2025-06-16}
}

@article{schmidDampingMechanismsSingleclamped2008,
  title = {Damping Mechanisms of Single-Clamped and Prestressed Double-Clamped Resonant Polymer Microbeams},
  author = {Schmid, S. and Hierold, C.},
  year = 2008,
  month = nov,
  journal = {Journal of Applied Physics},
  volume = {104},
  number = {9},
  pages = {093516},
  issn = {0021-8979},
  doi = {10.1063/1.3008032},
  urldate = {2025-12-14}
}

@book{schmidFundamentalsNanomechanicalResonators2016,
  title = {Fundamentals of {{Nanomechanical Resonators}}},
  author = {Schmid, Silvan and Villanueva, Luis Guillermo and Roukes, Michael Lee},
  year = 2016,
  publisher = {Springer International Publishing},
  address = {Cham},
  doi = {10.1007/978-3-319-28691-4},
  urldate = {2021-03-22},
  isbn = {978-3-319-28689-1 978-3-319-28691-4},
  langid = {english}
}

@article{sementilliLowDissipationNanomechanicalDevices2025,
  title = {Low-{{Dissipation Nanomechanical Devices}} from {{Monocrystalline Silicon Carbide}}},
  author = {Sementilli, Leo and Lukin, Daniil M. and Lee, Hope and Yang, Joshua and Romero, Erick and Vu{\v c}kovi{\'c}, Jelena and Bowen, Warwick P},
  year = 2025,
  month = apr,
  journal = {Nano Letters},
  publisher = {American Chemical Society},
  issn = {1530-6984},
  doi = {10.1021/acs.nanolett.4c06475},
  urldate = {2025-04-07}
}

@article{streckerFormationPropagationMatterwave2002,
  title = {Formation and Propagation of Matter-Wave Soliton Trains},
  author = {Strecker, Kevin E. and Partridge, Guthrie B. and Truscott, Andrew G. and Hulet, Randall G.},
  year = 2002,
  month = may,
  journal = {Nature},
  volume = {417},
  number = {6885},
  pages = {150--153},
  publisher = {Nature Publishing Group},
  issn = {1476-4687},
  doi = {10.1038/nature747},
  urldate = {2026-02-19},
  copyright = {2002 Macmillan Magazines Ltd.},
  langid = {english}
}

@article{suhMicroresonatorSolitonDualcomb2016,
  title = {Microresonator Soliton Dual-Comb Spectroscopy},
  author = {Suh, Myoung-Gyun and Yang, Qi-Fan and Yang, Ki Youl and Yi, Xu and Vahala, Kerry J.},
  year = 2016,
  month = nov,
  journal = {Science},
  volume = {354},
  number = {6312},
  pages = {600--603},
  publisher = {American Association for the Advancement of Science},
  doi = {10.1126/science.aah6516},
  urldate = {2026-02-06}
}

@article{thurstonCollisionsDarkSolitons1991,
  title = {Collisions of Dark Solitons in Optical Fibers},
  author = {Thurston, R. N. and Weiner, Andrew M.},
  year = 1991,
  month = feb,
  journal = {JOSA B},
  volume = {8},
  number = {2},
  pages = {471--477},
  publisher = {Optica Publishing Group},
  issn = {1520-8540},
  doi = {10.1364/JOSAB.8.000471},
  urldate = {2026-01-12},
  copyright = {\copyright{} 1991 Optical Society of America},
  langid = {english}
}

@article{trilloExperimentalObservationTheoretical2016,
  title = {Experimental {{Observation}} and {{Theoretical Description}} of {{Multisoliton Fission}} in {{Shallow Water}}},
  author = {Trillo, S. and Deng, G. and Biondini, G. and Klein, M. and Clauss, G. F. and Chabchoub, A. and Onorato, M.},
  year = 2016,
  month = sep,
  journal = {Physical Review Letters},
  volume = {117},
  number = {14},
  pages = {144102},
  publisher = {American Physical Society},
  doi = {10.1103/PhysRevLett.117.144102},
  urldate = {2026-03-03}
}

@article{wangHexagonalBoronNitride2019,
  title = {Hexagonal {{Boron Nitride Phononic Crystal Waveguides}}},
  author = {Wang, Yanan and Lee, Jaesung and Zheng, Xu-Qian and Xie, Yong and Feng, Philip X.-L.},
  year = 2019,
  month = dec,
  journal = {ACS Photonics},
  volume = {6},
  number = {12},
  pages = {3225--3232},
  publisher = {American Chemical Society},
  doi = {10.1021/acsphotonics.9b01094},
  urldate = {2025-09-24}
}

@article{weinerExperimentalObservationFundamental1988,
  title = {Experimental {{Observation}} of the {{Fundamental Dark Soliton}} in {{Optical Fibers}}},
  author = {Weiner, A. M. and Heritage, J. P. and Hawkins, R. J. and Thurston, R. N. and Kirschner, E. M. and Leaird, D. E. and Tomlinson, W. J.},
  year = 1988,
  month = nov,
  journal = {Physical Review Letters},
  volume = {61},
  number = {21},
  pages = {2445--2448},
  publisher = {American Physical Society},
  doi = {10.1103/PhysRevLett.61.2445},
  urldate = {2026-01-12}
}

@article{wellerExperimentalObservationOscillating2008,
  title = {Experimental {{Observation}} of {{Oscillating}} and {{Interacting Matter Wave Dark Solitons}}},
  author = {Weller, A. and Ronzheimer, J. P. and Gross, C. and Esteve, J. and Oberthaler, M. K. and Frantzeskakis, D. J. and Theocharis, G. and Kevrekidis, P. G.},
  year = 2008,
  month = sep,
  journal = {Physical Review Letters},
  volume = {101},
  number = {13},
  pages = {130401},
  publisher = {American Physical Society},
  doi = {10.1103/PhysRevLett.101.130401},
  urldate = {2025-12-16}
}

@article{wuControlledGenerationDark2002,
  title = {Controlled {{Generation}} of {{Dark Solitons}} with {{Phase Imprinting}}},
  author = {Wu, Biao and Liu, Jie and Niu, Qian},
  year = 2002,
  month = jan,
  journal = {Physical Review Letters},
  volume = {88},
  number = {3},
  pages = {034101},
  publisher = {American Physical Society},
  doi = {10.1103/PhysRevLett.88.034101},
  urldate = {2026-02-01}
}

@article{xiSoftclampedTopologicalWaveguide2025,
  title = {A Soft-Clamped Topological Waveguide for Phonons},
  author = {Xi, Xiang and Chernobrovkin, Ilia and Ko{\v s}ata, Jan and Kristensen, Mads B. and Langman, Eric and S{\o}rensen, Anders S. and Zilberberg, Oded and Schliesser, Albert},
  year = 2025,
  month = jun,
  journal = {Nature},
  volume = {642},
  number = {8069},
  pages = {947--953},
  publisher = {Nature Publishing Group},
  issn = {1476-4687},
  doi = {10.1038/s41586-025-09092-x},
  urldate = {2025-07-07},
  copyright = {2025 The Author(s), under exclusive licence to Springer Nature Limited},
  langid = {english}
}

@article{yangBroadbandDispersionengineeredMicroresonator2016,
  title = {Broadband Dispersion-Engineered Microresonator on a Chip},
  author = {Yang, Ki Youl and Beha, Katja and Cole, Daniel C. and Yi, Xu and Del'Haye, Pascal and Lee, Hansuek and Li, Jiang and Oh, Dong Yoon and Diddams, Scott A. and Papp, Scott B. and Vahala, Kerry J.},
  year = 2016,
  month = may,
  journal = {Nature Photonics},
  volume = {10},
  number = {5},
  pages = {316--320},
  publisher = {Nature Publishing Group},
  issn = {1749-4893},
  doi = {10.1038/nphoton.2016.36},
  urldate = {2026-01-21},
  copyright = {2016 Springer Nature Limited},
  langid = {english}
}

@article{yiImagingSolitonDynamics2018,
  title = {Imaging Soliton Dynamics in Optical Microcavities},
  author = {Yi, Xu and Yang, Qi-Fan and Yang, Ki Youl and Vahala, Kerry},
  year = 2018,
  month = sep,
  journal = {Nature Communications},
  volume = {9},
  number = {1},
  pages = {3565},
  publisher = {Nature Publishing Group},
  issn = {2041-1723},
  doi = {10.1038/s41467-018-06031-5},
  urldate = {2025-12-16},
  copyright = {2018 The Author(s)},
  langid = {english}
}

@article{yuenNonlinearDeepWater1975,
  title = {Nonlinear Deep Water Waves: {{Theory}} and Experiment},
  shorttitle = {Nonlinear Deep Water Waves},
  author = {Yuen, Henry C. and Lake, Bruce M.},
  year = 1975,
  month = aug,
  journal = {The Physics of Fluids},
  volume = {18},
  number = {8},
  pages = {956--960},
  issn = {0031-9171},
  doi = {10.1063/1.861268},
  urldate = {2026-01-21}
}

@article{zabuskyInteractionSolitonsCollisionless1965,
  title = {Interaction of ``{{Solitons}}" in a {{Collisionless Plasma}} and the {{Recurrence}} of {{Initial States}}},
  author = {Zabusky, N. J. and Kruskal, M. D.},
  year = 1965,
  month = aug,
  journal = {Physical Review Letters},
  volume = {15},
  number = {6},
  pages = {240--243},
  issn = {0031-9007},
  doi = {10.1103/PhysRevLett.15.240},
  urldate = {2026-01-12},
  copyright = {http://link.aps.org/licenses/aps-default-license},
  langid = {english}
}

@article{zakharovExactTheoryTwoDimensional1972,
  title = {Exact {{Theory}} of {{Two-Dimensional Self-Focusing}} and {{One-Dimensional Self-Modulation}} of {{Waves}} in {{Nonlinear Media}}},
  author = {Zakharov, V.E. and Shabat, A.B.},
  year = 1972,
  month = jan,
  journal = {Soviet Journal of Theoretical and Experimental Physics},
  volume = {34},
  number = {1}
}

@article{zakharovKineticEquationsSolitons1971,
  title = {Kinetic {{Equations}} for {{Solitons}}},
  author = {Zakharov, V. E.},
  year = 1971,
  month = sep,
  journal = {Soviet Journal of Theoretical and Experimental Physics},
  volume = {33},
  number = {3},
  pages = {538--541},
  publisher = {Soviet Physics}
}

@article{zhangGigahertzTopologicalValley2022,
  title = {Gigahertz Topological Valley {{Hall}} Effect in Nanoelectromechanical Phononic Crystals},
  author = {Zhang, Qicheng and Lee, Daehun and Zheng, Lu and Ma, Xuejian and Meyer, Shawn I. and He, Li and Ye, Han and Gong, Ze and Zhen, Bo and Lai, Keji and Johnson, A. T. Charlie},
  year = 2022,
  month = mar,
  journal = {Nature Electronics},
  volume = {5},
  number = {3},
  pages = {157--163},
  publisher = {Nature Publishing Group},
  issn = {2520-1131},
  doi = {10.1038/s41928-022-00732-y},
  urldate = {2022-04-01},
  copyright = {2022 The Author(s), under exclusive licence to Springer Nature Limited},
  langid = {english}
}

@article{zhangProgrammableRobustStatic2019,
  title = {Programmable and Robust Static Topological Solitons in Mechanical Metamaterials},
  author = {Zhang, Yafei and Li, Bo and Zheng, Q. S. and Genin, Guy M. and Chen, C. Q.},
  year = 2019,
  month = dec,
  journal = {Nature Communications},
  volume = {10},
  number = {1},
  pages = {5605},
  publisher = {Nature Publishing Group},
  issn = {2041-1723},
  doi = {10.1038/s41467-019-13546-y},
  urldate = {2026-01-12},
  copyright = {2019 The Author(s)},
  langid = {english}
}

@article{shapir_imaging_2019,
	title = {Imaging the electronic {Wigner} crystal in one dimension},
	volume = {364},
	url = {https://www.science.org/doi/10.1126/science.aat0905},
	doi = {10.1126/science.aat0905},
	number = {6443},
	urldate = {2026-03-06},
	journal = {Science},
	publisher = {American Association for the Advancement of Science},
	author = {Shapir, I. and Hamo, A. and Pecker, S. and Moca, C. P. and Legeza, \"O. and Zarand, G. and Ilani, S.},
	month = may,
	year = {2019},
	pages = {870--875},
}

@article{deshpande_one-dimensional_2008,
	title = {The one-dimensional {Wigner} crystal in carbon nanotubes},
	volume = {4},
	copyright = {2008 Springer Nature Limited},
	issn = {1745-2481},
	url = {https://www.nature.com/articles/nphys895},
	doi = {10.1038/nphys895},
	number = {4},
	urldate = {2026-03-06},
	journal = {Nature Physics},
	publisher = {Nature Publishing Group},
	author = {Deshpande, Vikram V. and Bockrath, Marc},
	month = apr,
	year = {2008},
	pages = {314--318},
}

@article{redor2019experimental,
  title={Experimental evidence of a hydrodynamic soliton gas},
  author={Redor, Ivan and Barth{\'e}lemy, Eric and Michallet, Herv{\'e} and Onorato, Miguel and Mordant, Nicolas},
  journal={Physical review letters},
  volume={122},
  number={21},
  pages={214502},
  year={2019},
  publisher={APS}
}

@misc{neely_melting_2024,
	title = {Melting of a vortex matter {Wigner} crystal},
	url = {http://arxiv.org/abs/2402.09920},
	doi = {10.48550/arXiv.2402.09920},
	urldate = {2026-03-11},
	publisher = {arXiv},
	author = {Neely, Tyler W. and Gauthier, Guillaume and Glasspool, Charles and Davis, Matthew J. and Reeves, Matthew T.},
	month = sep,
	year = {2024},
	note = {arXiv:2402.09920 [cond-mat]},
}

@article{wigner_interaction_1934,
	title = {On the {Interaction} of {Electrons} in {Metals}},
	volume = {46},
	copyright = {http://link.aps.org/licenses/aps-default-license},
	issn = {0031-899X},
	url = {https://link.aps.org/doi/10.1103/PhysRev.46.1002},
	doi = {10.1103/PhysRev.46.1002},
	number = {11},
	urldate = {2026-03-06},
	journal = {Physical Review},
	author = {Wigner, E.},
	month = dec,
	year = {1934},
	pages = {1002--1011},
}

@book{hansen_theory_2007,
	address = {Amsterdam Boston},
	edition = {3rd ed},
	title = {Theory of simple liquids},
	isbn = {978-0-12-370535-8},
	publisher = {Elsevier / Academic Press},
	author = {Hansen, Jean-Pierre and McDonald, Ian Ranald},
	year = {2007},
}


\clearpage

\onecolumngrid  

\renewcommand{\thetable}{S\arabic{table}}%
\renewcommand{\figurename}{Fig.}
\renewcommand{\thefigure}{S\arabic{figure}}
\renewcommand{\thepage}{S\arabic{page}}
\renewcommand{\theequation}{S\arabic{equation}} 

\begin{center}
    {\Large\bfseries Supplementary Information for ``Solitary waves in a phononic integrated circuit"}
    \vspace{3mm}
    
    Timothy M.F. Hirsch, Xiaoya Jin, Nicolas P. Mauranyapin, Nishta Arora, Erick Romero, Matthew Reeves, Glen I. Harris, Warwick P. Bowen, and Christopher G. Baker
\end{center}

\setcounter{equation}{0} 
\setcounter{section}{0}
\setcounter{table}{0}
\setcounter{page}{1}
\setcounter{figure}{0}

\section{Onset of nonlinearity}
\label{sec:Supp-onset-of-nonlinearity}

\begin{figure*}
    \centering
    \includegraphics[width=1.0\linewidth]{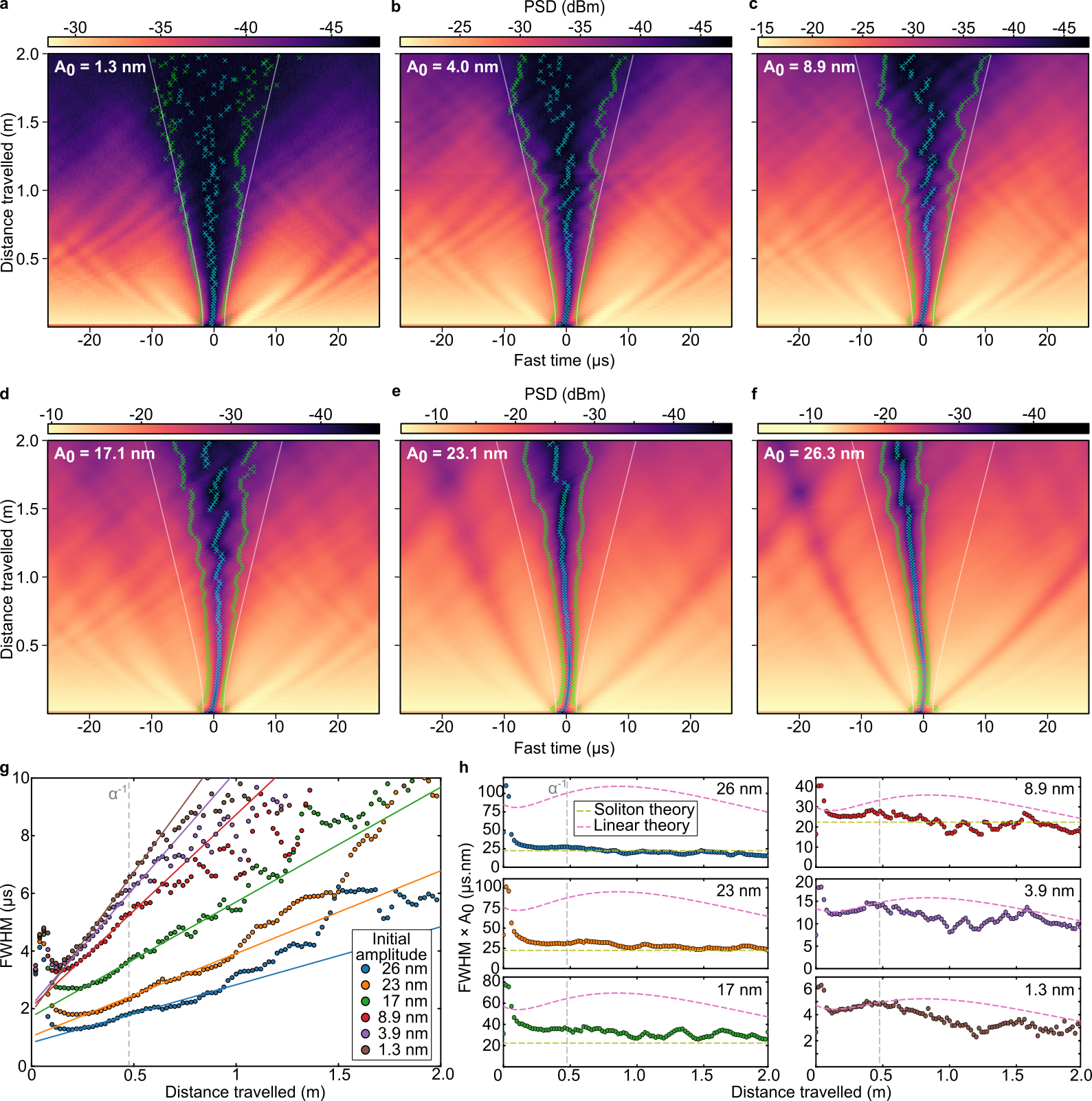}
    \caption{\textbf{Gradual onset of nonlinearity and solitary wave behaviour: identical tanh input pulses at different drive amplitudes.} \textbf{a-f} Experimentally measured evolution of tanh dark pulses of identical width and depth, with varying background amplitude ($A_0$ values in order: 1.3 nm; 3.9 nm;  8.9 nm; 17.1 nm;  23.1 nm and  26.3 nm). Blue markers: minimum photocurrent for each round trip. Green markers: $-6\,\mathrm{dB}$ from median photocurrent for each round trip.  White lines: expected pulse width assuming linear dispersive theory (Eq.~\eqref{eqn:linear-width-increase}). First and last panels ($A_0=1.3\,\mathrm{nm}$ and $26.3\,\mathrm{nm}$) are the same data as in Fig.~\ref{fig:Fig3}\textbf{c-d} of the main text. \textbf{g}  Circles: measured full width at half maximum (FWHM) versus propagation distance for pulses with different initial background amplitudes, shown in panels \textbf{a} through \textbf{f}. Lines: linear fits to the data between $8\,\,\mathrm{cm}$ and $60\,\mathrm{cm}$ of distance travelled. \textbf{h} Product of the measured FWHMs and the measured instantaneous background amplitude, versus propagation distance, for different initial background amplitudes. Pink and yellow dashed line respectively represent linear dispersive wave theory (Eq.\eqref{Eqamplitudetimeswidthlineartheory}) and soliton theory (Eq.~\eqref{eqamplitudetimeswisthsoliton}).}
    \label{fig:ExtFig-DarkSolAllVoltages+widthxamp}
\end{figure*}

In Fig.~\ref{fig:ExtFig-DarkSolAllVoltages+widthxamp} we show the same pulse evolution data as in Fig.~\ref{fig:Fig3} of the main text, alongside additional measurements at intermediate drive voltages. These data illustrate the gradual onset of nonlinear behaviour, such as pulse compression, as the initial background wave amplitude ramps up from $1.3\,\mathrm{nm}$ to $26.3\,\mathrm{nm}$ (Fig.~\ref{fig:ExtFig-DarkSolAllVoltages+widthxamp}\textbf{a} through \textbf{f}). From the measured  solitary wave trajectories we measure the pulse full width at half maximum (FWHM) after each round trip: green crosses in each row of the heatmaps mark the locations where the photocurrent PSD (proportional to the acoustic PSD) decreases by $6\,\mathrm{dB}$ from the median amplitude. Fig.~\ref{fig:ExtFig-DarkSolAllVoltages+widthxamp}\textbf{g} plots the measured FWHM versus propagation distance for the pulses with  different initial background amplitudes shown in panels \textbf{a} through \textbf{f}. The initial increase in FWHM observed within the first 10 cm of travel shown in Fig~\ref{fig:ExtFig-DarkSolAllVoltages+widthxamp}\textbf{g} is caused by transient effects (shedding of radiative components and/or grey solitary waves), and not indicative of the true width of the central solitary wave. To guide the eye, we provide lines of best fit for the data after this transient period and before loss and self-interference with shedded wave components significantly increase the noisiness of the data (specifically between $8\,\mathrm{cm}$ and $60\,\mathrm{cm}$ of travel distance). Under purely linear dispersion one would expect the amplitude to have no effect on the rate of change of the pulse width, so the markedly different slopes in Fig~\ref{fig:ExtFig-DarkSolAllVoltages+widthxamp}\textbf{a} are indicative of nonlinear effects. Specifically, the slopes decrease as the wave amplitude increases, corresponding to the self phase modulation (SPM) arising from Duffing nonlinearity counteracting the group velocity dispersion (GVD) and slowing the rate of pulse broadening.

In Fig.~\ref{fig:ExtFig-DarkSolAllVoltages+widthxamp}\textbf{h} we plot the experimentally measured  pulse width-background amplitude product $T_0A_0$ for each round trip of the solitary wave (measuring $A_0$ as the median amplitude outside of the fast time interval $(-20\,\upmu s,+20\,\upmu s)$), and compare it to both soliton and linear wave theory. In soliton theory this product is an important conserved quantity~\cite{agrawalNonlinearFiberOptics2019, kivsharDarkOpticalSolitons1998}:
\begin{equation}
    {T_0A_0}_{\mathrm{\,soliton\, theory}}=\sqrt{\frac{|k_2|}{\xi}}
    \label{eqamplitudetimeswisthsoliton}
\end{equation}
This width-amplitude product evaluates to approximately $22.4\,\mathrm{\upmu s.nm}$ for our values of $k_2$ and $\xi$ (see sections~\ref{supp:k2-k3-derivation} and~\ref{supp:xi-derivation}), and is represented by the horizontal yellow dashed line in Fig.~\ref{fig:ExtFig-DarkSolAllVoltages+widthxamp}\textbf{h}. For comparison, in the pink dashed line we plot the width-amplitude product expected from purely linear wave propagation. In that case the temporal width of the pulse would increase as~\cite{agrawalNonlinearFiberOptics2019}:
\begin{equation}
    \label{eqn:linear-width-increase}
    {T_0}_{\mathrm{\,linear\, theory}}(y) = T_0(y=0)\times\sqrt{1 + \left(\frac{y}{L_D}\right)^2}
\end{equation}
where the dispersive length $L_D=T_0^2/|k_2|$ is evaluated using the initial width. Moreover, the instantaneous amplitude $A(y)$ follows $A(y)=A_0\exp(-\frac{\alpha}{2}y)$ where $A_0$ is the initial background amplitude and $\alpha$ is the energy loss coefficient appearing in the NLSE. Combined, these two contributions lead to a different width-amplitude product, which in the linear (dispersive damped) wave theory takes the form:

\begin{equation}
    {T_0A_0 }_{\mathrm{\,linear\, theory}}(y) = T_0A_0(y=0)\times\sqrt{1+\left(\frac{y}{L_D}\right)^2}\times\exp\left(-\frac{\alpha}{2}y\right)
    \label{Eqamplitudetimeswidthlineartheory}
\end{equation}

In Fig.~\ref{fig:ExtFig-DarkSolAllVoltages+widthxamp}\textbf{h} the linear theory (pink dashed line) uses Eq.~\eqref{Eqamplitudetimeswidthlineartheory} for the width-amplitude product, taking the initial measured width and amplitude for the $y=0$ values. There is an illustrative trend in Fig.~\ref{fig:ExtFig-DarkSolAllVoltages+widthxamp}\textbf{h} in the region of propagation distances less than one decay length ($\sim48$cm): as the amplitude increases from 1.3nm from 17nm the width-amplitude product also increases; however, as the amplitude increases further from 17nm to 26nm, the width-amplitude product decreases and converges to the theoretical value of $\sqrt{|k_2|/\xi}$. This changing response at increased amplitude indicates the change from the linear regime to the nonlinear regime. Supporting this conclusion, we observe that a growing deviation from the linear theory begins at the same amplitudes. We note that at the highest amplitude our injected solitary wave matches soliton theory out to approximately $1.5\,\mathrm{m}$,  a propagation distance equal to  approximately $26\,000$ carrier wavelengths (Fig.~\ref{fig:ExtFig-DarkSolAllVoltages+widthxamp}\textbf{h}-top left panel). The constant width-amplitude product over this broad range highlights  how the solitary wave adiabatically evolves throughout the propagation, gradually broadening as the background amplitude decreases due to acoustic losses. This evolution ensures that the precise balance between dispersion and nonlinearity is continuously maintained, such that, at any given position $y$ throughout the first 1.5 meters of travel, the solitary wave properties precisely match that of a black soliton with background amplitude $A_0(y)$. This excellent matching to soliton theory is further evidenced in Fig.~\ref{fig:SuppFig-soliton-fission-linecuts}, where we show the solitary waves emerging from a fission process can be well fitted to analytical black and grey solitons profiles. To confirm that the nonlinearity remains sufficient to produce soliton behaviour, we compare the dispersive and nonlinear length scales at $y=1.5\,\mathrm{m}$ for the case of the high amplitude initialisation (Fig.~\ref{fig:ExtFig-DarkSolAllVoltages+widthxamp}\textbf{f}). We measure a background amplitude of $A_0\approx4\,\mathrm{nm}$, corresponding to a nonlinear lengthscale of $L_{NL}=1/(\xi A_0^2)\approx 0.86\,\mathrm{m}$, which is less than the dispersive length scale of $L_\mathrm{D}=T_0^2/|k_2|=1.29\,\mathrm{m}$. Therefore, the nonlinearity remains sufficient to produce soliton behaviour over metre-scale propagation distances.

\section{Bright pulse evolution}
\label{sec:Supp-bright-pulse}

\begin{figure}
    \centering
    \includegraphics[width=1.0\linewidth]{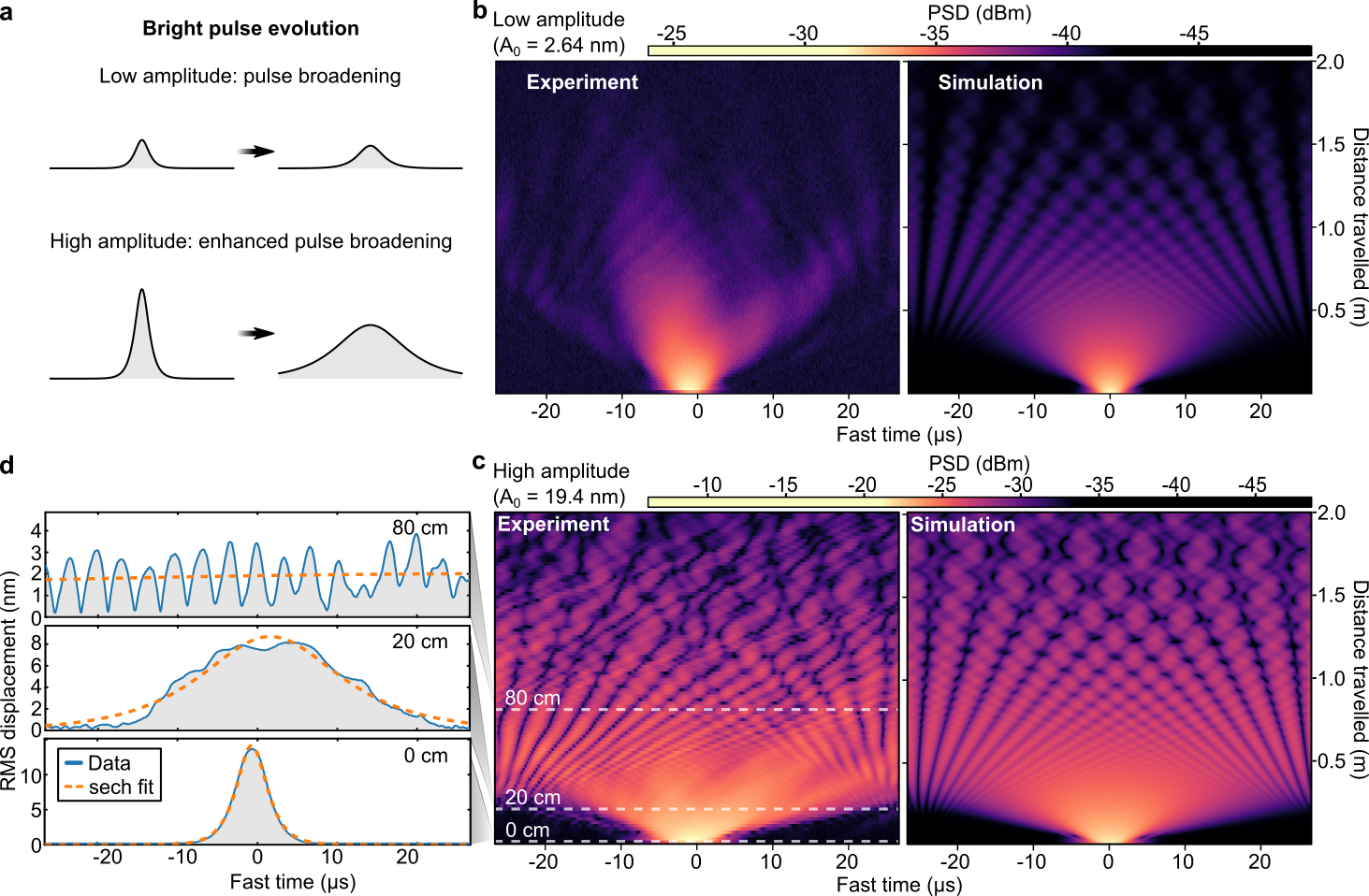}
    \caption{\textbf{Evolution of a bright $\sech$ pulse.} \textbf{a} Sketch of predicted bright pulse behaviour in the low amplitude (linear) and high amplitude (nonlinear) regimes. The addition of a defocussing Duffing nonlinearity to the anomalous GVD causes enhanced pulse broadening. \textbf{b} Experimental pulse evolution (left panel) and NLSE simulation (right panel) for a low-amplitude bright pulse in the linear regime ($A_0=2.64\,\mathrm{nm}$). The initial pulse FWHM is $1\,\mathrm{\upmu s}$. \textbf{c} Experimental pulse evolution (left panel) and NLSE simulation (right panel) in the high amplitude regime ($A_0=26.3\,\mathrm{nm}$). The same initial pulse width is used. \textbf{d} Cross-sections of the measured RMS displacement at the coordinates shown in \textbf{c}, along with a $\sech$ model fit, highlighting how bright solitary waves do not form in our system.}
    \label{fig:ExtFig-BrightPulse}
\end{figure}

In Fig.~\ref{fig:ExtFig-BrightPulse}, we contrast the dark pulse behaviour discussed above with that of a bright pulse in the same nonlinear medium. For a bright pulse, we expect the anomalous dispersion and defocussing nonlinearity to act in concert to enhance the pulse broadening, such that a higher amplitude pulse (experiencing greater nonlinear self-phase modulation) broadens faster than a lower amplitude pulse (Fig.~\ref{fig:ExtFig-BrightPulse}\textbf{a}). To demonstrate this, we repeat the same experiment as in Fig.~\ref{fig:Fig3} of the main text and Fig.~\ref{fig:ExtFig-DarkSolAllVoltages+widthxamp} of the supplementary material, using a bright `$\sech$' pulse with $\mathrm{FWHM=1\,\mathrm{\upmu s}}$, varying the AC actuation voltage to achieve different initial displacement amplitudes. The results are shown in Figs.~\ref{fig:ExtFig-BrightPulse}\textbf{b-c} alongside NLSE simulations. As expected, we see the higher-amplitude pulse broadens faster: it fills the waveguide and self-interferes after approximately $25\,\mathrm{cm}$ of distance travelled (Fig.~\ref{fig:ExtFig-BrightPulse}\textbf{c}), while the low amplitude pulse fills the waveguide after around $40\,\mathrm{cm}$ travelled (Fig.~\ref{fig:ExtFig-BrightPulse}\textbf{b}). This difference is not an artefact of having greater photocurrent signal at higher amplitudes, as we observe the same difference in the pulse broadening rates in the NLSE simulations where the dynamic range is infinite (right panels). In both measurements we see the formation of interference fringes, but in the high amplitude experiment we additionally see in the collisions of these fringes the temporal shift indicative of repulsive dark solitary wave interactions. For example, in Fig.~\ref{fig:ExtFig-BrightPulse}\textbf{c} at fast time $=0\,\mathrm{\upmu s}$ and $80\,\mathrm{cm}$ of travel distance, the crossings display the hexagonal pattern observed in Fig.~\ref{fig:Fig5}\textbf{f} of the main text and supplementary Fig.~\ref{fig:tanh-in-exp}. The comparison of the low- and high-amplitude data provides insight into the process of solitary wave formation: dark fringes form even in the linear regime due to interference and dispersion; it is the nonlinearity that stabilises them into dark solitary waves with mutually repulsive interactions.

\section{Soliton fission and fitting}
\label{sec:Supp-solitons-fission-and-fitting}

\begin{figure*}
    \centering
    \includegraphics[width=\linewidth]{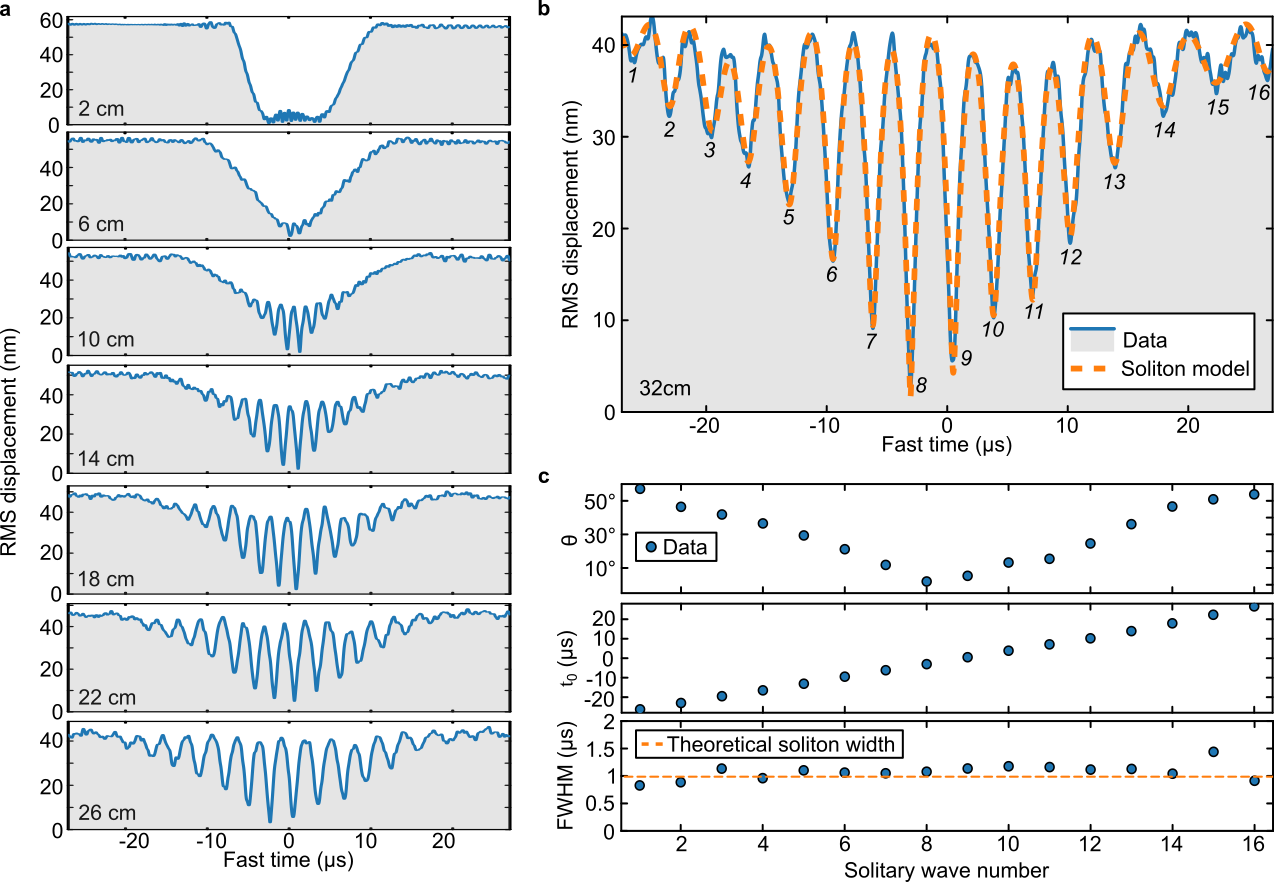}
    \caption{\textbf{Analysis of solitary wave fission.} \textbf{a} Measured RMS displacements at different propagation distances for a broad dark pulse (same data as Fig.~\ref{fig:Fig4} of the main text), showing breakdown of the pulse into its solitary wave components. \textbf{b} Solid blue line: Experimentally measured RMS displacement after $32\,\mathrm{cm}$ of pulse propagation. Dashed orange line: fit using Eq.~\eqref{eqn:soliton-fission-N-soliton-fit} for $N=16$ grey solitons. \textbf{c} Extracted fitting parameters: greyness parameter $\theta$,  time location $t_0$, and temporal FWHM. Dashed orange line in bottom panel: theoretical soliton width predicted from Eq.~\eqref{eqamplitudetimeswisthsoliton} for a background amplitude $A=40\,\mathrm{nm}$.}
    \label{fig:SuppFig-soliton-fission-linecuts}
\end{figure*}

In this section we provide additional analysis of the solitary wave fission data shown in Fig.~\ref{fig:Fig4} of the main text. Figure~\ref{fig:SuppFig-soliton-fission-linecuts}\textbf{a} provides line cuts showing the experimentally measured displacement $A_0(t)$ as the initial dip undergoes soliton fission into a predicted total of $N_\mathrm{fiss}=\sqrt{L_D/L_{NL}}\simeq18$ solitary waves~\cite{agrawalNonlinearFiberOptics2019}. In Fig.~\ref{fig:SuppFig-soliton-fission-linecuts}\textbf{b} we show the recorded RMS displacement after $32\,\mathrm{cm}$, which is approximately the furthest propagation distance at which the solitary waves can be observed before they are affected by self-interference (because of their combined temporal width exceeding the round trip time of the waveguide). At this distance the fission process is well underway and has produced $N=16$ solitary waves (solid blue line). We fit this experimental trace with the following analytical equation for $N$ grey solitons~\cite{onoratoRogueShockWaves2016}:
\begin{equation}
    \label{eqn:soliton-fission-N-soliton-fit}
    |A(t)|=A_0\times\prod_{k=1}^{N}\bigg|\sin(\theta_k)+i\cos(\theta_k)\tanh\left(\frac{\cos(\theta_k)(t-t_{0,k})}{w_k}\right) \bigg|.
\end{equation}
Here $A_0$ is the common background displacement amplitude, and each soliton is characterised by the following three independent parameters: $\theta_k$ for depth, $t_{0,k}$ for time location, and $w_k$ for width. These fit parameters are shown in subfigure $\textbf{c}$
 for each of the $N=16$ solitons, where $\theta=0^\circ$ corresponds to a black soliton with full extinction and $\theta=90^\circ$ to a soliton of vanishing depth. The fit results (dashed orange) indicate soliton FWHMs in the range of $1-1.2\,\upmu s$.
 
 The observed widths can be compared to NLSE theory (see Eq.~\eqref{eqamplitudetimeswisthsoliton}), which predicts that the product of background amplitude and soliton width will equal a constant~\cite{agrawalNonlinearFiberOptics2019}: $T_0A_0=\sqrt{|k_2|/\xi}$. Plugging in $k_2=-36.6\,\mathrm{\upmu s^{2}\cdot m^{-1}}$ from Eq.~\eqref{eqn:GVD-formula} and $\xi=7.30\times10^{-2}\,\mathrm{nm^{-2} \cdot m^{-1}}$ from Eq.~\eqref{eqn:xi-formula}, and a background amplitude of $A_0=40\,\mathrm{nm}$ (see Fig.~\ref{fig:SuppFig-soliton-fission-linecuts}\textbf{b}), we predict a soliton full width at half maximum of $0.99\,\mathrm{\upmu s}$, in good agreement to the fitted widths (dashed orange line in Fig.~\ref{fig:SuppFig-soliton-fission-linecuts}\textbf{c}, buttom panel). The fit parameters additionally show the solitary wave depth (greyness) is most pronounced in the centre of the pulse (top panel of Fig.~\ref{fig:SuppFig-soliton-fission-linecuts}\textbf{c}), becoming gradually greyer on either side, and that the solitary waves achieve a nearly uniform spacing under their repulsive interactions (middle panel of Fig.~\ref{fig:SuppFig-soliton-fission-linecuts}\textbf{c}).

\section{Bright exponential envelopes}

\begin{figure*}
    \centering
    \includegraphics[width=1.0\linewidth]{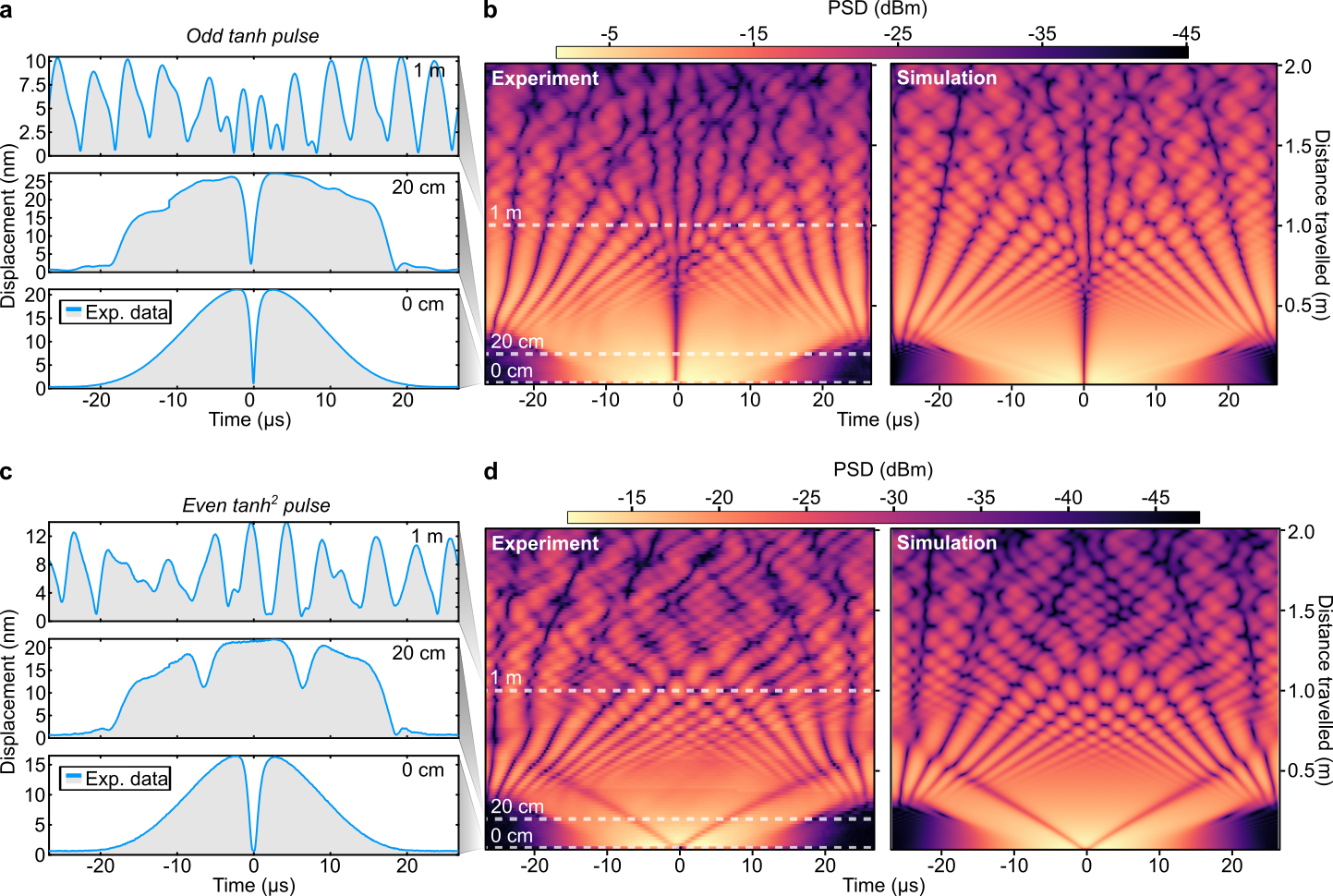}
    \caption{\textbf{Influence of pulse parity: odd tanh and even tanh$^2$ input pulses on a Gaussian envelope.} \textbf{a} Experimentally measured evolution of an odd $\tanh$ pulse with $\mathrm{FWHM}=1\,\mathrm{\upmu s}$ embedded in a Gaussian envelope with $\mathrm{FWHM}=20\,\mathrm{\upmu s}$, after propagation distance of 0 cm; 20 cm and 1 m. Initial background amplitude  $A_0= 36$ nm. \textbf{b} Surface plots showing the full pulse evolution (left: experimental data; right: NLSE simulation). \textbf{c} Experimentally measured evolution of an even $\tanh^2$ pulse with $\mathrm{FWHM}=1\,\mathrm{\upmu s}$ embedded in a Gaussian envelope with $\mathrm{FWHM}=20\,\mathrm{\upmu s}$, after propagation of 0 cm; 20 cm and 1 m. Initial background amplitude  $A_0= 36$ nm. \textbf{d} Surface plots showing the full pulse evolution (left: experimental data; right: NLSE simulation), showing the splitting of the even pulse into two grey solitary waves and eventual repulsive collisions between those waves and others formed by soliton fission.}
    \label{fig:tanh-in-exp}
\end{figure*}

To highlight the advantages of our recombining approach to initialising a dark soliton---whereby we imprint half of a hyperbolic tangent ($\tanh$) notch onto either end of an AC burst of length equal to the round trip time---here we investigate the more conventional approach of superimposing a dark notch onto a time-limited bright pulse~\cite{weinerExperimentalObservationFundamental1988}. Specifically, we use a broad Gaussian envelope for the bright background, and multiply that envelope by an odd parity $\tanh$ function with $\mathrm{FWHM}=\mathrm{1\,\upmu s}$ (Fig.~\ref{fig:tanh-in-exp}\textbf{a} bottom panel). Additionally, to experimentally verify the influence of the pulse parity, we prepare the same bright Gaussian envelope multiplied by an even parity $\tanh^2$ function (Fig.~\ref{fig:tanh-in-exp}\textbf{c} bottom panel).

Using a bright pulse for the envelope has a clear advantage over the initialisation method outlined in Fig.~\ref{fig:Fig2} of the main text, in that it allows the notch to be programmed directly into the arbitrary waveform (instead of having to recombine two halves of a $\tanh$ notch). This allows us to provide the ideal $\tanh$ profile as the initial condition, and test whether it is a stationary solution---we find that it is, since we observe no shedding of grey solitons or non-soliton radiation in Fig.~\ref{fig:tanh-in-exp}. We can also cleanly test the theory of soliton parity by inputting a $\tanh^2$ profile. Despite the odd and even input envelopes looking broadly similar (bottom panels of Fig.~\ref{fig:tanh-in-exp}\textbf{a,\,c}), the behaviour is drastically different. Instead of being a stationary solution, the even pulse immediately bifurcates into two grey solitons as discussed in the main text. This illustrates a fundamental difference between bright and dark solitons, namely a constant phase across bright solitons, and a discontinuous $\pi$ phase shift across black ($\theta\simeq0$) dark solitons~\cite{kivsharDarkOpticalSolitons1998}, responsible for their topologically protected nature. (In the case of grey solitons where $0<\theta<\pi/2$, the phase shift across the dip is continuous and of a magnitude less than $\pi$~\cite{kivsharDarkOpticalSolitons1998}).

Bright pulse envelopes have a significant downside, however: once the widening envelope fills the waveguide and begins to self-interfere, the resulting dark fringes (which stabilise into dark solitary waves) quickly fill the waveguide and make it impossible to track the centrally initialised wave (Figs.~\ref{fig:tanh-in-exp}\textbf{a} and~\textbf{c}). This is a significant downside as it prohibits the observation of slow-moving soliton interactions. For example, the slow overtaking collision in Fig.~\ref{fig:Fig5}\textbf{d} would be difficult to fit into the $\sim70\,\mathrm{cm}$ of propagation that it takes here for the background to become chaotic.

We can observe in Figs.~\ref{fig:tanh-in-exp}\textbf{a} and~\textbf{c} that as the bright Gaussian pulse broadens, it acquires a more square, `shelf-like' profile under the combined effects of nonlinearity and dispersion, in a manner similar to experiments with optical fiber solitons \cite{weinerExperimentalObservationFundamental1988}. This highlights another downside of the Gaussian pulse initialisation, which is that the background amplitude is non-uniform. This makes it difficult to test theoretical predictions under which a uniform background amplitude is a common assumption. The consequences of this are demonstrated below in section~\ref{sec:Supp-Predicted-separation-speed}, where we show that the soliton separation speed is impacted by the non-uniform background.

\section{Solitary wave Wigner crystal}

Here we provide additional analysis on the data shown in Fig.~\ref{fig:Fig5} of the main text, describing the particular configuration of co-propagating waves in the acoustic waveguide as a finite temperature Wigner-like crystal of dark solitary waves. 
While a Wigner crystal originally describes the crystalline phase of electrons interacting via Coulomb repulsion~\cite{wigner_interaction_1934,deshpande_one-dimensional_2008, shapir_imaging_2019}, it also more generally refers to the crystal phase formed by (quasi)particles experiencing mutually repulsive interactions at low densities---for example, quantised vortices in Bose-Einstein condensates~\cite{neely_melting_2024} or dark solitary waves as is the case here. 

\begin{figure}
    \centering
    \includegraphics[width=1.0\linewidth]{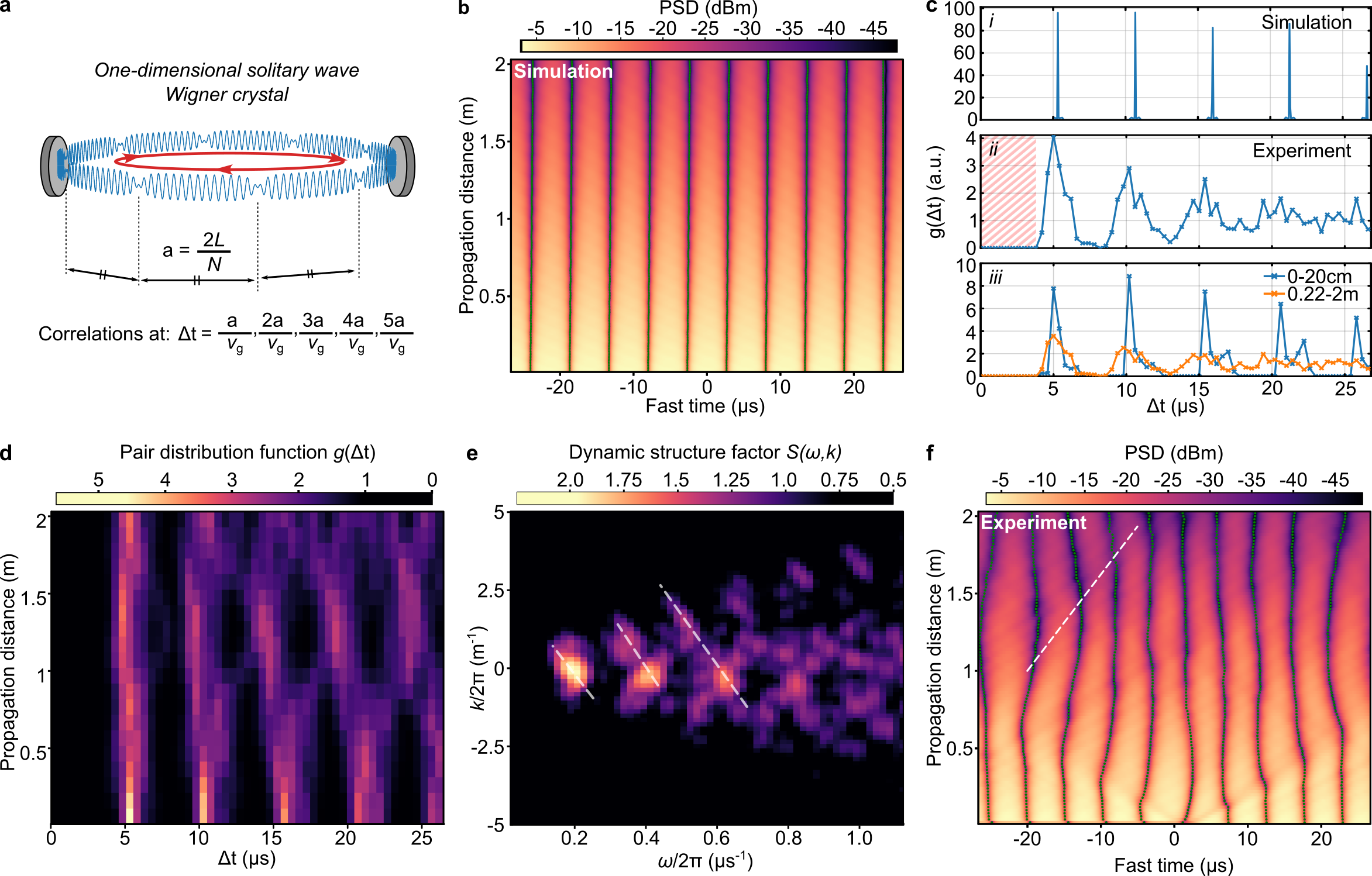}
    \caption{\textbf{One-dimensional solitary wave Wigner crystal.} \textbf{a} Sketch of the waveguide initialised with a one-dimensional Wigner crystal of equally spaced mutually repulsive solitary waves. \textbf{b} NLSE simulation of ten co-propagating dark solitary waves, each initialised in the ideal $\tanh$ solution such that none shed dispersive radiation. Faint green dots: extracted solitary wave locations used to calculate the pair distribution function $g(\Delta t)$. \textbf{c} \textit{i)} $g(\Delta t)$ for the simulation in subfigure $\textbf{b}$. \textit{ii)} $g(\Delta t)$ for our experimentally measured ten co-propagating solitary waves over the entire $2\,\mathrm{m}$ propagation distance. Red hatched region highlights the solitary wave exclusion zone. \textit{iii)} Blue: $g(\Delta t)$ integrated over the first $20\,\mathrm{cm}$ of propagation. Orange: Same data integrated from $22\,\mathrm{cm}$ to $2\,\mathrm{m}$. \textbf{d} $g(\Delta t)$ using a sliding window of length $10\,\mathrm{cm}$. A Gaussian blur with standard deviation of 0.7 pixels has been applied to smooth the data. \textbf{e} Dynamic structure function $S(\omega,k)$ from the experimental data. White lines: linear fits used to estimate the group velocity of vibrations in the crystal. A Gaussian blur with standard deviation of 0.8 pixels has been applied to smooth the data. \textbf{f} Experimental data used to calculate $g(\Delta t)$ in the previous subplots (same as top panel of Fig.~\ref{fig:Fig5}\textbf{b} of the main text). Faint green dots: extracted solitary wave locations. White line: suggested trajectory of a wave travelling in the solitary wave lattice at the group velocity estimated in subfigure \textbf{e}.}
    \label{fig:SuppFig-coprop-crystal}
\end{figure}

Here our initialisation can be understood as a finite-size lattice of $N=10$ dark solitary waves travelling around a closed ring (periodic boundary) topology of circumference $2L=2\,\mathrm{cm}$. The lowest energy, zero-temperature configuration corresponds to $N$ dark solitary waves equally spaced at a distance $a=\frac{2L}{N}$ from each other (see Fig.~\ref{fig:SuppFig-coprop-crystal}\textbf{a}). In this configuration the solitary waves will be spatially correlated at integer multiples of $a$, up to a maximum of $5a$ due to the closed ring topology. These correlations in space correspond to correlations of separation times $\Delta t$ by the relation $a\mapsto \Delta t\,v_g$.

A measure of the organisation of the system is provided by the 
pair distribution (or pair correlation) function $g(\Delta t)$~\cite{hansen_theory_2007,coleSolitonCrystalsKerr2017,neely_melting_2024}. This function is proportional to the conditional probability density that a solitary wave is found at time $t_0+\Delta t$, given that a different solitary wave is present at time $t_0$. In the crystal phase with long range order, $g(\Delta t)$ exhibits sharp peaks at multiples of the lattice spacing $a/v_g$; in the liquid phase with medium range order, the oscillations in $g(\Delta t)$ decay exponentially and correlations are lost at larger multiples of the lattice spacing; in the gas phase, $g$ is uniform with value $\sim~1$ in the ideal gas limit~\cite{hansen_theory_2007}.

To obtain $g(\Delta t)$ from our simulations and experimental data, we track the temporal locations $t_i(y)$ of all solitary waves across each of the propagation distances $y$ (faint green dots centred on the solitary wave minima in Fig.~\ref{fig:SuppFig-coprop-crystal}\textbf{b} and \textbf{f}). For each propagation distance the code finds the $N=10$ minimum values of the photocurrent PSD, enforcing a minimum spacing of $1.5\,\mathrm{\upmu s}$ to ensure that no two minima come from within the same solitary wave. From this table of separation times $\{t_i\}$ for each $y$, we compute $g(\Delta t)$ as the histogram of all pairwise temporal distances, normalised such that $g(\Delta t)=1$ indicates an absence of correlation. Specifically, the discrete calculation is performed as:
\begin{equation}
    g(\Delta t) = \frac{\sum_{y=1}^{N_y}\sum_{i=1}^{N}\sum_{j=i+1}^{N}\textbf{1}_{\Delta t}(d_P(t_i,t_j))}{\left(\frac{N_yN(N-1)}{2}\right)\left(\frac{\Delta t}{T_\mathrm{max}}\right)}
\end{equation}
The numerator counts every unique solitary wave pair separation that falls within the bin centred around $\Delta t$, summing over every propagation distance in the data. Here $N_y$ is the number of propagation distances in the data (typically $N_y=101$ for our experimental data where we track solitary waves over $2\,\mathrm{m}$, and $N_y=20,000$ in our NLSE simulations over the same distance). $N=10$ is the number of solitary waves, and $\textbf{1}_\mathrm{\Delta t}(t)$ is the indicator function that equals $1$ if $t$ is in the bin $[\Delta t - \frac{\text{bin size}}{2}, \Delta t+\frac{\text{bin size}}{2}]$, and 0 otherwise. Because of the differences in $N_y$, we use a narrow bin size of $0.05\,\mathrm{\upmu s}$ for the NLSE simulations, and larger bin size of $0.4\,\mathrm{\upmu s}$ for the sparser experimental data. $d_P$ is the periodic distance between two solitons, defined as:
\begin{equation}
    d_P(t_i,t_j)=\min\left(|t_i-t_j|, \;T_\mathrm{max}-|t_i-t_j|\right),
\end{equation}
where $T_\mathrm{max}=L/v_g$ is the maximum temporal separation between two solitary waves under the periodic boundary conditions. The denominator normalises $g(\Delta t)$: the left parentheses represent the total number of pair separations across all bins; the right parentheses represent the probability of a separation distance falling into that bin assuming no correlations.

We first plot the ideal, zero-temperature pair correlation function by simulating the waveguide with ten co-propagating solitary waves, where each wave is initialised using the ideal $\tanh$ soliton solution (Fig.~\ref{fig:SuppFig-coprop-crystal}\textbf{b}). In this configuration, no dispersive radiation is produced and the solitary waves maintain their even spacing over the full propagation distance (even in the presence of loss and third-order dispersion). The corresponding pair distribution function is shown in Fig.~\ref{fig:SuppFig-coprop-crystal}\textbf{c}\textit{i}, exhibiting sharp peaks at multiples of the crystal lattice as expected for a Wigner crystal.

We next study our experimentally realised solitary wave Wigner crystal (same data as Fig.~\ref{fig:Fig5}\textbf{b} of the main text, and here shown again in Fig.~\ref{fig:SuppFig-coprop-crystal}\textbf{f}). As discussed in the main text, in our experiment one of the solitary waves is initialised imperfectly, representing an energetic defect that ultimately results in a disordered, out of equilibrium lattice. The solitary waves continually adjust their positions in response to movement introduced by the defect, leading to a gradual loss of order. In this case, the disorder in the lattice manifests as a smearing and broadening of the peaks of the correlation function (Fig.~\ref{fig:Fig5}\textbf{c}\textit{ii})~\cite{neely_melting_2024,coleSolitonCrystalsKerr2017, karpovDynamicsSolitonCrystals2019}. However, despite the presence of the defect, the order is still faintly visible up to the maximum pairwise spacing of 5 lattice sites. The repulsive interactions are also clearly visible, through the vanishing likelihood of finding two solitary waves separated by less than $\sim4\,\mathrm{\upmu s}$. This region where $g(\Delta t)\sim 0$ (illustrated by the red hatched box in Fig.~\ref{fig:Fig5}\textbf{c}\textit{ii}), represents the width of the solitary wave exclusion zone, i.e. the temporal width of the dark solitary waves' repulsion (corresponding to a spatial width of $\sim4\,\mathrm{\upmu s}\cdot v_g=1.5\,\mathrm{mm}$).

In Fig.~\ref{fig:SuppFig-coprop-crystal}\textbf{c}\textit{iii} we calculate $g(\Delta t)$ separately over the first 20 cm (blue trace) and the subsequent 180 cm of propagation (orange trace). This indicates that the solitary waves are initially able to maintain the Wigner-crystal like correlation with which they were initialised, with clearly visible peaks at multiples of the lattice spacing. The subsequent loss of order represents a melting of the Wigner crystal into a semi-ordered liquid state. To better observe this melting, in Fig.~\ref{fig:SuppFig-coprop-crystal}\textbf{d} we show $g(\Delta t)$ across the full experimental data using a sliding window of length 10cm. The correlations initially start as bright, narrow peaks separated by exclusion zones, but over time the peaks at larger lattice spacings blur together. There is a slight recurrence of the correlations around 1.3m; we tentatively interpret this as associated with the lengthscale over which the bulk of the dispersive radiation completes a round trip around the waveguide in the moving frame, at which point the system may achieve a transient revival of the initial conditions.

Finally we calculate the dynamic structure factor $S(\omega, k)$, which represents the Fourier transform over time and space of the pair correlation function~\cite{hansen_theory_2007}. We calculate $S(\omega,k)$ using the list of measured time coordinates $\{t_{ij}\}$ (green dots in Fig.~\ref{fig:SuppFig-coprop-crystal}\textbf{b} and \textbf{f}), where $i\in[1,\ldots,N=10]$ indicates soliton number and $j\in[1,\ldots,N_y]$ indicates the propagation distance. In the first step we Fourier transform in the time domain to obtain a density fluctuation in frequency space:
\begin{equation}
    \rho(\omega,y_j)=\sum_{i=1}^N e^{i\omega t_{ij}}
\end{equation}
where $\omega\in[\pi m/T_\mathrm{max}:\,1\leq m\leq60]$. Because there are far fewer steps in $y$ (limited by the 2cm round trip distance of the waveguide) than number of time bins, to obtain a smoother, interpolated final plot we pad the spatial dimension with zeros out to $N_\mathrm{padded}=4N_y$:
\begin{equation}
    \tilde{\rho}(\omega,y_j)=\begin{cases}
    \rho(\omega,y_j) & 1\leq j \leq N_y\\
    0 & N_y<j\leq N_\mathrm{padded}
    \end{cases}
\end{equation}
We then obtain $S(w,k)$ as the fast Fourier transform of $\tilde{\rho}$ along the $y$ axis.

Figure~\ref{fig:SuppFig-coprop-crystal}\textbf{e} shows $S(\omega,k)$ for the experimental data. We first identify a bright spot at $\omega/2\pi\approx0.2\,\mathrm{\upmu{s}^{-1}}$ and $k=0$, corresponding to the first peak of $g(\Delta t)$, i.e. the fundamental lattice spacing unit. The location of this point at $k=0$ indicates this separation is globally uniform, which is to be expected since it is caused by the fundamentally repulsive interactions of the dark solitary waves. Similar bright points appear at multiples of the same frequency along the $k\approx0$ line, albeit with a slight downwards drift of $\sim-0.6\,\mathrm{\upmu s/m}$ that indicates a small error in our estimate of the round-trip time $T_\mathrm{max}$. Around several of the bright points we can see vaguely `X'-shaped structures, indicative of a vibrational mode within the crystal~\cite{landigMeasuringDynamicStructure2015}. The diagonal from top left to bottom right is more strongly visible (indicating our data more clearly exhibits the mode travelling backwards in the moving frame), so we tentatively perform linear fits to that orientation to obtain an estimate of the group velocity of the mode. We obtain a relative velocity of $\sim16\,\mathrm{\upmu s.m^{-1}}$, which is plotted on top of the soliton trajectories in Fig.~\ref{fig:SuppFig-coprop-crystal}\textbf{f} for reference, and appears to qualitatively match the speed as which disturbances propagate in the solitary wave lattice. 
Going forward, both the initial conditions of the solitary waves (for example their number $N$ and initial velocities), as well as characteristics of the background (such as localised or distributed acoustic energy of a chosen power spectrum) can be precisely controlled by programming the arbitrary waveform~\cite{wuControlledGenerationDark2002}. These results mean it is possible to systematically study non-equilibrium soliton thermodynamics using a phononic waveguide such as ours. Future measurements could study the phase transitions between crystal~\cite{coleSolitonCrystalsKerr2017}, liquid, and gas phases~\cite{redor2019experimental}, and quantify the phonon dispersion relation by introducing engineerable defects into the lattice structure. This could be relevant to studies of Kerr solitons in microresonators~\cite{karpovDynamicsSolitonCrystals2019} and Wigner lattices in one-dimensional semiconductors~\cite{meyerWignerCrystalPhysics2008} and quantum wires~\cite{schulzWignerCrystalOne1993}.

\section{Predicted separation speed}
\label{sec:Supp-Predicted-separation-speed}

\begin{figure}
    \centering
    \includegraphics[width=1.0\linewidth]{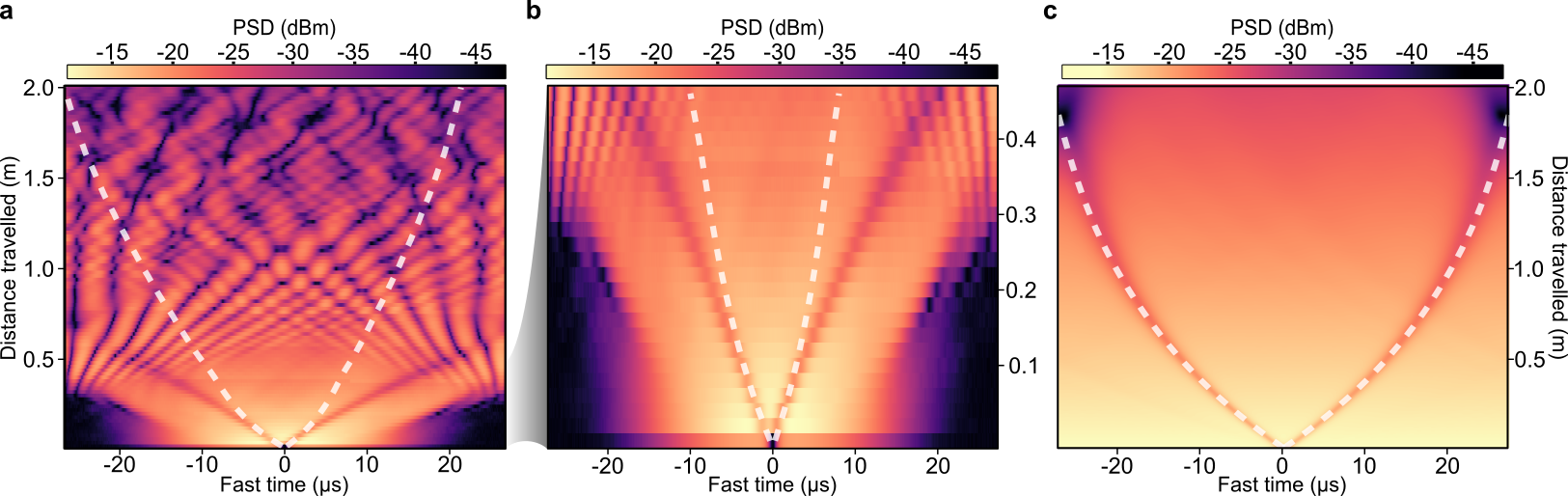}
    \caption{\textbf{Dark solitary wave separation speed.} \textbf{a} Experimentally measured evolution of a $\tanh^2$ dark pulse initialised inside a Gaussian envelope (identical data as Fig.~\ref{fig:tanh-in-exp}\textbf{d}). White: expected trajectories of the two solitary waves moving at speeds $c_s\sin(\theta_f)$ and $c_s\sin(\theta_f)$. \textbf{b} Zoom-in on the first 48 cm of propagation. \textbf{c} NLSE simulations using a uniform background amplitude instead of the Gaussian envelope. White lines: expected trajectories on the uniform background, assuming equal depths of $\theta=(\theta_f+\theta_b)/2$.}
    \label{fig:ExtFig-csep-tanhsq-test}
\end{figure}

In Fig.~\ref{fig:ExtFig-csep-tanhsq-test} we further analyse the experimental data of Fig.~\ref{fig:tanh-in-exp}\textbf{d}, which shows the separation of two grey solitary waves formed from a single $\tanh^2$ notch within a bright Gaussian envelope. This experiment provides a good setting to verify whether the two formed grey solitary waves move with velocity $c_s\sin\theta$, as the ideal $\tanh^2$ notch cleanly separates with minimal radiation.

In the white lines of Figs.~\ref{fig:ExtFig-csep-tanhsq-test}\textbf{a-b} we plot the predicted solitary wave trajectories overlaid upon the experimental data, calculating $c_s$ at each round trip from Eq.~\eqref{eqn:c_s} (we take the instantaneous background amplitude as the median amplitude recorded over the round trip). We achieve almost perfectly equal bifurcation of the initial $\tanh^2$ notch, with the forward-propagating (i.e. moving left on the heatmap) solitary wave having depth $\theta_f\approx26^\circ$, and the backward-propagating solitary wave having depth $\theta_b\approx34^\circ$. As in the main text discussion of Fig.~\ref{fig:Fig5}\textbf{d}, we estimate these depths from the relationship $\sin(\theta)=A_\mathrm{dip}/A_0$. In Fig.~\ref{fig:ExtFig-csep-tanhsq-test}\textbf{b} we can see that the solitary waves initially follow the predicted trajectory, but after around $10\,\mathrm{cm}$ of propagation they begin to deviate, separating faster than expected. Our hypothesis is that this deviation from theory is because of the Gaussian envelope, which forms a nonuniform background.

To test this hypothesis, we take the NLSE simulation that closely matches these data (right panel in Fig.~\ref{fig:tanh-in-exp}\textbf{d}) and change the initial conditions from a Gaussian envelope to a uniform background. The simulation results, shown in Fig.~\ref{fig:ExtFig-csep-tanhsq-test}\textbf{c}, show the two grey solitary waves separating at the expected rate, validating the previous hypothesis that the measured speed differs from $c_s\sin\theta$ because of the nonuniform background. Because the simulated $\tanh^2$ notch separates into exactly equal grey solitary waves, in Fig.~\ref{fig:ExtFig-csep-tanhsq-test}\textbf{c} the predicted separations are calculated assuming the two solitary waves have identical depths equal to the average of $\theta_f$ and $\theta_b$.

\section{Observing solitary waves in the middle of the waveguide}
\begin{figure}
    \centering
    \includegraphics[width=1.0\linewidth]{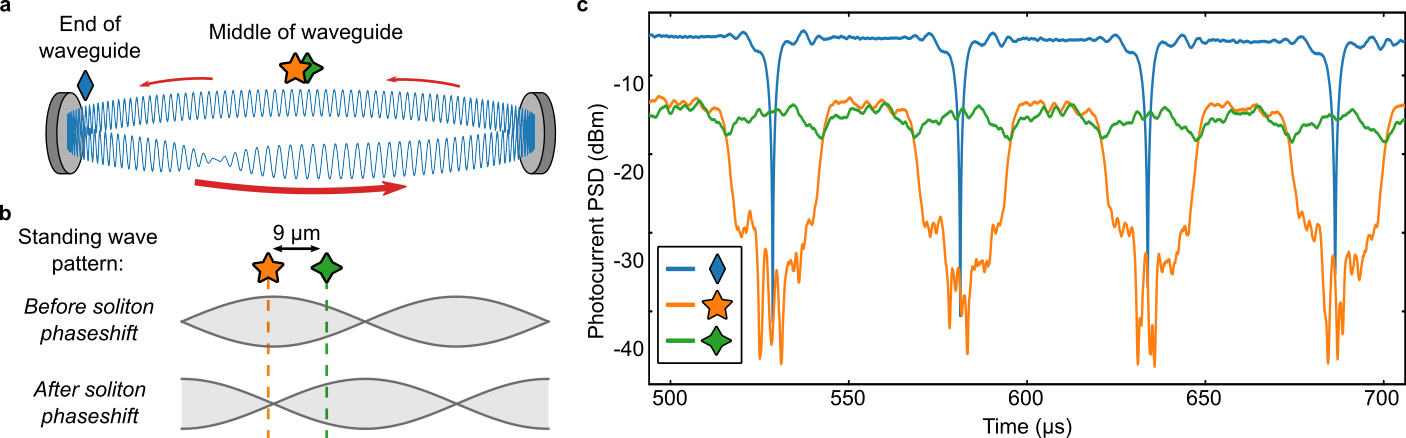}
    \caption{\textbf{Observing solitary waves at different positions along the waveguide.} \textbf{a} We initialise a black ($\theta\simeq0$) solitary wave with carrier frequency $\Omega=2\pi\times15.18\,\mathrm{MHz}$ and measure the photocurrent PSD in three locations: near the actuation electrode at the end of the waveguide (blue diamond); and two locations near the middle of the waveguide, separated by $9\,\mathrm{\upmu m}$ (orange star and green cross). \textbf{b} The two locations in the middle of the waveguide are $9\mathrm{\upmu m}$ apart, so that before passage of the solitary wave (top sketch), one (orange star) lies on a local maximum of the standing wave pattern and the other (green cross) lies on an intermediate point of the pattern. \textbf{c} Excerpt of the resulting photocurrent power spectral densities measured from the time since pulse launch, showing inversion of the standing wave pattern (Fig.~\ref{fig:SuppFig-PhaseFlip}\textbf{b} bottom sketch) with a $\pi$ phase flip caused by the dark solitary wave.}
    \label{fig:SuppFig-PhaseFlip}
\end{figure}

As discussed in the main text and illustrated in Fig.~\ref{fig:Fig2}, we find dark solitary waves are best observed by positioning the optical fibre at either end of the waveguide. In this configuration the detected acoustic power drops to zero when the wave passes below the fiber (provided the fiber is positioned at a distance $l_\mathrm{edge}$ from the edge of the waveguide less than half that of the solitary wave's spatial extent, i.e. $l_\mathrm{edge}<\frac{1}{2}\, T_0v_g$). In contrast, away from the edges of the waveguide the amplitude drops in just one of two counter-propagating wave trains, as illustrated in Fig.~\ref{fig:SuppFig-PhaseFlip}\textbf{a}. However, because a black ($\theta\simeq0$) solitary wave is associated with a $\pi$ phase shift across its central dip, it can be indirectly observed by how its passage affects the pattern of standing waves formed by the two counterpropagating wave trains---local maxima become local minima, and vice versa (Fig.~\ref{fig:SuppFig-PhaseFlip}\textbf{b}). Stated differently, if the forward and backward components were constructive before passage of the solitary wave, they become destructive afterwards due to the $\pi$ phase shift, and vice versa. 

We experimentally observe this phase shift for the case of a single black solitary wave propagating back and forth in the waveguide. Using the same actuation settings, we measure the photocurrent PSD (which is proportional to the membrane RMS displacement) at three locations, the first being $\sim150\,\mathrm{\upmu m}$ away from the electrode at the end of the waveguide (blue diamond in Fig.~\ref{fig:SuppFig-PhaseFlip}). This location satisfies the criterion for total extinction of the detected acoustic power: $l_\mathrm{edge}=150\,\mathrm{\upmu m}$, and $\frac{1}{2}T_0v_g=\frac{1}{2}\times1\,\mathrm{\upmu s}\times376\,\mathrm{m/s}\approx188\,\mathrm{\upmu m}$. The other two measurement locations are near the middle of the waveguide, spaced $\sim9\,\mathrm{\upmu m}$---approximately $15\%$ of the carrier wavelength---apart along the longitudinal axis of the waveguide, such that one (orange star) is situated near a local maximum of the standing wave pattern and the other (green cross) is situated on an intermediate point between the local maximum and adjacent local minimum (Fig.~\ref{fig:SuppFig-PhaseFlip}\textbf{b}). Figure~\ref{fig:SuppFig-PhaseFlip}\textbf{c} shows an extract of the resulting measured photocurrent over time. At the end of the waveguide (blue trace) we observe sharp dips in the acoustic power when the soliton passes by, with spacing between the dips indicating the soliton round-trip travel time is approximately $50\,\mathrm{\upmu s}$ (as expected with $v_g\approx376\,\mathrm{m/s}$). Near the middle of the waveguide, the orange trace shows that local minima become local maxima and vice versa every time the solitary wave passes by, leading to abrupt transitions in the magnitude of the photocurrent PSD. This transition occurs twice per round trip, corresponding to the solitary wave passing by in the $+y$ and $-y$ propagation directions. Because we positioned the fibre approximately in the middle of the waveguide, the inversion also appears with a roughly equal period---if we repeat this experiment closer to either end of the waveguide, we measure a duty cycle departing from 50:50 (not shown here). 
Finally, the green trace demonstrates how the solitary wave appears when the fiber position is not precisely located either over a node or an antinode of the standing wave. In this configuration the acoustic powers measured before and after passage of the solitary wave are very similar, although we do still observe a $\sim3\,\mathrm{dB}$ dip in the acoustic power as the standing wave undergoes the $\pi$ phase shift. In the pulic repository associated with this paper we include an animated visualisation of this phase shift; while the initial and final standing wave patterns appear only a quarter-wavelength apart, over the course of the phase shift the standing waves appear to translate several wavelengths, leading to the observed transient change in the standing wave amplitude. 

\section{Nonlinear Schr\"odinger Equation theory}

\begin{figure}
    \centering
    \includegraphics[width=1.0\linewidth]{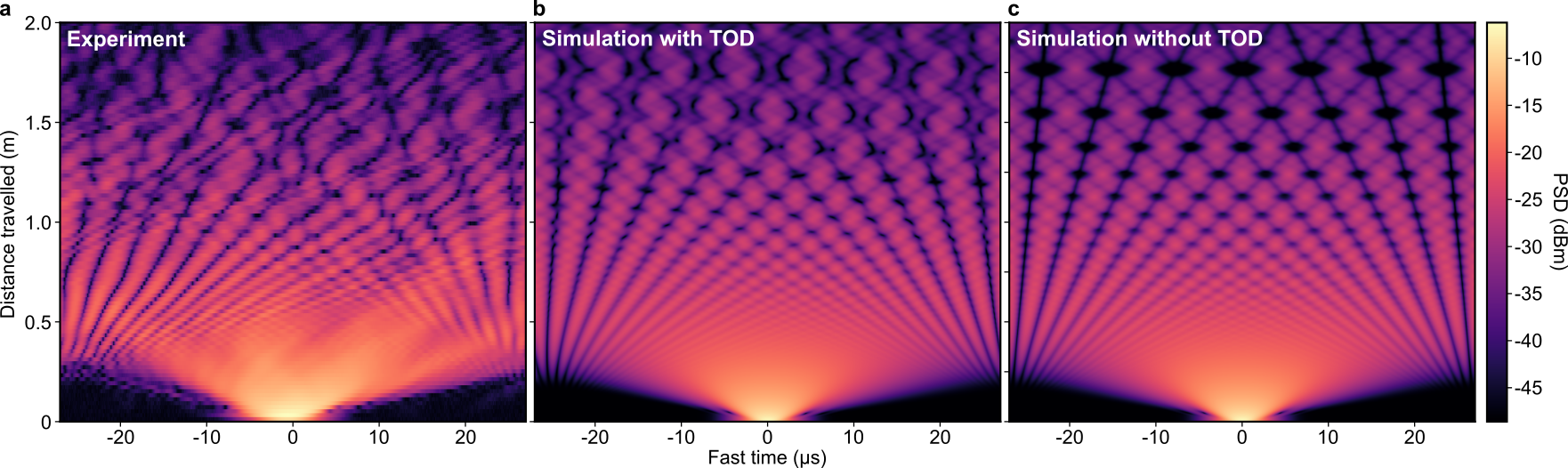}
    \caption{\textbf{Effect of third-order dispersion.} \textbf{a} and \textbf{b} respectively show experimental data and NLSE simulations for the bright pulse with $\mathrm{FWHM}=1\,\mathrm{\upmu s}$ and $A_0=19.4\,\mathrm{nm}$ (same results as in Fig.~\ref{fig:ExtFig-BrightPulse}\textbf{c}). \textbf{c} Results of performing the same NLSE simulation without the TOD term ($k_3=0$).}
    \label{fig:SuppFig_sim-wwout-TOD}
\end{figure}

As mentioned in the main text, we model the evolution of a pulse in our phononic waveguide using the nonlinear Schr\"odinger equation (NLSE), which is the standard treatment for modelling wave envelopes in nonlinear, dispersive media~\cite{nayfehNonlinearOscillations2004,kurosuMechanicalKerrNonlinearity2020}. In Eq.~\eqref{eqn:NLSE} of the main text we write the NLSE in its canonical form with up to second-order dispersion. However, we find our simulations provide the best agreement to the data when accounting for up to third-order dispersion (TOD):
\begin{equation}
    \label{eqn:NLSE-with-TOD}
    \frac{\partial A}{\partial y}=-\frac{\alpha}{2}A-\frac{i}{2}k_2\frac{\partial^2A}{\partial t^2}+\frac{1}{6}k_3\frac{\partial A^3}{\partial t^3}+i\xi|A|^2A.
\end{equation}
The TOD term $(k_3)$ is an additional correction that results in asymmetric broadening of the pulse and the formation of ripples~\cite{agrawalNonlinearFiberOptics2019}. Because of its inclusion, Eq.~\eqref{eqn:NLSE-with-TOD} differs from the canonical NLSE and no longer accepts idealised $\tanh$-shaped solitons as stationary solutions. However, in our case the TOD term is small compared with the second-order GVD term, which allows us to efficiently analyse our solitary waves using conventional soliton theory (as shown for instance in Figs.~\ref{fig:ExtFig-DarkSolAllVoltages+widthxamp} and~\ref{fig:SuppFig-soliton-fission-linecuts}). To quantify the importance of the TOD term we consider a bandwidth-limited $\tanh$-shaped pulse with $\mathrm{FWHM}=1\,\mathrm{\upmu s}$ and carrier frequency of $15\,\mathrm{MHz}$. In this case we can compare the relative importance of second and third-order dispersion by evaluating the ratio: $(\mathrm{bandwidth}\times k_3)/|k_2|\approx2.3\%$ 
(this calculation uses formulae for $k_2$ and $k_3$ that we will derive in the following subsections). Because the TOD term is relatively small, it only produces a noticeable correction for long propagation distances, over which the TOD-induced dispersive phase shift has sufficiently accumulated. This is shown in Fig.~\ref{fig:SuppFig_sim-wwout-TOD} where we compare experimentally measured bright pulse data from Fig.~\ref{fig:ExtFig-BrightPulse}\textbf{c} to NLSE simulations performed both with (\textbf{b}) and without (\textbf{c}) TOD. The initial pulse evolution appears identical in either case, but the `butterfly-like' or `x'-shaped patterns visible around 1.5 m of propagation in the experimental data (Fig.~\ref{fig:SuppFig_sim-wwout-TOD}\textbf{a}) are better reproduced when accounting for TOD (Fig.~\ref{fig:SuppFig_sim-wwout-TOD}\textbf{b}), in contrast to the no-TOD case where they appear as ellipses (Fig.~\ref{fig:SuppFig_sim-wwout-TOD}\textbf{c}).

\subsection{Heuristic derivation of the NLSE for our system}

The NLSE is a paradigmatic equation for dispersive and weakly nonlinear waves, and can be derived for practically any dispersive and weakly nonlinear wave system, i.e., a system with a dispersion relation of the form $\omega = \Omega(k_y,|A|^2)$, where $\omega$ is the frequency, $k_y$ the wavenumber, $A$ the complex wave amplitude, and $\Omega$ the dispersion relation between those variables. The dispersion relation has an equivalent inverse formulation of $k_y = K(\omega,|A|^2)$ for a function $K$. The choice of formulation depends on whether the natural independent variable of the carrier is $k_y$ or $\omega$. In our system it is the latter and we consider $k_y = K(\omega,|A|^2)$.  

Assume the wave is localised on the dispersion relation such that $k_y = k_0 + \delta k_y$, with $\delta k_y /k_y \sim \mathcal{O}(\beta)\ll 1$ and $\omega = \omega_0 + \delta \omega$ with $\delta \omega / \omega_0 \sim \mathcal{O}(\beta) \ll 1$, where $\beta$ is a parameter that measures the modulation. Additionally, measure the wave amplitude as $|A| \sim \mathcal{O}(\varepsilon)$.
In a quasi-1D waveguide, such as in our system, we may then consider a modulated harmonic wave of the form:
\begin{equation}
u(y,t) = A(y,t)  e^{i(k_0 y - \omega_0 t)} + \rm{c.c},
\end{equation}
where c.c. stands for complex conjugate. Here $A(y,t)$ is slowly varying, such that the Fourier space representation $A(k_y,\omega)$ is strongly localised near the carrier at $(k_0,\omega_0)$.
We may then expand the dispersion relation around the carrier at $(\omega_0,k_0)$ and about the linear regime, $A=0$. Expanding and collecting terms in order of increasing powers of $(\beta,\varepsilon)$ gives
\begin{equation}
k_y = K(\omega_0,0) + \frac{\partial K}{\partial\omega}\bigg|_{0} \delta \omega + \frac{\partial K}{\partial |A|^2} \bigg|_{0} |A|^2 +  \frac{1}{2}  \frac{\partial^2 K}{\partial \omega^2}\bigg|_{0} (\delta \omega )^2 +   \frac{1}{6}  \frac{\partial^3 K}{\partial \omega^3}\bigg|_{0} (\delta \omega )^3 + \frac{\partial^2 K}{\partial \omega \partial |A|^2}\bigg|_{0} (\delta \omega)  |A|^2 +  \mathcal{O}(\beta^4,\varepsilon^4, \beta^2 \epsilon^2) 
\end{equation}
where the shorthand $|_0$ means evaluating the expressions at $\omega=\omega_0$ and $A=0$. We retain terms up to order $\varepsilon^2, \beta^3$; keeping these terms and rearranging, this can be written as 
\begin{equation}
    \label{eqn:NLSE-deriv-fourier-expansion}
  \delta k_y - \frac{1}{v_g} \delta \omega - \frac{1}{2}\frac{\partial^2 K }{\partial\omega^2} \bigg|_0 (\delta \omega)^2  - \frac{1}{6}\frac{\partial^3 K }{\partial\omega^3} \bigg|_0 (\delta \omega)^3- \frac{\partial K}{\partial |A|^2} \bigg|_0 |A|^2 -  \dots = 0.
\end{equation}
We consider up to third-order dispersion as higher-order effects are negligible when the bandwidth is much less than the carrier frequency~\cite{agrawalNonlinearFiberOptics2019}. Indeed, in this experiment we employ $\tanh$ and $\sech$ pulses as short as $1\,\mathrm{\upmu s}$ (corresponding to bandwidths approximately $\lesssim 1\,\mathrm{MHz}$), compared with carrier frequencies of over $10\,\mathrm{MHz}$.

The multiplicative factors of $\delta k_y$, $\delta \omega$, acting on $A(k_y,\omega)$ may then be promoted to operators acting on $A(y,t)$ through the usual correspondences through the Fourier transform, $\delta \omega \rightarrow -i \partial _t$ and $\delta k_y \rightarrow i \partial_y$, yielding~\cite{dingemansWaterWavePropagation1997,karjantoNonlinearSchrodingerEquation2019}:
\begin{equation}
i \left (\partial_y + \frac{1}{v_g} \partial_t\right) A - \frac{1}{2} \frac{\partial^2 K }{\partial\omega^2} \bigg|_0 \partial_t^2 A + \frac{i}{6}\frac{\partial^3 K }{\partial\omega^3} \bigg|_0 \partial_t^3 A -  \frac{\partial K}{\partial |A|^2} \bigg |_0 |A|^2 A = 0.
\end{equation}
Recalling the fast time $T = t - y/v_g$, we obtain the NLSE without loss:
\begin{align}
    \label{eqn:NLSE-deriv-lossless-eqn}
i \frac{\partial A}{\partial y} A &=  \frac{1}{2} k_2 \frac{\partial^2 A}{\partial T^2} + \frac{i}{6} k_3 \frac{\partial^3 A}{\partial T^3} +  \xi |A|^2 A; 
\end{align}
where 
\begin{align}
    \label{eqn:NLSE-coeffs-as-derivatives}
  k_2&= \frac{\partial^2K}{\partial\omega^2} \bigg|_{0}; & k_3&= \frac{\partial^3K}{\partial\omega^3} \bigg|_{0}; & \xi & = \frac{\partial K }{\partial |A|^2} \bigg|_0.    
\end{align}
The loss term $-\frac{\alpha}{2} A$ can be added now to obtain our complete NLSE, Eq.~\eqref{eqn:NLSE-with-TOD}, recognising that since the loss is independent of frequency and amplitude, it would not affect the above derivation had it been included from the start.

The coefficients in Eq.~\eqref{eqn:NLSE-coeffs-as-derivatives} can be readily evaluated from $K(\omega,|A|^2)$, or, equivalently, through the inverse derivative relations for its inverse function, $\Omega(k_y,|A|^2)$. In our 1D waveguide with Duffing nonlinearity, the dispersion relation takes the form~\cite{hirschTutorialMembranePhononic2026}:
\begin{align}
\Omega(k_y,|A|^2) &= \Omega_0(k_y) \left(1 + \frac{\alpha_{\rm{eff}}}{ 4 k_{\rm eff}} |A|^2\right)\label{eqn:nonlinearDispersion},
\end{align}
where $\Omega_0(k_y)$ is the dispersion relation for linear waves:
\begin{equation}
    \Omega_0(k_y) = c \sqrt{k_y^2 + k_x^2} .
\end{equation}
Here $k_x = \pi/W$ for the fundamental mode of a waveguide of width $W$~\cite{romeroPropagationImagingMechanical2019}. The linear dispersion relation is associated with modes of the form $u(x,y,t) = \sin(k_x x) e^{ik y} e^{-i \omega t}$.  

The inverse formulation of Eq.~\eqref{eqn:nonlinearDispersion} is:
\begin{equation}
\label{eqn:nonlinearDispersionInverted}
K(\omega,|A|^2) = \sqrt{\frac{\omega^2}{c^2(1 + \alpha_{\rm{eff}} |A|^2 /4 k_{\rm{eff}})} - k_x^2}.
\end{equation}
We can take appropriate derivatives of this expression (Eq.~\eqref{eqn:NLSE-coeffs-as-derivatives}) to obtain formulae for the nonlinear and dispersive NLSE coefficients.

\subsection{Determination of the loss coefficient}
\label{sec:Supp-alpha-vg-determination}

\begin{figure*}
    \centering
    \includegraphics[width=1.0\linewidth]{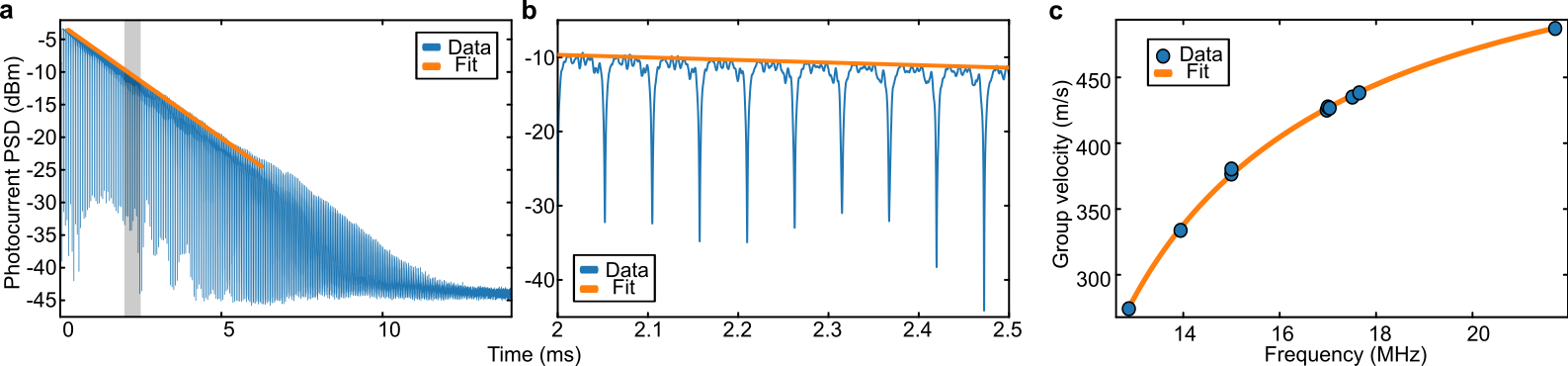}
    \caption{\textbf{Experimental determination of simulation parameters.} \textbf{a} Ringdown measurement of a dark solitary wave initialised at $\Omega=2\pi\times15\,\mathrm{MHz}$, providing a decay rate of $\alpha=2.1\,\mathrm{m^{-1}}$. Blue: Time trace of the photocurrent power spectral distribution (proportional to the acoustic PSD). Orange: Linear fit, plotted across the subset of data used for fitting. Grey shaded box defines zoom-in region. \textbf{b} Zoom-in on the time trace in the grey box in subfigure~\textbf{a}. \textbf{c} Circles: experimentally measured group velocities versus solitary wave frequency. Orange line: Fit using Eq.~\eqref{eqn:vg-analytic} with $c=570.4\,\mathrm{m/s}$ and $W=25.3\,\mathrm{\upmu m}$.}
    \label{fig:SuppFig_parameter-fitting}
\end{figure*}

The coefficient $\alpha$ in Eq.~\eqref{eqn:NLSE-with-TOD} defines the inverse decay length of the envelope. If we neglect all other terms, the equation simply reads $\partial A/\partial x=-\frac{\alpha}{2}A$, which has a solution of $A(x)=A_0\exp(-\frac{\alpha}{2}x).$ Because the acoustic energy is proportional to amplitude squared, $\alpha^{-1}$ is therefore the travel distance over which the envelope decreases in energy by a factor of $e$. We calculate this distance as:
\begin{equation}
    \frac{1}{\alpha}=v_g\times\tau_\mathrm{decay},
\end{equation}
where $\tau_\mathrm{decay}$ is the characteristic decay time of the waveguide at the carrier frequency of the envelope; $\tau_\mathrm{decay}=\gamma^{-1}$ where $\gamma$ is the damping rate of the waveguide~\cite{hirschTutorialMembranePhononic2026}. We experimentally determine $\gamma$ by measuring the ringdown of the background acoustic energy captured during a dark solitary wave measurement~\cite{sementilliLowDissipationNanomechanicalDevices2025}. An example is provided in Fig.~\ref{fig:SuppFig_parameter-fitting}\textbf{a} where we initialise a dark solitary wave at a carrier frequency of $15\,\mathrm{MHz}$, consistent with the main text figures. 
Under linear damping the acoustic energy should decay proportional to $\exp(-\gamma t)$~\cite{hirschTutorialMembranePhononic2026}, so by fitting the logarithmically-scaled acoustic power spectral density with a straight line we obtain the damping rate from the slope. Specifically:
\begin{equation}
    \gamma = \frac{\text{slope}}{-10\,\mathrm{dB}\times\log_{10}(e)},
\end{equation}
where the slope is in units of dB per unit time. At the measured carrier frequency this damping rate corresponds to a quality factor of $Q\sim120,000$---better or comparable to other waveguides fabricated from suspended membranes~\cite{kurosuOnchipTemporalFocusing2018,romeroPropagationImagingMechanical2019,wangHexagonalBoronNitride2019}, though still well below what is achievable using soft-clamped structures~\cite{xiSoftclampedTopologicalWaveguide2025}. A detailed comparison with the broader integrated phononic landscape is provided in Table~\ref{tab:Comparison-phononic-waveguide-landscape}.

To calculate the group velocity $v_g$, we experimentally measure the round trip period $T_{RT}$ of the dark solitary waves and use our knowledge that the round trip distance is $2L=2\,\mathrm{cm}$, as defined in the lithographic chip design (yielding $v_g=2L/T_{RT}$). The decay length can then be found as $\alpha^{-1}=v_g/\gamma$. For the measurements at $15\,\mathrm{MHz}$ shown in Fig.~\ref{fig:SuppFig_parameter-fitting}\textbf{a-b} we find $\alpha^{-1}\approx48\,\mathrm{cm}$, giving $\alpha\approx2.1\,\mathrm{m^{-1}}$. The quality factor and decay length varies with frequency: for example, by performing a similar ringdown measurement at $16.9769\,\mathrm{MHz}$ we find $Q\sim160,000$ and $\alpha\approx1.56\,\mathrm{m^{-1}}$.

\subsection{Determination of the dispersion coefficients}
\label{supp:k2-k3-derivation}

We calculate the GVD and TOD coefficients using the formulae in Eq.~\eqref{eqn:NLSE-coeffs-as-derivatives} using the linear dispersion relationship, Eq.~\eqref{eqn:nonlinearDispersionInverted} with $|A|^2\approx0$:
\begin{equation}
    \label{eqn:dispersion-relation}
    k_y(\Omega) = \sqrt{\left(\frac{\Omega}{c}\right)^2-\left(\frac{\pi}{W}\right)^2}.
\end{equation}
Here we use $\Omega$ for the frequency, $c=\sqrt{\sigma/\rho}$ the speed of sound, and $W$ the waveguide width.

Taking derivatives yields the formula for the GVD coefficient at some frequency $\Omega$:
\begin{equation}
    \label{eqn:GVD-formula}
    k_2(\Omega)=\frac{\partial^2k_y(x)}{\partial x^2}\bigg|_{x=\Omega}=\frac{-\left(\frac{\pi}{W}\right)^2}{c^2 k_y(\Omega)^3}.
\end{equation}
The corresponding formula for the TOD coefficient is:
\begin{equation}
    \label{eqn:TOD-formula}
k_3(\Omega)=\frac{\partial^3k_y(x)}{\partial x^3}\bigg|_{x=\Omega}=\frac{3\pi^2\Omega}{c^4W^2k_y(\Omega)^5}.
\end{equation}

Evaluating Eqs.~\eqref{eqn:GVD-formula} and~\eqref{eqn:TOD-formula} requires knowledge of $c$ and $W$. We know approximate values for both from the material properties and waveguide design: for LPCVD high-stress silicon nitride $\sigma\sim1\,\mathrm{GPa}$ and $\rho=3200\,\mathrm{kg/m^{3}}$~\cite{hirschTutorialMembranePhononic2026}, giving $c\approx559\,\mathrm{m/s}$, and the waveguide width is nominally intended to be $25\,\mathrm{\upmu m}$ after etching. However, the release holes used for sacrificial etching lead to a small correction in the effective density (making the membrane lighter) and in the tensile stress through stress relaxation~\cite{hirschTutorialMembranePhononic2026}. Therefore, instead of relying on the approximate values, we precisely determine $c$ and $W$ from measured group velocities captured in ringdown measurements over a range of frequencies (shown in Fig.~\ref{fig:SuppFig_parameter-fitting}\textbf{c}). To do this, we differentiate and rearrange Eq.~\eqref{eqn:dispersion-relation} to get an expression for the group velocity $v_g=(\partial k_y/\partial \Omega)^{-1}$:
\begin{equation}
    \label{eqn:vg-analytic}
    v_g(\Omega) = c\sqrt{1-\left(\frac{\Omega_c}{\Omega}\right)^2}.
\end{equation}
Here $\Omega_c=c\pi/W$ is the cutoff frequency of the fundamental travelling wave mode~\cite{romeroPropagationImagingMechanical2019}. By fitting Eq.~\eqref{eqn:vg-analytic} to the measured group velocities (orange line in Fig.~\ref{fig:SuppFig_parameter-fitting}\textbf{c}), we calibrate $c=570.4\,\mathrm{m/s}$ and $W=25.3\,\mathrm{\upmu m}$, close to the expected values.

\subsection{Determination of the nonlinear coefficient}
\label{supp:xi-derivation}

To evaluate the nonlinear NLSE coefficient $\xi$ from Eq.~\eqref{eqn:NLSE-coeffs-as-derivatives}, we note that the fixed carrier frequency $\Omega$ is independent of amplitude:
\begin{equation}
\frac{\upd\Omega}{\upd|A|^2} = \frac{\partial\Omega}{\partial K}\frac{\partial K}{\partial |A|^2} + \frac{\partial \Omega}{\partial |A|^2} = 0 
\end{equation}
giving
\begin{equation}
\frac{\partial K}{\partial |A|^2} = -\frac{1}{v_g} \frac{\partial \Omega}{\partial |A|^2}.
\end{equation}
By evaluating the right hand side using Eq.~\eqref{eqn:nonlinearDispersion} and noting on the left hand side that $\xi=\partial^2K/\partial|A|^2$, we obtain:
\begin{equation}
\label{eqn:xi-formula}
\boxed{
   \xi  = -\frac{1}{v_g} \frac{\alpha_{\rm{eff}}\Omega_0(k_0)}{4 k_{\rm{eff}}}
   }
\end{equation}

From Eq.~\eqref{eqn:xi-formula} one case see $\xi$ is different to the usual Duffing spring term used in lumped-element models, which are commonly  encountered in the case of string and membrane resonators~\cite{lifshitzNonlinearDynamicsNanomechanical2008,schmidFundamentalsNanomechanicalResonators2016,romeroAcousticallyDrivenSinglefrequency2024}. In previous phononic waveguide literature $\xi$ has been experimentally measured~\cite{kurosuOnchipTemporalFocusing2018,kurosuMechanicalKerrNonlinearity2020}, but not analytically derived from the acoustic modeshape as done here.

\begin{figure}
    \centering
    \includegraphics[width=0.6\textwidth]{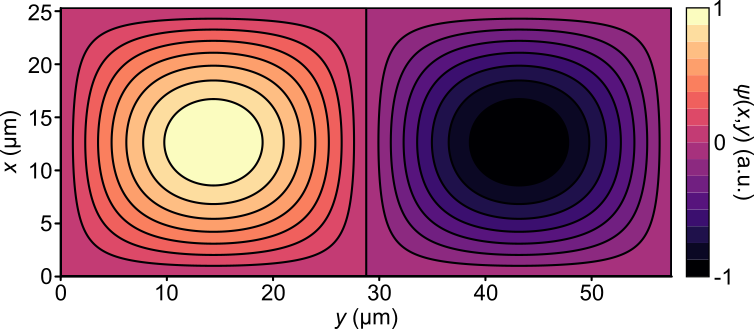}
    \caption{\textbf{Analytical modeshape used for calculating $k_\mathrm{eff}$ and $\alpha_\mathrm{eff}$.} We plot the normalised modeshape $\psi(x,y)$ from Eq.~\eqref{eqn:single-mode-modeshape} over the full width of the waveguide, $W$, and one wavelength (approximately $\sim58\,\mathrm{\upmu m}$). Here $W=25.3\,\mathrm{\upmu m}$; $k_y$ is calculated from Eq.~\eqref{eqn:dispersion-relation} assuming $c=570\,\mathrm{m/s}$ and $\Omega=2\pi\times15\,\mathrm{MHz}$.}
    \label{fig:SuppFig_modeshape-for-alphaeff}
\end{figure}

To numerically evaluate Eq.~\eqref{eqn:xi-formula}, we use analytic expressions for the effective spring and Duffing coefficients, derived assuming plane stress and purely out-of-plane displacement~\cite{hirschTutorialMembranePhononic2026}:
\begin{equation}
    k_{\mathrm{eff}}=\sigma h \int_0^{\tfrac{\pi}{k_y}}\int_0^W\left[\left( \frac{\partial \psi}{\partial x}\right)^2 + \left( \frac{\partial \psi}{\partial y}\right)^2 \right] \D x \, \D y
    \label{Eqkeff}
\end{equation}
\begin{equation}
    \alpha_{\mathrm{eff}}=\frac{1}{2}\frac{Y}{1-\nu^2} h \int_0^{\tfrac{\pi}{k_y}}\int_0^W\left[\left( \frac{\partial \psi}{\partial x}\right)^2 + \left( \frac{\partial \psi}{\partial y}\right)^2 \right]^2 \D x \, \D y
    \label{Eqalphaeff}
\end{equation}
Here $\sigma$, $Y$, $\nu$ and $h$  are respectively the silicon nitride membrane's tensile stress, Young's modulus, Poisson's ratio and thickness. The function $\psi$ describes the carrier wave modeshape and is dimensionless and normalised such that $\max|\psi|=1$, as shown in Fig.~\ref{fig:SuppFig_modeshape-for-alphaeff}. To first order the modeshape can be written as~\cite{romeroPropagationImagingMechanical2019}:
\begin{equation}
    \label{eqn:single-mode-modeshape}
    \psi(x,y)=\sin(k_x x)\sin(k_yy)
\end{equation}
where $k_x=\pi/W$. The integration in Eq.~\eqref{Eqalphaeff} is taken over a half-wavelength, $\pi/k_y=\lambda/2$, capturing the complete modeshape information through symmetry arguments. Integrating over different multiples of this length does not affect the ratio $k_\mathrm{eff}/\alpha_\mathrm{eff}$ that determines the nonlinear coefficient $\xi$ (Eq.\eqref{eqn:xi-formula}). However, it does change the specific numerical values of $k_\mathrm{eff}$ and $\alpha_\mathrm{eff}$ themselves. Our choice provides the value of effective stiffness $k_\mathrm{eff}$ relevant to the calculation of the electrostatic deflection outlined in section~\ref{sec:validation-through-electrostatic-deflection-estimate}, as it most closely resembles the static deflection produced by the electrode.

\section{NLSE simulations}

Our numerical simulation evolves a pulse envelope $A(t,y)$ using the NLSE in the moving frame with up to third-order dispersion (Eq.~\ref{eqn:NLSE-with-TOD}). The technique we use is the popular split-step Fourier method~\cite{agrawalNonlinearFiberOptics2019,kurosuMechanicalKerrNonlinearity2020}. To briefly explain this approach, consider that the NLSE can be written as $\frac{\partial A}{\partial y}=(\hat{L}+\hat{N})A$, where $\hat{L}$ is the linear differential operator:
\begin{equation}
    \hat{L}=-\frac{\alpha}{2}-\frac{i}{2}k_2\frac{\partial ^2}{\partial t^2}+\frac{1}{6}k_3\frac{\partial ^3}{\partial t^3},
\end{equation}
and $\hat{N}$ is the nonlinear operator:
\begin{equation}
    \hat{N}=-i\xi|A|^2.
\end{equation}

The linear equation $\frac{\partial A}{\partial y}=\hat{L}A$ has an analytic solution in the frequency domain:
\begin{equation}
    \label{eqn:Supp-NLSE-simulation-linear-op}
    \tilde{A}(\Omega,y+\Delta y)=\exp\left(\left(-\frac{\alpha}{2}+i\left(\frac{1}{2}k_2\Omega^2+\frac{1}{6}k_3\Omega^3\right)\right)\Delta y\right)\tilde{A}(\Omega, y).
\end{equation}
Here $\tilde{A}(\Omega,y)$ is the Fourier transform of $A(t,y)$. Similarly, the nonlinear equation $\frac{\partial A}{\partial y}=\hat{N}A$ has an analytic solution in the time domain:
\begin{equation}
    \label{eqn:Supp-NLSE-simulation-nonlinear-op}
    A(t,y+\Delta y)=\exp(-i\xi|A(t,y)|^2\Delta y)A(t,y).
\end{equation}

Equations~\eqref{eqn:Supp-NLSE-simulation-linear-op} and~\eqref{eqn:Supp-NLSE-simulation-nonlinear-op} are both easy to numerically implement, but one takes place in the frequency domain and the other in the time domain. To get around this issue, the key strategy of the split-step method is to perform the linear and nonlinear operations separately:
\begin{equation}
    \label{eqn:Supp-split-step-formula}
    A(t,y+\Delta y)\approx\exp\left(\frac{\Delta y}{2}\hat{N}\right)\exp(\Delta y\,\hat{L})\exp\left(\frac{\Delta y}{2}\hat{N}\right)A(t,y)+\mathcal{O}((\Delta y)^3).
\end{equation}
Separating the operators assumes they commute---they do not, but the error incurred is acceptable for small spatial steps. The cubic order of accuracy obtained here stems from the symmetric arrangement of the exponential operators~\cite{agrawalNonlinearFiberOptics2019}.

The simulation begins with an initial condition $A(t,y=0)$, which we represent numerically as a vector $A[t_1,t_2,\ldots,t_N;y=0]$ using $N=2^{13}$ time bins. To allow for overtaking collisions, we set $t_1=0$ and $t_N=2L/v_g$ and use the periodic boundary conditions that are automatically enforced by using a Fourier transform method. We use a spatial step of $\Delta y=0.1\,\mathrm{mm}$. The main simulation loop iterates Eq.~\eqref{eqn:Supp-split-step-formula} until the desired propagation distance has been reached.

The simulation code and corresponding \texttt{conda} environment is provided in the public respository associated with this paper.

\section{Choice of operating frequency}

The main figures in this work use data collected at a drive frequency of $\Omega\approx2\pi\times15\,\mathrm{MHz}$. This is just one frequency in the single-mode transmission band, which starts at the cutoff frequency of $\Omega_c=\pi c/W\approx2\pi\times11.3\,\mathrm{MHz}$, and finishes at the second-mode cutoff of $2\Omega_c$~\cite{romeroPropagationImagingMechanical2019}. We have verified that solitary waves can be initialised at a range of frequencies within this band---see, for example, the data in Fig.~\ref{fig:SuppFig_parameter-fitting}\textbf{c}---but to minimise the number of variables changing between experiments, we chose to maintain a consistent carrier frequency across the main figures.

\begin{figure}
    \centering
    \includegraphics[width=1.0\linewidth]{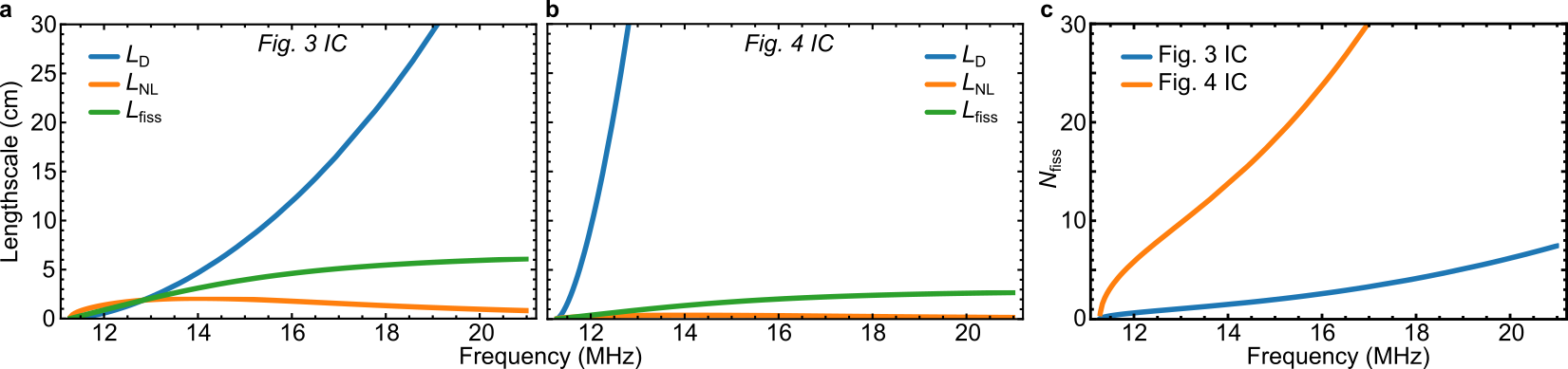}
    \caption{\textbf{Characteristic length scales at different frequencies.} \textbf{a} Characteristic dispersive ($L_D$), nonlinear ($L_{NL}$), and fission ($L_\mathrm{fiss}$) length scales at different frequencies, for width and amplitude parameters corresponding to the initial conditions (IC) of the experiment in Fig.~\ref{fig:Fig3}.
     \textbf{b} Same length scales, for width and amplitude parameters corresponding to the initial conditions (IC) of experiment in Fig.~\ref{fig:Fig4}. \textbf{c} Fission number $L_\mathrm{fiss}$ at different frequencies, for initial conditions corresponding to Figs.~3 and~4.}
    \label{fig:SuppFig_length-scales}
\end{figure}

The motivation for choosing a drive of $\Omega=2\pi\times15\,\mathrm{MHz}$ is illustrated in Fig.~\ref{fig:SuppFig_length-scales}, where we plot the characteristic length scales $L_D=T_0^2/|k_2|$, $L_{NL}=1/(\xi A_0^2)$, and $L_\mathrm{fiss}=\sqrt{L_DL_{NL}}$, and predicted number of solitons $N_\mathrm{fiss}=\sqrt{L_D/L_{NL}}$ emerging from the fission process, at different frequencies, assuming initial conditions (pulse width $T_0$ and background amplitude $A_0$) corresponding to the experiments in Figs.~\ref{fig:Fig3} and~\ref{fig:Fig4}. This lets us explore what those experiments would have looked like at different frequencies.

From Fig.~\ref{fig:SuppFig_length-scales}\textbf{a} we can see that for pulses initialised with an initial width $T_0$ near the ideal soliton width ($T_0\sim T_s=\sqrt{|k_2|/(\xi A_0^2)}$), $15\,\mathrm{MHz}$ is high enough (above $\sim\,\mathrm{13\,MHz}$) such that $L_{NL}<L_D$. This is necessary for solitary wave formation---technically $L_{NL}=L_D$ would produce a `perfectly' initialised solitary wave at $y=0$, but because of dissipation $L_{NL}$ will continuously increase over the course of the experiment, so we find it is better to initialise the experiments with $L_{NL}<L_D$. While sufficiently high frequencies thus ensure solitary wave formation, at even higher frequencies $L_D$ becomes much larger than $L_{NL}$ (Fig.~\ref{fig:SuppFig_length-scales}\textbf{a}). This corresponds to experiments where the linear pulse dispersion is so small that the pulse compression from the Duffing nonlinearity is difficult to observe within the dissipative lifetime (making the amplitude-dependent pulse compression shown in Fig.~\ref{fig:Fig3} less evident). We find $\Omega=2\pi\times15\,\mathrm{MHz}$ provides the appropriate level of dispersion to both enable and highlight solitary wave formation.

As an additional note, in Fig.~\ref{fig:SuppFig_length-scales}\textbf{b} and~\textbf{c} we see that $L_D$ increases with frequency faster than $L_{NL}$ decreases, so greater numbers of solitary waves are expected to be formed from the fission process at higher frequencies. This could be useful for future experiments aiming to maximise the number of solitary waves and overtaking collisions in the waveguide, for instance for the study of the soliton gas regime~\cite{zakharovKineticEquationsSolitons1971}.

\section{Comparison with existing bright/dark soliton landscape}

\begin{figure}[ht]
    \centering
    \includegraphics[width=0.7\linewidth]{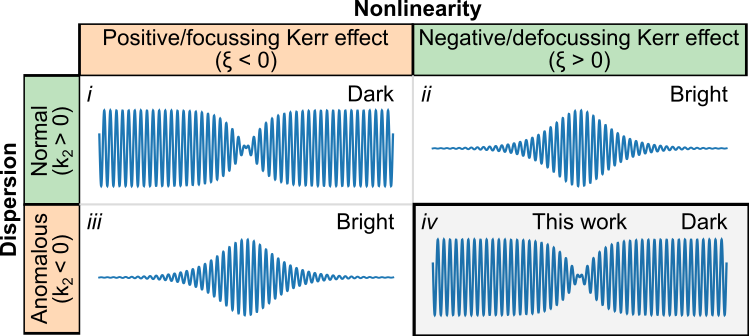}
    \caption{\textbf{Landscape of dark and bright soliton regimes}. (\textit{i}) Dark solitons occur in the case of positive Kerr effect ($\xi<0$) with normal dispersion ($k_2>0$). This case is representative of shallow water waves~\cite{chabchoubExperimentalObservationDark2013} and optical fibers operating in the normal dispersion regime~\cite{weinerExperimentalObservationFundamental1988,kivsharDarkOpticalSolitons1998}. (\textit{ii}) Bright solitons occur in the case of negative Kerr effect ($\xi>0$) with normal dispersion $(k_2>0$). This case is representative of deep water waves~\cite{yuenNonlinearDeepWater1975,chabchoubRogueWaveObservation2011}. (\textit{iii}) Bright solitons occur in the case of positive Kerr effect ($\xi<0$) with anomalous dispersion $(k_2<0)$. This case is representative of optical fibers and microresonators with anomalous dispersion~\cite{mollenauerExperimentalObservationPicosecond1980,herrTemporalSolitonsOptical2014}. (\textit{iv}) Dark solitons occur in the case of negative Kerr effect ($\xi>0$) with anomalous dispersion ($k_2<0$). This is the representative case for this work.}
    \label{fig:SuppFig-GVD+NL-landscape}
\end{figure}

NLSE envelope solitons occur in a wide variety of physical systems, whenever dispersion can be balanced by nonlinearity. If the GVD and nonlinear coefficients have the same sign, the system will support bright solitons; conversely, opposite-signed coefficients indicate the system supports dark solitons, as shown in Fig.~\ref{fig:SuppFig-GVD+NL-landscape}. This matrix representation allows us to place our phononic waveguide (where $k_2<0$ and $\xi>0$) within the broader solitary wave literature. Below we provide representative examples for each combination of positive and negative $k_2$ and $\xi$:
\renewcommand{\labelenumi}{\textit{\roman{enumi}}.} 
\begin{enumerate}
\itemsep-1.5em
    \item $k_2>0$ and $\xi<0$ corresponds to the case of optical fibers operating in the normal dispersion regime~\cite{weinerExperimentalObservationFundamental1988,kivsharDarkOpticalSolitons1998} and shallow water waves~\cite{chabchoubExperimentalObservationDark2013}.\\
    \item $k_2>0$ and $\xi>0$ corresponds to the case of deep water waves~\cite{yuenNonlinearDeepWater1975,chabchoubRogueWaveObservation2011}, BECs with attractive interactions~\cite{khaykovichFormationMatterWaveBright2002,streckerFormationPropagationMatterwave2002}, and phononic crystal waveguides operating in the anomalous dispersion regime~\cite{kurosuOnchipTemporalFocusing2018}. We note for the latter case that while numerical simulations predicted the ability to guide acoustic bright solitary waves, the experimentally achieved nonlinearities were insufficient to achieve this in practice, limiting observations of the acoustic Kerr effect to self- and cross-phase modulation measurements.\\
    \item $k_2<0$ and $\xi<0$ corresponds to the case of optical fibers operating in the anomalous dispersion regime~\cite{mollenauerExperimentalObservationPicosecond1980} and optical microresonators with anomalous cavity dispersion~\cite{herrTemporalSolitonsOptical2014,okawachiOctavespanningFrequencyComb2011,yangBroadbandDispersionengineeredMicroresonator2016}.\\
    \item $k_2<0$ and $\xi>0$ corresponds to the case of hard-clamped straight phononic waveguides (this work)
     and BECs with repulsive interactions~\cite{burgerDarkSolitonsBoseEinstein1999}.
\end{enumerate}

\section{Comparison with phononic waveguide landscape}

\begin{table}
    \centering
    \begin{tabularx}{1.0\textwidth}{|>{\raggedright\arraybackslash}p{15mm}|>{\raggedright\arraybackslash}p{30mm}|X|X|p{14mm}|X|p{15mm}|p{24mm}|X|}
    \hline
        Material & Geometry & Operating frequency (MHz) & Vibration amplitude (nm) & $v_g$ ($\mathrm{m/s})$ & $\alpha$ ($\mathrm{m^{-1}}$) & $k_2$ $(\mathrm{\upmu s^2.m^{-1}})$& $\xi$ ($\mathrm{nm^{-2}.m^{-1}}$) & Refs \\  \hline \hline
       
        $\mathrm{Si_3N_4}$ & Suspended membrane &  $11.3-22.2$ & $59.5$ & $376^\mathrm{a}$ & $2.1^\mathrm{a}$ & -36.6\footnote{at $15\,\mathrm{MHz}$} & $0.073$ & This work\\
         \hline
        GaAs/ AlGaAs & Suspended membrane, 1D PnC & $2.4-7.4$ & $1$\footnote{at $5.38\,\mathrm{MHz}$} & $100^\mathrm{b}$ & 66.8 & $-200^\mathrm{b}$ & $-16$ & \cite{kurosuOnchipTemporalFocusing2018,kurosuMechanicalKerrNonlinearity2020}\\
        \hline
        $\mathrm{SiN}_x$ & Suspended membrane, 1D PnC, tuning electrodes & $11.5-16$ & - & $50-150$ & $484$ & - & - & \cite{chaElectricalTuningElastic2018}\\
\hline
        $\mathrm{Si_3N_4}$ & Suspended membrane, soft clamped, topological waveguide & $1.18-1.3$ & - 
        & $280$ & $6.9\times10^{-4}$ & $\sim0$ 
        & - & \cite{xiSoftclampedTopologicalWaveguide2025}\\
\hline
        AlN & Suspended membrane, topological waveguide & 1015-1080 & - & $890$
        & - & $\sim 0$ & - & \cite{zhangGigahertzTopologicalValley2022}\\
\hline
        GaN on sapphire & Index-contrast slab & $\sim100$ & - & 5000 & 11.5 & - &-  & \cite{fuPhononicIntegratedCircuitry2019}\\
\hline
        $\mathrm{LiNbO_3}$ on sapphire & Index-contrast slab & $3420$ & - & $3320\pm30$ & $1128\pm368$ & $1.1\times10^{-3}$ & $7\,\mathrm{mW^{-1}.mm^{-1}}$\footnote{The dimensions of this nonlinear coefficient are for use with a mechanical amplitude $A_0$ normalised such that $|A_0|^2$ is the acoustic power. To compare with our value of $\xi$, we can compare the nonlinear phase shift $\varphi_\mathrm{NL}$ over one decay length. Using the values in the top row of the table, for our system $\varphi_\mathrm{NL}=\xi A_0^2 \alpha^{-1}\approx32.2\pi \,\,\mathrm{rad}$. Ref.~\cite{mayorGigahertzPhononicIntegrated2021} gives a typical power of $|A_0|^2\sim10\,\mathrm{\upmu W}$ and decay length $\alpha^{-1}\approx1\,\mathrm{mm}$, giving $\varphi_\mathrm{NL}\approx0.02\pi\,\mathrm{rad}$, which is roughly three orders of magnitude lower.}
        & \cite{mayorGigahertzPhononicIntegrated2021}\\
         
         \hline
    \end{tabularx}
    \caption{Comparison of the device in this work with phononic waveguide landscape. PnC = phononic crystal.}
    \label{tab:Comparison-phononic-waveguide-landscape}
\end{table}

Table~\ref{tab:Comparison-phononic-waveguide-landscape} compares this work with other phononic waveguiding platforms.

\section{Device fabrication}
\label{sectionsuppfabricationprocessflow}

The fabrication process begins with commercially obtained wafers from MicroChemicals GmbH. The wafers are coated on both sides with a $60\,\mathrm{nm}$ layer of high stress~\cite{gardeniersLPCVDSiliconrichSilicon1996,beliaevOpticalStructuralComposition2022} ($\sim1\,\mathrm{GPa}$) stoichiometric $\mathrm{Si_3N_4}$ deposited via low-pressure chemical vapour deposition on a $500\,\mathrm{nm}$ $\mathrm{SiO_2}$ layer. The wafer substrate is low-resistivity silicon.

\begin{figure}
    \centering
    \includegraphics[width=0.9\linewidth]{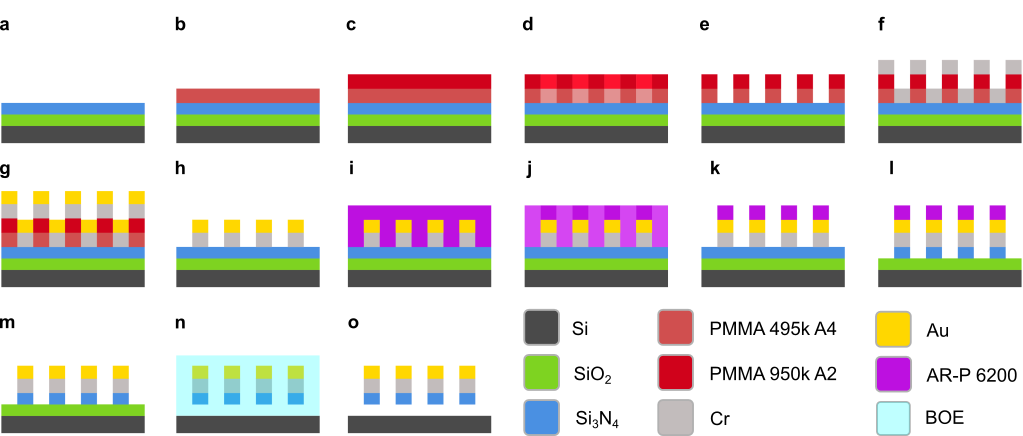}
    \caption{\textbf{Device fabrication process.} Steps \textbf{a} through \textbf{o} detail full device fabrication procedure, see text for details. Si: Silicon; SiO$_2$: Silica;  Si$_3$N$_4$: Silicon nitride;  Cr: Chromium; PMMA: Poly(methyl methacrylate); Au: Gold; BOE: Buffered Oxide Etchant.}
    \label{fig:SuppFig-fabrication-process}
\end{figure}

The fabrication process is summarised in Fig.~\ref{fig:SuppFig-fabrication-process}. There are two exposures, the first to define the metal electrodes and the second to define the release holes used to suspend the membrane. Both exposures are performed using electron beam lithography (RAITH EBPG 5150). For the metal exposure we spin-coat a double layer of 495K A4 beneath 950k A2 poly(methylmethacrylate) (PMMA). After exposure and development, this double layer produces a re-entrant profile suitable for liftoff. Metal deposition is performed with an e-beam evaporator (Temescal FC-2000), depositing $5\,\mathrm{nm}$ of chromium as an adhesion layer before $45\,\mathrm{nm}$ of gold as the main conductive layer. Liftoff is performed using acetone. For the next exposure we spin-coat Allresist AR-P 6200.09, which has a high resistance to plasma etching. After exposure and development of this resist, we form the release holes in the $\mathrm{Si_3N_4}$ membrane using reactive ion etching (Oxford Instruments PlasmaPro 80). The chip is then stripped and cleaned via sonication in n-methyl-2-pyrrolidone-based solvent (Microchem Remover PG) followed by acetone. To release the membrane we remove the oxide layer using buffered oxide etchant (BOE), which is a mixture of ammonium fluoride and hydrofluoric acid. A custom-made `turbulence shielding'~\cite{norteNanofabricationOnChipOptical2014} chip carrier protects the centimetre-scale membranes from tearing while they are in solution. We dry the released devices using a critical-point dryer (Leica EM CPD300).

\section{Sample mounting}

\begin{figure}
    \centering
    \includegraphics[width=0.7\linewidth]{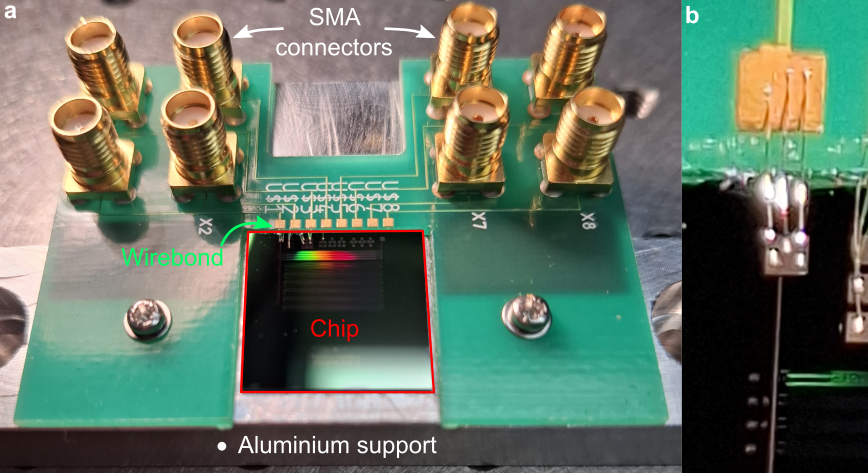}
    \caption{\textbf{Phononic integrated circuit mounting solution.} \textbf{a} Overview of the chip mounting solution, consisting of a custom aluminium support and printed circuit board for connecting to SMA cables. \textbf{b} Zoom-in on the wirebond connection between metallic pads on the PCB and large-scale pads on the phononic chip. These pads connect to the micrometer-sizes patterned electrodes on the waveguide used for electrostatic actuation (see Fig.~\ref{fig:Fig2} of the main text).}
    \label{fig:throne+PCB}
\end{figure}

The released chip is mounted onto a custom-made aluminium support (Fig.~\ref{fig:throne+PCB}), which supports a custom-made printed circuit board (PCB). Electrodes on the chip are wirebonded to pads on the PCB, which run to SMA cable connectors for interfacing with external electronics. The entire package screws into the breadboard of our vacuum chamber via through-holes in the aluminium support. We record the data in this work at pressures $\lesssim10^{-5}\,\mathrm{mbar}$ where gas damping is negligible~\cite{schmidDampingMechanismsSingleclamped2008}.

\begin{table}
    \centering
    \begin{tabular}{|l|c|c|c|c|}
    
    \hline
    \textbf{Name} & \textbf{Symbol} & \textbf{Unit} & \textbf{Value} \\
    \hline \hline
     Silicon nitride density     &  $\rho$& kg$\cdot$m$^{-3}$ & 3200   \\
     \hline
     Silicon nitride Young's modulus & $Y$ & GPa & 250 \\
     \hline
     Silicon nitride tensile stress & $\sigma$ &  GPa & 1  \\
     \hline
     Silicon nitride Poisson's ratio & $\nu$ & - & 0.27\\
     \hline
     Silicon nitride thickness & $h$ & nm  & 60  \\
     \hline
     Sound speed & $c=\sqrt{\sigma/\rho}$ & m/s& 570 \\
     \hline
     Electrode effective gap & $g$ & nm & 500\\
     \hline
     Waveguide width & $W$ & $\mu$m & 25.3\\
     \hline
    Waveguide cutoff frequency & $\Omega_c/2\pi$ &  MHz & 11.27\\
     \hline
     Pulse carrier frequency & $\Omega_0/2\pi$ &  MHz & 15\\
     \hline
     Waveguide acoustic energy loss rate (@15 MHz) & $\alpha$ & m$^{-1}$ & 2.1\\
     \hline
     Waveguide nonlinear coefficient  (@15 MHz) & $\xi$ & $\mathrm{nm^{-2}.m^{-1}}$  & $7.30\times10^{-2}$  \\
     \hline 
     Waveguide dispersive coefficient  (@15 MHz) & $k_2$ &  $\mathrm{\upmu s^{2}.m^{-1}}$  & -36.6  \\
     \hline 
          
    \end{tabular}
    \caption{\textbf{System parameters.} Experimentally measured parameters, and material parameters used in the simulations.}
    \label{tab:systemparametersrecaptable}
\end{table}

\section{Optical measurement}

We detect out-of plane motion using a custom-built laser Doppler interferometer with heterodyne detection~\cite{hirschDirectionalEmissionOnchip2024}. Free-space coherent $780\,\mathrm{nm}$ light is provided by a titanium-sapphire laser (M Squared SolTiS). We employ a half-wave plate and polarised beam splitter to tune the optical power in free space before collimating the beam and using fiber optics for the rest of the setup. A $2\times2$ fiber coupler splits the beam into a probe and local oscillator (see main text, Fig.~\ref{fig:Fig2}).  The local oscillator is frequency-shifted by $f_\mathrm{LO}=80\,\mathrm{MHz}$ through an acousto-optic modulator (AA Opto-Electronic MT80-NIR60-Fio). Meanwhile, the probe passes through a fiber polarisation controller before entering a circulator (Precision Micro-Optics FBCI-7811113323) which directs the light towards and away from the sample. A lensed optical fiber (Nanonics; working distance = $4\,\mathrm{\upmu m}$, spot size = $1\,\mathrm{\upmu m}$) is used to focus the light onto the surface of the membrane. To position the lensed fiber in three dimensions, we glue it to a custom aluminium mounting piece that is attached to three stacked single-axis piezoelectric nanopositioners (Smaract SLC-1720). After passing through the circulator for a second time, the reflected, phase-modulated probe recombines with the local oscillator at a variable coupler (KS Photonics FTDC-N-078-1-FA). The outputs of the variable coupler go into a balanced photodetector (Newport 1807-FC). The DC component of the photocurrent is sent to an oscilloscope for monitoring; we tune it to zero using the variable coupler. The AC component of the photocurrent is sent to a spectrum analyser (Agilent EXA N9010A). To observe the mechanical motion over time at frequency $f_d$, we measure the photocurrent at frequency $f_\mathrm{LO}-f_d$, in zero-span mode. We use the spectrum analyser's maximum bandwidth of $8\,\mathrm{MHz}$ to achieve the maximum temporal resolution. To increase our number of sampling points, we use the trigger delay function on the spectrum analyser to measure the pulse evolution over small times ($\sim1\,\mathrm{ms}$) with a high sampling rate, then construct larger datasets (e.g. 10 ms of observation) by stitching together those measurements. Drifts in photocurrent during this process produce the horizontal lines visible in some of our data, e.g. Fig.~\ref{fig:tanh-in-exp}\textbf{d}.

\subsection{Determination of RMS displacement}

\label{suppsubsectionopticaldisplacementcalibration}
This section describes the calibration of membrane displacement, enabling the conversion of the optically measured power spectral density (PSD) into a calibrated membrane RMS displacement.  We model the out-of-plane oscillating membrane as a moving, partially reflective mirror. Let the membrane displacement be $U_0\cos(\Omega t)$, where $U_0$ is the amplitude of displacement and $\Omega$  the vibration frequency. The reflected light travels a shorter distance when the displacement is positive (i.e. upwards/towards the lensed fiber) and a longer distance when the displacement is negative (i.e. downwards/away from the fiber). This produces a phase shift in the reflected light field $E_r$:
\begin{equation}
    E_r=E_{r,0}e^{i(\omega t-kU_0\cos(\Omega t))}=E_{r,0}e^{i\omega t}e^{i(-kU_0)\cos(\Omega t)}.
\end{equation}
Here $E_{r,0}$ is the amplitude of the reflected light field, $\omega$ is the light frequency, and $k=2\pi/(780\,\mathrm{nm}$) is the wavenumber of the light. We are neglecting amplitude modulation that would appear from a change in angle of the membrane, because we position the fiber at an antinode of the standing wave pattern where the membrane locally remains horizontal during an oscillation period. We express the reflected light using the Jacobi-Anger expansion:
\begin{equation}
    e^{iz\cos\theta}=J_0(z)+2\sum_{n=1}^\infty i^nJ_n(z)\cos(n\theta).
\end{equation}
Here $J_n$ is the $n^\mathrm{th}$ Bessel function of the first kind. In our case $z=-kU_0$ and $\theta=\Omega t$. Substituting the expansion into the expression for $E_r$, we obtain:
\begin{equation}
    E_r = E_{r,0}e^{i\omega t}\left[J_0(-kU_0)+2\sum_{n=1}^\infty i^n J_n(-kU_0)\cos(n\Omega_mt)\right].
\end{equation}
The oscillating phase has the effect of generating sidebands at multiples of the membrane oscillation frequency. We restrict our attention to the central frequency $(n=0)$ and the first $(n=1)$ sideband where the acoustic signal is greatest. We perform a Taylor expansion on the Bessel functions (valid as $U_0\sim10^{-8}\,\mathrm{m}$ and $k\simeq8\times10^{6}\,\,\mathrm{m}^{-1}$, such that $|kU_0|\ll1$). The expansions are $J_0(x)\approx1+\mathcal{O}(x^2)$ and $J_1(x)\approx \frac{x}{2}+\mathcal{O}(x^2)$. Substituting these in (and keeping only the $n=1$ term of the sum) the reflected light is:
\begin{equation}
    E_r=E_{r,0}e^{i\omega t}\left[1-ikU_0\cos(\Omega t)\right].
\end{equation}
The reflected light field consists of a carrier component and a component at a sideband of $\Omega$, with the ratio of sideband to carrier being $ikU_0\cos(\Omega t)$.
Because we are performing a heterodyne measurement, the photocurrent detected will be proportional to the absolute value of $|E_r|$ (not $|E_r|^2$). 
Therefore the ratio of power spectral distributions (PSDs) at the mechanical sideband and AOM frequency will be:
\begin{equation}
    \frac{\mathrm{PSD}_\mathrm{mech}}{\mathrm{PSD}_\mathrm{AOM}}=kU_0\cos(\Omega t).
\end{equation}
Because $|\cos(\Omega t)|$ oscillates faster than our measurement bandwidth it is averaged out to a value of $2/\pi\approx0.6366$. The relationship between amplitude and measured photocurrent is therefore:
\begin{equation}
     \frac{\mathrm{PSD}_\mathrm{mech}}{\mathrm{PSD}_\mathrm{AOM}} = \frac{2}{\pi}kU_0\implies U_0 = \frac{780\,\mathrm{nm}}{4}\cdot\frac{\mathrm{PSD}_\mathrm{mech}}{\mathrm{PSD}_\mathrm{AOM}}.
\end{equation}
During experiments, we measure $\mathrm{PSD}_\mathrm{AOM}$ immediately before recording a time trace of acoustic pulse propagation. In postprocessing we use this measurement to calibrate the displacement from $\mathrm{PSD}_\mathrm{mech}$.

\subsection{Validation through electrostatic deflection magnitude estimation}
\label{sec:validation-through-electrostatic-deflection-estimate}

In this section, we corroborate the RMS displacement obtained through optical calibration (see section~\ref{suppsubsectionopticaldisplacementcalibration}) through an independent order-of-magnitude estimate of the membrane deflection based on the applied capacitive force.  The capacitive force $F_{es}$ applied by the drive electrode takes the form~\cite{schmidFundamentalsNanomechanicalResonators2016,bakerHighBandwidthOnchip2016,hirschTutorialMembranePhononic2026, romeroAcousticallyDrivenSinglefrequency2024}:
\begin{equation}
    F_{\mathrm{es}}=-\frac{1}{2}\left(\frac{\partial C(g)}{\partial g} \right)V^2= \frac{1}{2}\frac{\varepsilon_0  A}{g^2}V^2,
\end{equation}
where $A=\pi\times(2.5\times 10^{-6})^2\simeq 2\times 10^{-11}$ m$^2$ is the circular actuation electrode area (see Fig.~\ref{fig:Fig1}) and $g$ is the capacitive gap distance. Here we have used the parallel-plate approximation for the displacement amplitude-dependent capacitance $C(g)$ formed by the top electrode deposited on the waveguide and the ground electrode below the chip:  $C(g)=\frac{\varepsilon_0 A}{g}$.  Indeed, we have shown in previous work~\cite{romeroAcousticallyDrivenSinglefrequency2024} that the capacitance is well approximated by a parallel-plate capacitor of gap $g$ equal to the undercut depth (here $g=500$ nm, i.e. the sacrificial silica layer thickness, see Table~\ref{tab:systemparametersrecaptable}).
In the regime of strong actuation characteristic of the data shown in Figs.~\ref{fig:Fig3},\ref{fig:Fig4}\&\ref{fig:Fig5} of the main text, typical values of applied voltages are $V_{\mathrm{DC}}=150$ V and $V_{\mathrm{AC}}=50$ V$_{pp}$. In this case, the applied voltage on the capacitor plate is correspondingly comprised between $V_{\mathrm{max}}=V_{\mathrm{DC}}+V_{\mathrm{AC}}/2=175$V and  $V_{\mathrm{min}}=V_{\mathrm{DC}}-V_{\mathrm{AC}}/2=125$V.
The associated static deflection $F_{\mathrm{es}}/k_{\mathrm{eff}}$ is calculated with  $k_{\mathrm{eff}}=397$ N$\cdot$m$^{-1}$ determined through Eq.\eqref{Eqkeff}. We find a downward deflection of 27 nm and 14 nm respectively (the capacitive force being always attractive), leading to a launched travelling wave amplitude $A_0$ of order 13 nm. When interfering with itself after a round-trip, the amplitudes are summed, leading to a wave amplitude of 26 nm  (in the antinodes of the generated standing wave) in this example, in broad agreement with the values obtained through optical calibration above.

\section{AWG excitation}

Actuation signals are provided by a function/arbitrary waveform generator (Siglent SDG5162), with the output signal fed into an amplifier (Minicircuits ZX60-100VH+) and bias-tee (Minicircuits ZFBT-6GW+) before running via SMA cables to the actuation electrodes on the chip. After amplification, the AC signal voltage typically ranges up to $50\,\mathrm{V_{pp}}$. The DC bias is provided by an open-loop piezo controller (Thorlabs MDT693B). The arbitrary waveform generator is set to deliver one waveform when triggered by a $50\,\mathrm{Hz}$ external trigger signal. This trigger is also fed to the spectrum analyser to allow for trace averaging. The acoustic power spectrum decays to the noise floor in approximately $10\,\mathrm{ms}$ (see Fig.~\ref{fig:SuppFig_parameter-fitting}), so we can safely consider the waveguide has fully thermalised between each trigger event.
Single dark solitary waves such as those shown in Figs.~\ref{fig:Fig3} and~\ref{fig:ExtFig-DarkSolAllVoltages+widthxamp} are initialised using arbitrary waveforms of the form:
\begin{equation}
    V(t) = \sin(\Omega_0 t)\times\tanh\left(\frac{t}{T_0}\right)\times\tanh\left(\frac{t-T_\mathrm{AWG}}{T_0}\right)
\end{equation}
Here $\Omega_0$ is the carrier frequency (typically around $15\,\mathrm{MHz}$), $T_0$ is the initial temporal width of the pulse, and $T_\mathrm{AWG}$ is the period of the arbitrary waveform. To create a clean solitary wave, $T_\mathrm{AWG}$ should equal the round-trip time of the waveguide, i.e. $T_\mathrm{AWG}=2L/v_g$, so that the beginning and end of the waveform recombine to form a smooth $\tanh$ profile (see Fig.~\ref{fig:Fig2} of the main text and accompanying discussion). To create the wide dark pulse in Fig.~\ref{fig:Fig4}, we deliberately set $T_\mathrm{AWG}<2L/v_g$ so that there is a pronounced period of zero excitation between the decreasing and rising half $\tanh$ profiles after recombination---thus forming a wide notch with $\tanh$ sidewalls and a flat, zero-amplitude centre.

The multiple dark solitary waves in Figs.~\ref{fig:Fig5}\textbf{b} and~\textbf{d} are initialised using waveforms of the form:
\begin{equation}
    V(t) = \sin(\Omega_0t)\times\tanh\left(\frac{t}{T_0}\right)\times\tanh\left(\frac{t-T_\mathrm{AWG}}{T_0}\right)\times\bigg[\prod_{k=1}^N\tanh\left(\frac{t-t_k}{T_0}\right)\bigg]
    \label{EqsuppNsolitonsAWG}
\end{equation}
where $N$ is the number of additional solitary waves (centred around times $t_k$) that will be created in addition to the one centred around $t=0$ formed by the stitching of the waveform's beginning and end as described above (to create the even-parity dark pulses leading to bifurcating solitary waves seen in Fig.~\ref{fig:Fig5}\textbf{f}, we replace the $\tanh$ function where it appears in Eq.~\eqref{EqsuppNsolitonsAWG} with the function $\tanh^2$). In Eq.~\eqref{EqsuppNsolitonsAWG} the $N$ additional dark solitary waves produced by the product in square brackets will have perfect $\tanh$ shape, unlike the one formed by the recombination of the beginning and end of the waveform. This difference can be seen in Fig.~\ref{fig:Fig5} where one out of the ten solitary waves---for example, solitary wave \#6 in Figs.~\ref{fig:Fig5}\textbf{b} and~\textbf{c}---sheds non-solitary wave components during its initial propagation. In the simulations, we replicate this imperfection by replacing one of the $\tanh$ functions with a similar but unequal function. Specifically, we use the replacement rules: 
\begin{equation}
    \tanh\left(\frac{t}{T_0}\right)\rightarrow\tanh^3\left(\frac{t}{1.2T_0}\right)
\end{equation}
and
\begin{equation}
    \tanh^2\left(\frac{t}{T_0}\right)\rightarrow0.25\times\bigg[1+\tanh\left(\frac{t-1.5T_0}{T_0}\right)*\tanh\left(\frac{t+1.5T_0}{T_0}\right)\bigg]^2.
\end{equation}
The dark solitary waves embedded in Gaussian envelopes in Figs.~\ref{fig:tanh-in-exp} and~\ref{fig:ExtFig-csep-tanhsq-test} are initialised using  waveforms of the kind:
\begin{equation}
    V(t) = \sin(\Omega_0t)\times\exp\left(-\frac{1}{2}\left(\frac{t-T_\mathrm{AWG}/2}{T_\mathrm{gauss}}\right)^2\right)\times\tanh\left(\frac{t-T_\mathrm{AWG}/2}{T_0}\right),
\end{equation}
where $T_\mathrm{gauss}$ is the width of the Gaussian pulse, and the $\tanh$ function can be replaced with $\tanh^2$ to create even parity dark pulses which split into two grey solitary waves, as shown in Fig.~\ref{fig:ExtFig-csep-tanhsq-test}. Finally, the bright pulses in Fig.~\ref{fig:ExtFig-BrightPulse} are initialised using arbitrary waveforms of the form:
\begin{equation}
    V(t) = \sin(\Omega_0t)\times\mathrm{sech}\left(\frac{t-T_\mathrm{AWG}/2}{T_0}\right).
\end{equation}
In this case the only requirement on the AWG period is that $T_\mathrm{AWG}\gg T_0$, which ensures the pulse smoothly decays to zero, in the wings, and fully fits inside the waveguide.

\begin{figure*}
    \begin{center}
        \includegraphics[width=1.0\linewidth]{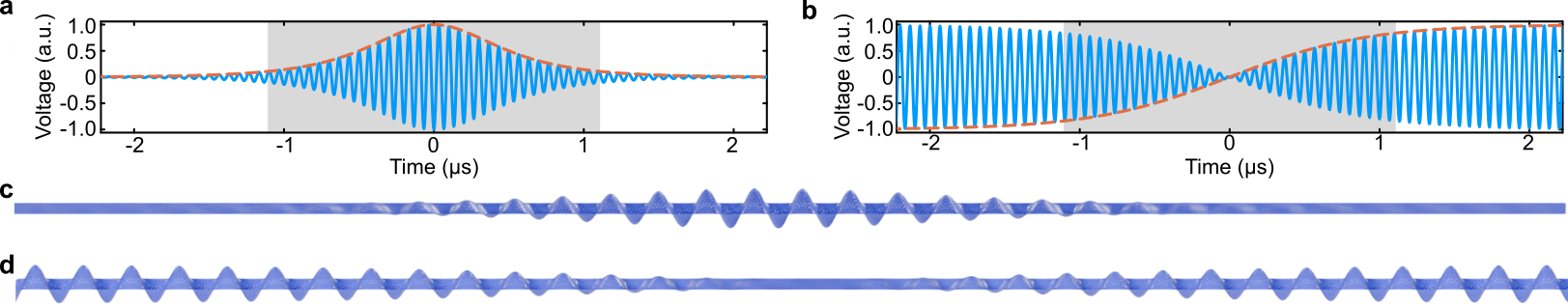}
        \caption{\textbf{Correspondence between drive waveforms and resulting acoustic waves.} \textbf{a-b} Examples of synthesised AC drive waveforms for bright (sech) and dark (tanh) pulses with carrier frequency of $15\,\mathrm{MHz}$ and $\mathrm{FWHM}=1\,\mathrm{\upmu s}$. Blue: product of $15\,\mathrm{MHz}$ sinusoid with envelope function. Orange: $\mathrm{sech}$/$\tanh$ envelopes. Grey shading corresponds to the regions illustrated in the subfigures below. \textbf{c-d} 3D illustrations of the corresponding waveguide displacement, assuming direct proportionality to the AC actuation voltage. Horizontal dimensions are to scale while transverse displacement is exaggerated.}
        \label{fig:ExtFig-AWG+pulses}
    \end{center}
\end{figure*}

To visualise what the resulting pulses look like in the spatial domain, in Fig.~\ref{fig:ExtFig-AWG+pulses} we illustrate arbitrary waveforms for bright (sech) and dark ($\tanh)$ pulses, alongside 3D illustrations of the corresponding membrane displacement. The translation from time to spatial domain assumes a direct proportionality between voltage and displacement (which is the case when the AC drive is applied to the electrode alongside a large DC bias~\cite{hirschTutorialMembranePhononic2026}), a carrier frequency of $2\pi\times15\,\mathrm{MHz}$ and a speed of sound of $c=570\,\mathrm{m/s}$. The waveguide width and length are drawn to scale while the vertical displacement is exaggerated.

\end{document}